\documentclass[11pt,english]{article}
\usepackage{amsfonts, amsmath, amssymb, verbatim, setspace, pdfsync, graphicx,comment,epsfig}
\usepackage{color}
\definecolor{black}{rgb}{.2,.2,.2}

\usepackage{hyperref}
\hypersetup{colorlinks,%
citecolor=black,%
filecolor=black,%
linkcolor=black,%
urlcolor=black,%
pdftex}

\begin{document}

\title{Onsager reciprocity principle for kinetic models and kinetic schemes}
\author{Ajit Kumar Mahendra\\ 
{\it Homi Bhabha National Institute,}\smallskip \\
{\it Anushaktinagar, Mumbai-400094,}\\
{\it India.}\\
\and Ram Kumar Singh \\
{\it Reactor Safety Division, BARC,}\smallskip \\
{\it Mumbai-400085,}\\
{\it India.}\\
}
\date{}
\maketitle
\setcounter{page}{1}

\begin{abstract}
Boltzmann equation requires some alternative simpler kinetic model like BGK to replace the collision term. Such a kinetic model which replaces the Boltzmann collision integral should preserve the basic properties and characteristics of the Boltzmann equation and comply with the requirements of non equilibrium thermodynamics. Most of the research in development of kinetic theory based methods have focused more on entropy conditions, stability and ignored the crucial aspect of non equilibrium thermodynamics. The paper presents a new kinetic model formulated based on the principles of non equilibrium thermodynamics. The new kinetic model yields correct transport coefficients and satisfies Onsager's reciprocity relationship. The present work also describes a novel kinetic particle method and gas kinetic scheme based on this linkage of non-equilibrium thermodynamics and kinetic theory. The work also presents derivation of kinetic theory based wall boundary condition which complies with the principles of non-equilibrium thermodynamics, and can simulate both continuum and rarefied slip flow in order to avoid extremely costly multi-scale simulation.
\end{abstract}
Presented in \emph{The Tenth International Conference for Mesoscopic Methods in Engineering and Science} (ICMMES-2013), 22-26 July 2013, Oxford.

\pagestyle{plain}

\setcounter{tocdepth}{1}

\tableofcontents

\begin{figure} \centering
{\includegraphics[width=.8\textwidth,bb=0 0 1202 414]{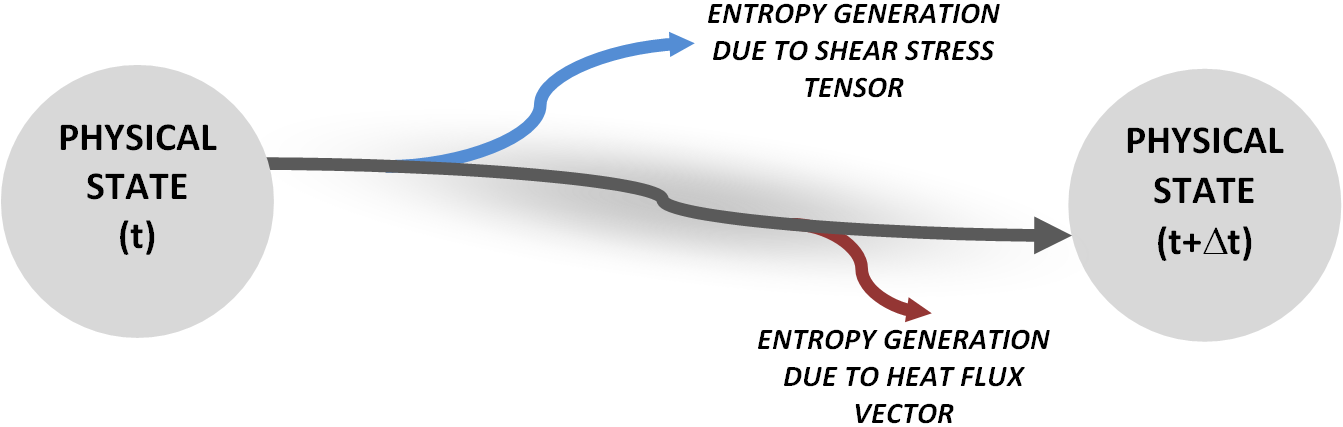}}
\caption{Two components of entropy as physical state evolves from time $t$ to time $t+\Delta t$.\label{entropy}}
\end{figure}

\begin{section}
{Introduction}\label{intro}
\end{section}
All the research in the development of upwind scheme based on macroscopic theories can be seen in terms of inclusion of physically consistent amount of entropy. In many case, a single solver operating from rarefied flow to hypersonic continuum flow requires corrections and tuning, as most of the time it is not known what is the correct amount of entropy generation for a particular regime and the correct distribution of entropy generation for each thermodynamic force. Figure \ref{entropy} shows schematic of entropy generation as physical state evolves from time $t$ to time $t+\Delta t$. The components of entropy due to thermodynamic forces associated with stress tensor and thermal gradient vector differ in magnitude and vary with locations in the flow domain. \emph{Genuine upwind scheme should resolve these different components of entropy generation due to its conjugate thermodynamic force in order to satisfy thermodynamics while the state update happens.} Most of the upwind schemes basically aim to add the correct dissipation or entropy but fail to resolve and ensure the correct distribution of the entropy associated with its conjugate thermodynamic force.  If the solver follows and mimics the physics then we can have a single monolithic solver serving the entire range from rarefied flow to continuum flow, creeping flow to flow with shocks. The entropy generation observed at the macroscopic level is a consequence of molecular collisions at the microscopic level. Mesoscopic method based on kinetic theory uses statistical description of a system of molecules and provides model for molecular collisions leading to non-equilibrium phenomena. Non-equilibrium thermodynamics (NET) being a phenomenological theory describes this non-equilibrium phenomena and provides linkage with kinetic theory (KT) based coefficients of transport and relaxation. The molecular description is provided by the kinetic theory while the relationship between the entropy generation due to thermodynamic forces associated with stress tensor and thermal gradient vector is a feature of non-equilibrium thermodynamics. Kinetic theory and non-equilibrium thermodynamics together become a powerful tool to model non-equilibrium processes of compressible gas.

This work introduces maximum entropy production principle and investigates its relationship with Onsager's reciprocity principle and Boltzmann equation. A new kinetic model based on Onsager's principle is proposed, which gives correct Prandtl number and also complies with the requirements of non-equilibrium thermodynamics.  The paper describes kinetic flux vector splitting and kinetic particle method based on the new kinetic model which incorporates features of  non-equilibrium thermodynamics. It also gives derivation of kinetic theory based wall boundary condition which complies with the principles of non-equilibrium thermodynamics, and can simulate both continuum and rarefied slip flow as an efficient and economical alternative to extremely costly multi-scale simulation. Finally, simulation and validation of continuum, and rarefied slip flow test cases are presented to illustrate the present formulation.

\section{Kinetic theory and non-equilibrium thermodynamics}
The kinetic theory of gases is a very vast field which successfully explains the irreversible laws of fluid mechanics through a statistical description of a system composed of large number of particles. Kinetic theory based method should preserve the basic properties and characteristics of the Boltzmann equation and also comply with the principles of non-equilibrium thermodynamics like i) positive entropy production, ii) satisfaction of Onsager's relation and maximum entropy production principle. Non-equilibrium thermodynamics being a phenomenological theory gives the symmetry relationship between kinetic coefficients as well as general structure of equations describing the non-equilibrium phenomenon. The Onsager's symmetry relationship is a consequence of microscopic reversibility condition due to the equality of the differential cross sections for direct and time reversed collision processes. For a prescribed irreversible force the actual flux which satisfies Onsager's theory also maximizes the entropy production. Maximum entropy production principle is an additional statement over the second law of thermodynamics telling us that the entropy production is not just positive, but tends to a maximum. Research in this area is yet to enter the domain of computational fluid dynamics, publications concerning maximum entropy principle is still in the realm of physics\footnote{ Refer to publication of Martyushev and Seleznev \cite{mart-sele}, and Beretta \cite{beretta} for more details.}. The solution of the Boltzmann equation is in accordance with the principle of maximum entropy production (MEP). Non-equilibrium thermodynamics provides a tool for checking the correctness of the kinetic theory based solutions \cite{zhdanov}. Distribution function derived using kinetic theory has to comply with requirements of non-equilibrium thermodynamics like following Onsager's principle and maximization of entropy production under constraint imposed due to conservation laws.  This section introduces kinetic theory and describes Boltzmann equation and its moments. The section also presents maximum entropy production (MEP) principle and brings out relationship between Onsager's variational principle and linearized Boltzmann equation. 
\subsection{Boltzmann equation}
The Boltzmann equation in Bogoliubov's generalized form is expressed as follows 
\begin{equation} \label{slkns1} 
\frac{\partial f}{\partial t} + \vec{a}\cdot\nabla _{\vec{v}} f +\nabla _{\vec{x}} \cdot (\vec{v}f) =J(f,f)+K(f,f,f)+L(f,f,f,f)+ \cdots 
\end{equation} 
where $\vec{x}$, the position vector, $\vec{a}$ is the acceleration vector and $\vec{v}$ is the velocity vector of molecules given in $\mathbb{R}^{D}$, here $D$ is number of directions a molecule is allowed to move. The left hand side describes the streaming operation as $\nabla \cdot \vec{v}$$=0$, thus it expresses advection of molecules written in conservative form. On the right hand side factor $J(f,f)$ is binary or two particle collision, $K(f,f,f)$ being the ternary or three particle collision and $L(f,f,f,f)$ is quaternary or four particle collision. Here in $J, K, L$ the difference in position between the colliding particles is taken into account. 

Consider dilute polyatomic gas with binary collisions, the Boltzmann transport equation in such a case describes the transient single particle molecular distribution $f(\vec{x},\vec{v},\mathbb{I},t):\mathbb{R}^{D} \times \mathbb{R}^{D} \times \mathbb{R}^{+} \times \mathbb{R}^{+} \to \mathbb{R}^{+} $ where $D$ is the degree of freedom. An additional internal energy variable $\mathbb{I}\in \mathbb{R}^{+} $ is added as polyatomic gas consists of particles with additional degree of freedom \cite{morse} required for conservation of total energy instead of translational energy alone.  Thus a molecule of a polyatomic gas  is characterized by a $(2D+1)$ dimensional space given by its position $\vec{x}$ $\in$ $\mathbb{R}^{D}$, molecular velocity vector $\vec{v}$ $\in$ $\mathbb{R}^{D}$ and internal energy $\mathbb{I}$ $\in$ $\mathbb{R}$. Distribution function expresses the probability of finding the molecules in the differential volume $d^{D}x d^{D}v d\mathbb{I}$\footnote{For example when D=3  the polyatomic gas  is characterized by a $7$ dimensional space and the differential volume in phase space is $d^{3}x d^{3}v d\mathbb{I}$ where $d^{3}x$ is $dxdydz$ and $d^{3}v$ is $dv_{x}dv_{y}dv_{z}$.}  of the phase space. The equilibrium or Maxwellian distribution function for the polyatomic gas is given by
\begin{equation} \label{slkns1a} 
f_{0} =\frac{\rho}{\mathbb{I}_{o}}\left({\frac{\beta}{\pi}}\right)^{\frac{D}{2}}exp\left(-\beta{\left(\vec{v}-\vec{u}\right)}^{2}-\frac{\mathbb{I}}{\mathbb{I}_{o}}\right)
\end{equation}
where $\beta=1/2RT$ with $R$ as the specific gas constant and $\mathbb{I}_{o}$ is given as
\begin{equation} \label{slkns1b} 
\mathbb{I}_{o}=\frac{<\mathbb{I},f_{0} >}{<f_{0} >}= \frac{1}{\rho}\int _{\mathbb{R}^{+} }\int _{\mathbb{R}^{D} }\mathbb{I}f_{0}(\vec{x},\vec{v},\mathbb{I},t)  d\vec{v} d\mathbb{I}=\frac{2-(\gamma-1)D}{4(\gamma-1)\beta}
\end{equation}
where $\gamma$ is the specific heat ratio. Many polyatomic gases are calorically imperfect i.e. specific heat varies with temperature and in most of the engineering applications translational, rotational and vibrational partition functions contribute to the thermodynamic properties. For such a case distribution function can be represented as the probability of dominant macrostate with specific heat ratio, $\gamma$ as
\begin{equation}\label{slkns1d}
\begin{array}{l}
\gamma = 1+ \left[{3 + \sum_{i=1}^{i=n_v}\left\{ {g_{i}\left(\frac{\theta_{i}}{T}\right)^{2}exp(-\frac{\theta_{i}}{T})}{\left(1-exp(-\frac{\theta_{i}}{T})\right)^{-2}}\right\}}\right]^{-1}
\end{array}
\end{equation}
where $g_{i}$ is the degeneracy of the $i^{th}$ vibrational mode, $\theta_{i}$ is the  characteristic temperature and $n_v$ is number of vibrational modes.
\subsection{Moments and hyperbolic conservation equations}
The moment of a function, $\boldsymbol{\Psi} =\boldsymbol{\Psi} (\vec{v},\mathbb{I},t):\mathbb{R}^{D} \times \mathbb{R}^{+} \times \mathbb{R}^{+} \to \mathbb{R} $ is defined as Hilbert space of functions generated by the inner product
\begin{equation} \label{OnsBGK_3_} 
\boldsymbol{\bar{\Psi}} (\vec{x},t)=\left\langle \boldsymbol{\Psi},f\right\rangle \equiv \int _{\mathbb{R}^{+} }\int _{\mathbb{R}^{D} }\Psi  (\vec{v},\mathbb{I},t) f(\vec{x},\vec{v},\mathbb{I},t) d\vec{v} d\mathbb{I} 
\end{equation} 
The five moments function defined as  $\boldsymbol{\Psi} =\left[1,\vec{v},\mathbb{I}+\frac{1}{2} v^{2}\right]^{T} $ gives the macroscopic mass, momentum, and energy densities i.e.  $\left\langle {\rm \boldsymbol{\Psi}},f \right\rangle$$=$ $ [\rho ,\rho \vec{u},\rho E]^{T}$, where $E=RT/(\gamma-1)$$+$$\frac{1}{2} u^{2}$, $\vec{u}$ is the fluid velocity vector. When we take moments of the Boltzmann equation we get the hyperbolic conservation equation. For example with $f=f_{0} $ we get Euler equations that are set of inviscid compressible coupled hyperbolic conservation equations written as 
\begin{equation} \label{OnsBGK_4_} 
\int _{\mathbb{R}^{+} }\int _{\mathbb{R}^{D} }\boldsymbol{\Psi}   \left(\frac{\partial f_{0} }{\partial t} +\nabla _{\vec{x}} .(\vec{v}f_{0} )=0\right)d\vec{v}d\mathbb{I}\equiv \frac{\partial \boldsymbol{U}}{\partial t} +\frac{\partial \boldsymbol{GX}_I}{\partial x} +\frac{\partial \boldsymbol{GY}_I}{\partial y} +\frac{\partial \boldsymbol{GZ}_I}{\partial z} =0 
\end{equation} 
where $\boldsymbol{U}=[\rho ,\rho \vec{u},\rho E]^{T} =\left\langle {\rm \boldsymbol{\Psi} \; },f_{0} \right\rangle \equiv \int _{\mathbb{R}^{+} }  \int _{\mathbb{R}^{D} } \boldsymbol{\Psi}   f_{0} (\vec{x},\vec{v},\mathbb{I},t)d\vec{v}d\mathbb{I}$ is the vector of conserved variable and $(\boldsymbol{GX}_{I},\boldsymbol{GY}_{I},\boldsymbol{GZ}_{I})$ are the Cartesian components of the inviscid flux vector defined as
\begin{equation} \label{OnsBGK_5_} 
\boldsymbol{GX}_{I}={\left< \boldsymbol{\Psi} v_x  f_{0} \right>}=\int _{\mathbb{R}^{+} }\int _{\mathbb{R}^{D} }\boldsymbol{\Psi} v_{x}   f_{0} (\vec{x},\vec{v},\mathbb{I},t)d\vec{v}d\mathbb{I}\equiv \left[\begin{array}{c} {\rho u_{x} } \\ {p+\rho u_{x}^{2} } \\ {\rho u_{x} u_{y} } \\ {\rho u_{x} u_{z} } \\ {(\rho E+p)u_{x} } \end{array}\right] 
\end{equation} 
\begin{equation} \label{OnsBGK_6_} 
\boldsymbol{GY}_{I}={\left< \boldsymbol{\Psi} v_y  f_{0} \right>}=\int _{\mathbb{R}^{+} }\int _{\mathbb{R}^{D} }\boldsymbol{\Psi} v_{y}   f_{0} (\vec{x},\vec{v},\mathbb{I},t)d\vec{v}d\mathbb{I}\equiv \left[\begin{array}{c} {\rho u_{y} } \\ {\rho u_{y} u_{x} } \\ {p+\rho u_{y}^{2} } \\ {\rho u_{y} u_{z} } \\ {(\rho E+p)u_{y} } \end{array}\right] 
\end{equation} 
\begin{equation} \label{OnsBGK_7_} 
\boldsymbol{GZ}_{I}={\left< \boldsymbol{\Psi} v_z  f_{0} \right>}=\int _{\mathbb{R}^{+} }\int _{\mathbb{R}^{D} }\boldsymbol{\Psi} v_{z}   f_{0} (\vec{x},\vec{v},\mathbb{I},t)d\vec{v}d\mathbb{I}\equiv \left[\begin{array}{c} {\rho u_{z} } \\ {\rho u_{z} u_{x} } \\ {\rho u_{z} u_{y} } \\ {p+\rho u_{z}^{2} } \\ {(\rho E+p)u_{z} } \end{array}\right] 
\end{equation} 
For an ideal law we have $p=\rho RT$ where $R$ is the specific gas constant and $T$ is the absolute temperature. No real gas follows the ideal gas law for all temperatures and pressures, in such a case a good engineering approach is to take the equation of state of the thermally imperfect gas.

The distribution function $f$ can be expressed as the Chapman-Enskog expansion\cite{chapman} in terms Knudsen number, $\text{Kn}$ as follows 
\begin{equation} \label{c2-ce2}  
f_{} =f_{0}^{} +\text{Kn}\bar{f}_{1}^{} +\text{Kn}^{2} \bar{f}_{2}^{} +\cdots  
\end{equation}
The perturbation terms satisfies the additive invariants property $<\boldsymbol{\Psi} ,\text{Kn}^{k} \bar{f}_{k} >_{\forall k\ge 1} =0$. Using the non-dimensionless Boltzmann equation and Chapman-Enskog expansion, higher order distribution is generated by virtue of iterative refinement as follows:
\begin{equation} \label{c2-ce3c} 
\bar{f}_{k} =-\frac{t_{R} }{\text{Kn}} \left[\frac{\partial \bar{f}_{k-1} }{\partial t} +\nabla _{\vec{x}}\cdot(\vec{v}\bar{f}_{k-1} )\right] 
\end{equation} 
 where $\text{Kn}$ is the Knudsen number with $\bar{f}_{0} =f_{0}$. With first order expansion $f=f_{0}^{} +\text{Kn}\bar{f}_{1}^{}$ we get Navier-Stokes-Fourier equations, and with second order expansion $f=f_{0}^{} +\text{Kn}\bar{f}_{1}^{} +\text{Kn}^{2} \bar{f}_{2}^{}$ we get a set of Burnett equations.  It is evident that Navier-Stokes-Fourier and Burnett equations are all obtained using the five moments and hence they do not have equations for evolution of shear stress tensor and thermal gradient vector. The moments of the Boltzmann equation satisfy an infinite hierarchy of balance laws such that from the continuum mechanics perspective the flux in an equation becomes the density at the next hierarchical level \cite{mullerruggeri}. There can be different set of moment equations based on entropy based closure or closure due to equilibrium solution.
\subsection{Maximum entropy production and Onsager's principle}
Any fluid flow moving from one conserved non-equilibrium state to another conserved non-equilibrium state will generate entropy $\sigma(\boldsymbol{J}_i,\boldsymbol{X}_i)$ due to its thermodynamic force $\boldsymbol{X}_i$ and associated conjugate thermodynamic flux $\boldsymbol{J}_i$. From Onsager's point of view thermodynamic force is defined as derivative of entropy density $\sigma$  with respect to thermodynamic variable $\boldsymbol{a}_i$  
\begin{equation}\label{a1-mep-tf}
\boldsymbol{X}_i = \left(\frac{\partial \sigma}{\partial \boldsymbol{a}_i}\right)_{\boldsymbol{a}_j \neq \boldsymbol{a}_i}
\end{equation}
The subscript "$i$" is not the index notation of the tensor, it signifies the type of thermodynamic force e.g $\boldsymbol{X}_{\tau}$ or $\boldsymbol{X}_{q}$ associated due to stress tensor or thermal gradient vector. 
 \emph{Maximum entropy production principle states that this entropy is not only positive it is also maximum.} The maximization of entropy takes place for a prescribed irreversible force under constraint imposed on entropy production due to conservation laws \cite{ziegler} written as
\begin{equation}\label{a1-mep}
\begin{array}{ll}
\text{Maximize} {\rm \; \; \;  }&\sigma_J(\boldsymbol{J}_i,\boldsymbol{J}_{k})\\
\text{subject to}{\rm \; \; \;  }& \sigma_J(\boldsymbol{J}_i,\boldsymbol{J}_{k}) - \sigma(\boldsymbol{J}_i,\boldsymbol{X}_i)   =0
\end{array}
\end{equation}
where $\sigma(\boldsymbol{J}_i,\boldsymbol{X}_i)$$=$ $\sum_i{\boldsymbol{J}_i\odot \boldsymbol{X}_i}$ is the entropy production density based on conservation law and $\sigma_J(\boldsymbol{J}_i,\boldsymbol{J}_{k})$ is the entropy production density in terms of fluxes.  Operator $\odot$ denotes full tensor contraction of forces and fluxes, which are of the same tensorial order following Curie principle. As described earlier, the term $\boldsymbol{J}_i$ signifies flux and the term $\boldsymbol{X}_i$ signifies the thermodynamic force. Expansion of the entropy production density $\sigma_J$   in terms of $N$ number of fluxes for a system close to equilibrium state gives
\begin{equation}\label{a1-mep-a}
\sigma_J(\boldsymbol{J}_i,\boldsymbol{J}_k)= \alpha+\sum_{i=1}^{N} \alpha_{i} \boldsymbol{J}_{i}+\sum_{i,k=1}^{N}\boldsymbol{\alpha}_{ik}\odot \boldsymbol{J}_{i}\odot\boldsymbol{J}_{k}+\cdots
\end{equation}
where coefficients $\alpha,\alpha_{i},\boldsymbol{\alpha}_{ik},\cdots,$ are properties  of the system in equilibrium state. The first term on the right hand side vanishes since there is no entropy production in the equilibrium state. Coefficients associated with odd power of fluxes also vanish i.e. $\alpha_{i}=0$ as entropy production is independent of the direction of flux flow. For linear irreversible thermodynamics (LIT) higher terms can be neglected and the entropy production density $\sigma_J(\boldsymbol{J}_i,\boldsymbol{J}_k)$  for LIT can be approximated as a bilinear function of fluxes as follows
\begin{equation}\label{a1-mep-a}
\sigma_J(\boldsymbol{J}_i,\boldsymbol{J}_k)\approx \sum_{i,k=1}^{N}\boldsymbol{\alpha}_{ik}\odot \boldsymbol{J}_{i}\odot\boldsymbol{J}_{k}
\end{equation}

The coefficients $\boldsymbol{\alpha}_{rs}$, $\boldsymbol{\alpha}_{sr}$ vanish only if flux $\boldsymbol{J}_{r}$ and $\boldsymbol{J}_{s}$ do not couple. The constrained maximization of $\sigma_J(\boldsymbol{J}_i,\boldsymbol{J}_k)$ leads to Lagrangian $\mathcal{L}(\boldsymbol{X},\boldsymbol{J},\lambda)$ with $\lambda$ as a  Lagrangian multiplier. The optimality conditions leads to KKT (Kharush-Kuhn-Tucker) equations $[\nabla_{\boldsymbol{J}} \mathcal{L}(\boldsymbol{X},\boldsymbol{J},\lambda)]_{\boldsymbol{X},\lambda}=0$ and $[\nabla_{\lambda} \mathcal{L}(\boldsymbol{X},\boldsymbol{J},\lambda)]_{\boldsymbol{X},\boldsymbol{J}}=0$. The first KKT condition $[\nabla_{\boldsymbol{J}_i} \mathcal{L}(\boldsymbol{X},\boldsymbol{J},\lambda)]_{\boldsymbol{X},\lambda}=0$ gives
$\boldsymbol{X}_{i}=\frac{2(\lambda-1)}{\lambda}\sum_{k=1}^{N}\boldsymbol{\alpha}_{ik}\odot \boldsymbol{J}_{k}
$. Using the second KKT  condition we obtain $\lambda=2$ and the Lagrangian can be written as
\begin{equation}\label{a1-mep-c}
\mathcal{L}(\boldsymbol{X},\boldsymbol{J})=\left[\frac{ \partial \left ( \sum_{i}^{N}  \boldsymbol{J}_{i}\odot\boldsymbol{X}_{i} -\frac{1}{2} \sum_{i,k=1}^{N}\boldsymbol{\alpha}_{ik}\odot \boldsymbol{J}_{i} \odot \boldsymbol{J}_{k} \right)}{\partial \boldsymbol{J}}\right]_{\boldsymbol{X}} =0
\end{equation}
This  Lagrangian  can be recast in a form similar to Onsager's variational principle
\begin{equation}\label{a1-mep-d}
\mathcal{L}(\boldsymbol{X},\boldsymbol{J})=\left[\frac{ \partial \left (\sigma(\boldsymbol{J}_{i},\boldsymbol{X}_{i}) -\sigma_J(\boldsymbol{J}_{i},\boldsymbol{J}_{k})\right)}{\partial \boldsymbol{J}}\right]_{\boldsymbol{X}} =0
\end{equation}
where derived entropy production term $\sigma_J(\boldsymbol{J}_{i},\boldsymbol{J}_{k})$ $=$ $\frac{1}{2} \sum_{i,k=1}^{N}\boldsymbol{\alpha}_{ik}\odot \boldsymbol{J}_{i} \odot \boldsymbol{J}_{k}$ for linear irreversible thermodynamics (LIT) is similar to Onsager's dissipative function density $\Phi (\boldsymbol{J}_{i},\boldsymbol{J}_{k}) $ for domain $\Omega$ in the flux space
\begin{equation}\label{a1-mep-e}
\Phi (\boldsymbol{J}_{i},\boldsymbol{J}_{k}) = \frac{1}{2}\int_{\Omega} \sum_{i,k=1}^{N} \boldsymbol{R}^{J}_{ik} \odot \boldsymbol{J}_{i}\odot \boldsymbol{J}_{k}d\Omega  
\end{equation}
and coefficients $\boldsymbol{\alpha}_{ik}$ is equivalent of Onsager's phenomenological symmetric tensor $\boldsymbol {R}^{J}_{ik}$. Onsager variational principle \cite{onsager1,onsager2} is one of the corner stone of linear non-equilibrium thermodynamics. It states that each flux is a linear homogeneous function of all the forces of the same tensorial order following Curie principle such that flux $\boldsymbol{J}_{i}=\sum_{j} \boldsymbol{L}_{ij} \odot \boldsymbol{X}_{j}$. In isotropic media $\boldsymbol{L}_{ij}$ vanish if forces couple with fluxes of different tensor types. \emph{For a prescribed irreversible force $\boldsymbol{X}_i$ the actual flux $\boldsymbol{J}_i$ which satisfies Onsager's principle  also maximizes the entropy production,  $\sigma(\boldsymbol{J}_i,\boldsymbol{X}_i)=\sum_i{\boldsymbol{J}_i \odot \boldsymbol{X}_i}$.} 

An alternative Gyarmati \cite{gyarmati} formulation in force space  can be written as
\begin{equation}\label{a1-mep-f}
\left[\frac{ \partial \left (\sigma(\boldsymbol{J}_{i},\boldsymbol{X}_{i}) -\sigma_X(\boldsymbol{X}_{i},\boldsymbol{X}_{k})\right)}{\partial \boldsymbol{X}}\right]_{\boldsymbol{J}} =0
\end{equation}
The derived entropy in terms of thermodynamic force, $\sigma_X(\boldsymbol{X}_{i},\boldsymbol{X}_{k})$ is similar to dissipation function density $\phi_X(\boldsymbol{X}_{i},\boldsymbol{X}_{k})$ in the force space expressed as
\begin{equation}\label{a1-mep-e}
\phi_X(\boldsymbol{X}_{i},\boldsymbol{X}_{k}) = \frac{1}{2}\int_{\Omega} \sum_{i,k=1}^{N} \boldsymbol{R}^{X}_{ik} \odot \boldsymbol{X}_{i}\odot \boldsymbol{X}_{k}d\Omega
\end{equation}
where $\boldsymbol{R}^{X}_{ik}$ is the  phenomenological symmetric tensor in the force space. Gyarmati formulation also leads to the same conclusion that for a prescribed thermodynamic fluxes $\boldsymbol{J}_i$ the actual irreversible forces $\boldsymbol{X}_i$  maximize entropy production.
\subsection{Onsager's principle and linearized Boltzmann equation}
Consider linearized distribution $f_{1}=f_{0}[1+\Phi]$  with the further assumption that $|\Phi|$ $\ll 1$  and both Maxwellian, $f_{0}$ and unknown $\Phi$ vary slowly in space and time. With this assumption we can neglect the product of $\Phi$ with derivatives of Maxwellian $f_{0}$ as well as derivatives of $\Phi$ . The linearized Boltzmann equation in terms of linearized collision operator $\mathbb{J}\Phi$  can be expressed as
\begin{equation} \label{b1-ons-1}
\frac{1}{f_{0}}\left(\frac{\partial f_{0}}{\partial t} +\nabla _{\vec{x}} \cdot (\vec{v}f_{0})\right)=-\mathbb{J} \Phi
\end{equation} 
Consider another trial linearized distribution $f_{1,T}=f_{0}[1+\Phi_{T}]$ which is not the solution of Boltzmann equation but it satisfies additive invariants property and produces entropy. Martyushev and Seleznev \cite{mart-sele} proved that the  distribution  $f_{1}=f_{0}[1+\Phi]$ which is the solution of Boltzmann equation also maximizes the entropy production
\begin{equation} \label{b1-ons-2}
\int _{\mathbb{R}^{+} } \int_{\mathbb{R^{D}}}f_{0} \Phi \mathbb{J}\Phi d\vec{v}d\mathbb{I} \geq \int _{\mathbb{R}^{+} } \int_{\mathbb{R^{D}}}f_{0} \Phi_{T} \mathbb{J}\Phi_{T} d\vec{v}d\mathbb{I}
\end{equation} 
Thus the \emph{solution of the Boltzmann equation is in accordance with the principle of maximum entropy production (MEP)}. To carry the investigation further on the subject it is essential to analyze linearized Boltzmann equation. Wang Chang and Uhlenbeck \cite{wangchang}  , Grad\cite{grad}, Ikenberry and Truesdell \cite{ikenberry}, Gross and Jackson \cite{gross},  and others have done extensive investigations on linearized Boltzmann equation. Researchers have tried to interpret the linear collision operator by  i) either considering its spectrum that includes eigenvalues for which $\mathbb{J}\Phi$ $=$ $\lambda \Phi$ has eigensolutions within the Hilbert space, ii) or decomposing it in terms of fluid dynamic gradients, iii) or expanding it in terms of thermodynamic forces. For example Wang Chang and Uhlenbeck \cite{wangchang}  interpreted $ \Phi$ in terms of  orthonormalized set of eigenfunctions written in separable form as tensor spherical harmonic and radial eigenfunctions, these eigenfunctions for Maxwell molecules can be written in terms of Laguerre-Sonine polynomials. Grad \cite{grad} interpreted $ \Phi$ in terms of Hermitian tensor polynomials whereas Gross and Jackson \cite{gross} used eigenvalue-theory of Wang Chang and Uhlenbeck \cite{wangchang} to construct kinetic model by replacing higher order eigenvalues by a suitable constant at each lower order approximation.  Loyalka \cite{loyalka71} used linearized Boltzmann equation with perturbation based on pressure and temperature gradients to investigate the Onsager reciprocal relationship for slip flows. Lang \cite{lang} decomposed the perturbation into three parts and made use of variational technique to calculate symmetric Onsager's matrix for slip flows. Zhdanov and  Roldughin \cite{zhdanov} have investigated linkage between kinetic theory and non-equilibrium thermodynamics by expanding $\Phi$ in terms of tensor spherical harmonic and Sonine polynomials while McCourt \textit{et al.} \cite{mccourt} interpreted $\Phi$ in term of flux and its conjugate thermodynamic force. Sharipov\cite{sharipov2011} investigated Onsager's reciprocal relationship for nonlinear irreversible phenomena by expanding Boltzmann equation in power series with respect to thermodynamic force. 

\begin{figure} \centering
{\includegraphics[width=0.8\textwidth,bb=0 0 523 268]{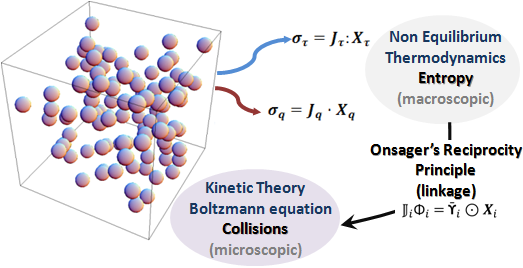}}
\caption{Onsager reciprocity principle linking non-equilibrium thermodynamics and kinetic theory.\label{kt-net-ons}}
\end{figure}

The present research investigates the  linkage between kinetic theory and non equilibrium thermodynamics by expanding unknown $\Phi$ $=$ $\sum_{i} \Phi_{i}$  as a sum of component of perturbation $\Phi_{i} =(\Phi)_{\boldsymbol{X}_{j}=0, \forall j \neq i}$ appearing due to thermodynamic force $\boldsymbol{X}_{i}$ such that all other thermodynamic forces are absent  i.e. $\boldsymbol{X}_{j}=0$ $, \forall j \neq i$. The linearized Boltzmann equation corresponding to thermodynamic force $\boldsymbol{X}_{i}$ becomes
\begin{equation} \label{b1-ons-4}
\frac{1}{f_{0}}\left(\frac{\partial f_{0}}{\partial t} +\nabla _{\vec{x}} \cdot (\vec{v}f_{0})\right)_{\boldsymbol{X}_{j}=0, \forall j \neq i}=-\mathbb{J}_i \Phi_{i}
\end{equation} 
The collision operator $\mathbb{J}_i \Phi_{i}$ will exists only when the system is disturbed from the state of equilibrium due to some thermodynamic forces $\boldsymbol{X}_{i}$. Such a thermodynamic force will lead to its associated conjugate microscopic flux tensor i.e. microscopic flux due to thermal gradient vector or stress tensor. Adopting the approach of McCourt \textit{et al.} \cite{mccourt} we can  express the linearized collision operator  in term of flux and its conjugate thermodynamic force such that $\mathbb{J}_i \Phi_{i}$ $=$ $\bar{\boldsymbol{\Upsilon}}_{i} \odot \boldsymbol{X}_{i}$, where $\bar{\boldsymbol{\Upsilon}}_{i}$ is the reduced microscopic flux tensor associated with its conjugate thermodynamic force, $\boldsymbol{X}_{i}$. As already described operator $\odot$ denotes full tensor contraction of forces and fluxes, which are of the same tensorial order following Curie principle. The subscript "$i$" is not the index notation of the tensor, it signifies the type of thermodynamic force e.g $\boldsymbol{X}_{\tau}$ or $\boldsymbol{X}_{q}$ associated due to stress tensor or thermal gradient vector. The inverse operator $\mathbb{J}^{-1}_i$ is defined on the non-hydrodynamic subspace orthogonal to the collisional invariants such that  
\begin{equation} \label{b1-ons-6}
\Phi=\sum_i \left(\mathbb{J}^{-1}_i\bar{\boldsymbol{\Upsilon}}_{i} \odot \boldsymbol{X}_{i}\right)=\sum_i \left((\mathbb{J}^{-1}\bar{\boldsymbol{\Upsilon}})_{i} \odot \boldsymbol{X}_{i}\right)
\end{equation}
Linearized Boltzmann equation leads to thermodynamic flux $\boldsymbol{J}_{i}$ in a  linear phenomenological form as 
\begin{equation} \label{b1-ons-7}
\begin{array}{ll}
\boldsymbol{J}_{i}&=\displaystyle  \int_{\mathbb{R}^{+}} \int_{\mathbb{R}^{D}} \bar{\boldsymbol{\Upsilon}}_{i} f_{0}\Phi   d\vec{v} d\mathbb{I}={\left<\bar{\boldsymbol{\Upsilon}}_{i} , f_{0}\Phi \right>}\\
&={\left<\bar{\boldsymbol{\Upsilon}}_{i} ,f_{0} \sum_{j} \left((\mathbb{J}^{-1} \bar{\boldsymbol{\Upsilon}})_{j} \odot \boldsymbol{X}_{j}\right)\right>}=\sum_{j} \boldsymbol{L}_{ij} \odot \boldsymbol{X}_{j}
\end{array}
\end{equation} 
where $\boldsymbol{L}_{ij}$ is the phenomenological tensor of transport coefficients defined as
\begin{equation} \label{b1-ons-8}
\boldsymbol{L}_{ij}={\left<\bar{\boldsymbol{\Upsilon}}_{i}, f_{0} (\mathbb{J}^{-1} \bar{\boldsymbol{\Upsilon}})_{j} \right>}
\end{equation} 
The phenomenological tensor obeys Onsager's reciprocal relationship  $\boldsymbol{L}_{ij}$ $=$ $\boldsymbol{L}_{ji}$. Casimir \cite{casimir} generalized this reciprocal relationship for thermodynamic force of arbitrary parity to include larger class of irreversible phenomena such that $\boldsymbol{L}_{ij}$ $=$$\eta_j$$\eta_i$$\boldsymbol{L}_{ji}$ where parity $\eta_j$$=$$-1$ when flux changes sign under microscopic motion reversal. If the reduced microscopic flux tensors $\bar{\boldsymbol{\Upsilon}}_{r}$ and $\bar{\boldsymbol{\Upsilon}}_{s}$ do not couple then  no cross effects will be present and phenomenological tensor of transport coefficients $\boldsymbol{L}_{rs}$ $=$ $\boldsymbol{L}_{sr}$ will vanish. For example in case of fluid flow described by Navier-Stokes-Fourier equations we have $\bar{\boldsymbol{\Upsilon}}_{\tau}$ due to thermodynamic force associated with stress tensor and $\bar{\boldsymbol{\Upsilon}}_{q}$  due to thermodynamic force associated with thermal gradient vector.  For such a case $\boldsymbol{L}_{\tau q}$ $=$ $\boldsymbol{L}_{q \tau}$  vanish as $\bar{\boldsymbol{\Upsilon}}_{\tau}$ and $\bar{\boldsymbol{\Upsilon}}_{q}$ are of different tensorial order and hence do not couple. We get only two tensors $\boldsymbol{L}_{\tau \tau}$ and $\boldsymbol{L}_{qq}$ of transport coefficients which are equivalent to scalars because of isotropy due to the rotational invariance of the collision operator\footnote{The operator $\mathbb{J}$  has rotational invariance if $\mathbb{J}=\mathcal{R}^{-1} \mathbb{J}\mathcal{R}$ for any rotational operator $\mathcal{R}$.}. Viscosity and thermal conductivity coefficients can be extracted from the reduced matrix element ${L}_{\tau \tau}$ and ${L}_{qq}$ respectively, where reduced matrix element ${L}_{ii}$ for any tensor $\boldsymbol{L}_{ii}$ of rank $\mathcal{T}$ is defined as
\begin{equation} \label{b1-ons-9}
L_{ii} = \frac{{\left<\bar{\boldsymbol{\Upsilon}}_{i}\odot, f_{0} (\mathbb{J}^{-1} \bar{\boldsymbol{\Upsilon}})_{i} \right>}}{2\mathcal{T} +1}
\end{equation}
Consider a binary mixture of non-reacting gas where macroscopic fluxes are generated because of thermodynamic forces associated due to stress tensor, thermal gradient vector and chemical potential gradient vector due to species distribution. Since the thermodynamic forces due to thermal gradient vector, $\boldsymbol{X}_q$ and chemical potential gradient vector, $\boldsymbol{X}_d$ are of the same tensorial order we get equality in Soret and Dufour coefficients due to Onsager symmetry. \emph{Onsager-Casimir symmetry relationship} is a consequence of positive semi-definiteness and self-adjoint property of the linearized collision operator $\mathbb{J}$ arising from the microscopic reversibility condition due to the equality of the differential cross sections for direct and time reversed collision processes. Entropy production density can be derived using moment $\Psi_{e}=-ln f$ for density based distribution function, $f$  as
\begin{equation} \label{b1-ons-10a}
\begin{array}{ll}
\sigma(\boldsymbol{J}_{i},\boldsymbol{X}_{i})=\frac{\partial \rho_{s}}{\partial t} +\nabla _{\vec{x}} \cdot (\vec{j}_{s}) &=  {\left< ln(f), f_{0}\mathbb{J} \Phi  \right>} 
\end{array}
\end{equation} 
where $\rho_{s}$ is the entropy density, $\vec{j_{s}}$ is the flux of entropy density and $\sigma(\boldsymbol{J}_{i},\boldsymbol{X}_{i})$ is the entropy production density. Since we deal with the linearized distribution, hence $ln(f)$ can be approximated as
\begin{equation} \label{b1-ons-10b}
ln(f) = ln(f_{0}) + ln(1+\Phi) \approx ln(f_{0}) + \Phi
\end{equation} 
There is no contribution from the term $\left< ln(f_{0}), f_{0} \mathbb{J} \Phi  \right>$ as it is a collisional invariant. Positive semi-definiteness of the linearized collision operator $\left<\Phi \mathbb{J} \Phi  \right>$ $\geq 0$ leads to non-negative entropy production as $\sigma(\boldsymbol{J}_{i},\boldsymbol{X}_{i})$ $\approx$ $  {\left< f_{0} \Phi \mathbb{J} \Phi  \right>}$ $\geq 0$ establishing the connection with linear irreversible thermodynamics as follows
\begin{equation} \label{b1-ons-10}
\begin{array}{ll}
\sigma(\boldsymbol{J}_{i},\boldsymbol{X}_{i})\approx  {\left< f_{0} \Phi \mathbb{J} \Phi  \right>} &= {\left<f_{0} \sum_{j} (\mathbb{J}^{-1} \bar{\boldsymbol{\Upsilon}})_{j} \odot \boldsymbol{X}_{j}  \sum_{i} \bar{\boldsymbol{\Upsilon}}_{i} \odot \boldsymbol{X}_{i}\right>}\geq 0\\\
&= \sum_{i}\left(\sum_{j}\boldsymbol{L}_{ij} \odot \boldsymbol{X}_{j}\right) \odot \boldsymbol{X}_{i}\geq 0=\sum_{i}\boldsymbol{J}_{i} \odot \boldsymbol{X}_{i} \geq 0
\end{array}
\end{equation} 
The present approach of defining linearized collision operator as $\mathbb{J}_i \Phi_{i}$ $=$ $\bar{\boldsymbol{\Upsilon}}_{i} \odot \boldsymbol{X}_{i}$ leads to non-equilibrium flux $\boldsymbol{J}_i$ in a  linear phenomenological form complying with  Onsager's principle. \emph{The flux follows Onsager's form hence entropy production for linearized Boltzmann equation described by Equation (\ref{b1-ons-10}) will also be in accordance with maximum entropy production (MEP) principle}. Figure \ref{kt-net-ons} shows the schematic of Onsager reciprocity principle linking the macroscopic non-equilibrium thermodynamics of entropy production due to microscopic collisions described by kinetic theory.  This exercise gives us the following guidance and directions:
\begin{itemize}
\item Kinetic model replacing  linearized collision operator $\mathbb{J}_i \Phi_{i}$ should be formulated based on the principles of non-equilibrium thermodynamics.
\item The perturbation term $\Phi$ in such a case can be written as a sum of perturbation components $\Phi_{i}$ for each thermodynamic forces $\boldsymbol{X}_{i}$. The perturbation components $\Phi_{i}$ can be expressed as tensor contraction of reduced microscopic  flux tensors with its conjugate thermodynamic force following Onsager's relationship and maximum entropy production principle.
\item Once the distribution function is formulated in the Onsager's form at the microscopic level it will also comply with the principles on non-equilibrium thermodynamics when it is projected to macroscopic Navier-Stokes-Fourier level.
\end{itemize}
\begin{section}
{Onsager's variational principle based kinetic model}\label{kmodel}
\end{section}
Boltzmann equation being a nonlinear integro-differential equation becomes difficult to handle. This requires some alternative simpler model to replace the collision term. In kinetic models, the Boltzmann collision term $J(f,f)$ is substituted  by a relaxation expression $J_{m}(f,f_{ref})$ in terms of suitable reference distribution function, $f_{ref}$ and mean collision frequency, $\nu$ or relaxation time, $t_{R}$. These models should preserve following basic properties and characteristics of the Boltzmann equation \cite{mahvrkfvs} enumerated as
\begin{enumerate}
 \item Locality and Galilean invariance\\
Since the Boltzmann equation is invariant under Galilean transformation hence the collision term $J_{m}(f,f_{ref})$ should depend only on peculiar velocity $\vec{c}=\vec{v}-\vec{u}$. 
\item Additive invariants of the collision integral\\
This property ensures conservation of mass, momentum and energy and is represented as
\begin{equation} \label{slkns1-2km2} 
 \int _{\mathbb{R}^{+} }\int _{\mathbb{R}^{D} }J_{m}(f,f_{ref}) \boldsymbol{\Psi}   d\vec{v}d\mathbb{I} =0
\end{equation} 
\item Uniqueness of equilibrium \\
The zero point of kinetic model $J_{m}(f,f_{ref})= 0$ representing collision term implies uniqueness of equilibrium.
\item Local entropy production inequality and Boltzmann H-theorem \\
This property represents  non-negative Boltzmann entropy production obtained using moment $\Psi_{e}=-ln f$.\footnote{When distribution function is expressed in terms of number density as $\breve{f}$ then entropy production per molecule is $\displaystyle \sigma(\vec{v},t)=\displaystyle -k_{B} \int _{\mathbb{R}^{+} }\int _{\mathbb{R}^{D} } ln \breve{f} J_{m}(\breve{f},\breve{f}_0) d\vec{v} d\mathbb{I} =\displaystyle -k_{B}\frac{dH}{dt} \geq 0$ where $k_{B}$ is the Boltzmann's constant. } by the kinetic model representing collision term 
\begin{equation} \label{slkns1-2km3}
\sigma(\vec{v},t)=-\int _{\mathbb{R}^{+} }\int _{\mathbb{R}^{D} } ln f J_{m}(f,f_0) d\vec{v} d\mathbb{I} = -\frac{dH}{dt} \geq 0
\end{equation} 
where H-function is given by
\begin{equation} \label{slkns1-2km4}
H=\int _{\mathbb{R}^{+} }\int _{\mathbb{R}^{D} } f ln f  d\vec{v} d\mathbb{I}
\end{equation} 
\item Positive distribution\\
The H-function of the kinetic model should decay monotonically such that Boltzmann equation gives positive distribution leading towards the unique equilibrium solution.
\item Correct transport coefficients in the hydrodynamic limit \\
In the hydrodynamic limit the kinetic model should generate correct transport coefficients such as viscosity, $\mu$ and thermal conductivity, $\kappa$ and Prandtl number, $Pr$ should be close to $2/3$. 
\item Onsager's relation for entropy production\\
Close to equilibrium for a prescribed irreversible force $\boldsymbol{X}_i$ a non-equilibrium flux $\boldsymbol{J}_i$  is generated due to collisions  such that  $\boldsymbol{J}_{i}= \sum_{j}\boldsymbol{L}_{ij} \odot \boldsymbol{X}_{j}$ leading to maximization of the entropy production given by  $\sigma(\boldsymbol{J}_i,\boldsymbol{X}_i)=\sum_i{\boldsymbol{J}_i\odot\boldsymbol{X}_i}$.
\end{enumerate}
The simplest of all the kinetic models which satisfies the Boltzmann H-theorem is the non-linear Bhatnagar-Gross-Krook (BGK) model \cite{bgk}. Here the reference distribution function is simply Maxwellian i.e. $f_{ref}=f_{0}$. With this the BGK kinetic model can be represented as
\begin{equation} \label{slkns1-2bgk1} 
J_{m}(f,f_{0})= -\nu (f-f_{0})= -\frac{(f-f_{0})}{t_{R}}
\end{equation} 
where $\nu$ is the collision frequency and $t_{R}$ is the relaxation time. BGK is a non-linear function of the moments of $f$ whereas the Boltzmann collision integral is non-linear in the distribution function itself. BGK model preserves most of the property of the collision integral but its evaluation in the hydrodynamic limit generates transport coefficients which cannot be adjusted to give the correct Prandtl number of $2/3$. A more generalized form of BGK model is Gross and Jackson\cite{gross} constant collision frequency kinetic model  which retains sufficient eigenfunctions of the collision operator, while doing so it introduces scaling or free parameter due to  certain cut-off in the intermolecular potential \cite{soga} that affects the thermal creep slip. The incorrect value of the Prandtl number due to BGK like model, can be corrected by kinetic models like Shakhov's model \cite{shakhov}, the ellipsoidal statistical BGK ( ES-BGK) model \cite{holway, perthame}. The ellipsoidal statistical BGK ( ES-BGK) model and the Shakhov's model is a generalization of the BGK model equation with correct relaxation of both the heat flux and stresses, leading thus to the correct continuum limit in the case of small Knudsen numbers. Both the models are computationally expensive in comparison with BGK model. Literature review also reveals Liu model \cite{liug}, the BGK model with velocity dependent collision frequency $\nu $(c)$-$BGK model of Mieussens and Struchtrup \cite{mieustruch} which yield the proper Prandtl number. Zheng and Struchtrup \cite{zhengstruch} have carried out detailed study on kinetic models. Gorban and Karlin \cite{gorban} have used specific thermodynamic parametrization of an arbitrary approximation of reduced description to construct kinetic model.

Kinetic model in lattice Boltzmann (LB) method is represented as scattering matrix between various discrete-velocity distributions e.g. in lattice BGK (LBGK) scattering matrix is in diagonal form with single relaxation parameter.  Kinetic model associated with multiple relaxation time (MRT) LB\cite{humrt,humrtluo} uses multiple relaxation to address the issue of fixed Prandtl number and fixed ratio between the kinematic and bulk viscosity while providing stability. Revised matrix LB model \cite{karlin} uses a two-step relaxation BGK like model to strike a balance between enhanced stability and simplicity. Yong \cite{yong} proposed Onsager like relation 
as a requirement and guide to construct stable LB models.

Almost all the kinetic models have focused on Prandtl number fix and satisfaction of H-theorem or stability, and have ignored the crucial aspect of  \emph{non-equilibrium thermodynamics i.e. satisfaction of Onsager's variational principle.} In order to derive a thermodynamically correct distribution function the kinetic model itself requires its foundation based on the principles of non-equilibrium thermodynamics. Task of development of such a kinetic model will require casting it in the Onsager's form at the microscopic level i.e. representing it terms of microscopic flux and its associated thermodynamic forces. The first step towards the development of such a kinetic model will  require identification of thermodynamic forces and microscopic flux tensors associated with the polyatomic gas.
\subsection{Identification of thermodynamic forces and microscopic tensors}
Important linkages with non-equilibrium thermodynamics can be drawn from the expression of the first order velocity distribution function, $f_{1}$  based on Morse-BGK model \cite{morse} of a polyatomic gas. In Morse's model \cite{morse} relaxation time for elastic and inelastic collision is considered separately. Due to inelastic collision particles relax to equilibrium distribution in internal and translational state at same temperature as there is equipartition of energy between the internal and translational degree of freedom.  Consider inelastic collisions and assumption of molecular level thermodynamic equilibrium with internal and translational mode\footnote{For polyatomic gas, vibrational relaxation may require several thousand collisions for relaxation to achieve molecular level thermodynamic equilibrium. For simulating effect of high frequency sound waves the present assumption of thermodynamic equilibrium with internal and translational mode may not hold.} such that the particles in non-equilibrium  are replaced exponentially by particles in equilibrium with characteristic time $t_{R(f)}$ and $t_{R(f_0)}$ respectively. Morse-BGK \cite{morse} kinetic model for polyatomic gas with an assumption $t_{R(f)}$$=$$t_{R(f_0)}$$=$$t_{R}$  is given by
\begin{equation} \label{slkns1x} 
J_m(f,f_0)=-\frac{f(\vec{x},\vec{v},\mathbb{I},t)}{t_{R(f)}(\vec{x},t)} + \frac{f_0(\vec{x},\vec{v},\mathbb{I},t)}{t_{R(f_0)}(\vec{x},t)}=-\frac{f(\vec{x},\vec{v},\mathbb{I},t)-f_{0}(\vec{x},\vec{v},\mathbb{I},t)}{t_{R}(\vec{x},t)}
\end{equation} 
Using Chapman-Enskog expansion, velocity distribution function $f_{1}$ is derived as
\begin{equation} \label{df-mbgk}
f_{1}=f_{0} - \sum_{j}{  \boldsymbol{\Upsilon}_{j} \odot \boldsymbol{X}_{j}}=f_{0}-(\boldsymbol{\Upsilon}_{\tau}\boldsymbol{:}\boldsymbol{X}_{\tau}+\boldsymbol{\Upsilon}_{q}\boldsymbol{\cdot} \boldsymbol{X}_{q})
\end{equation} 
where $\boldsymbol{\Upsilon}_{j}$ is the microscopic flux tensor and $ \boldsymbol{X}_{j}$ is the conjugate thermodynamic force tensor.  Using the definition of  perturbation term $\Phi$ the microscopic tensor $\boldsymbol{\Upsilon}_j$ can be expressed in terms of reduced  microscopic tensor as follows
\begin{equation} \label{df-mbgk-1}
\boldsymbol{\Upsilon}_j= -f_{0} (\mathbb{J}^{-1} \bar{\boldsymbol{\Upsilon}})_j=-f_{0} t_{R} \bar{\boldsymbol{\Upsilon}}_j
\end{equation} 
The linearized collision operator $\mathbb{J}$ is described by a single relaxation time $t_{R}$ which is not affected by thermodynamic force. The microscopic reduced tensor associated with the stress tensor for $D$ degree of freedom in terms of Morse-BGK  model is derived as
\begin{equation} \label{df-mbgk-2}
\bar{\boldsymbol{\Upsilon}}_{\tau}=-\left[\vec{c} \otimes\vec{c} +\frac{1}{2}\{ \frac{(2+D) \gamma -(4+D)}{2 \beta }-\frac{\mathbb{I} (\gamma -1)}{\mathbb{I}_o \beta }-c^2 (\gamma -1)\} \boldsymbol{I}  \right]
\end{equation}
where $\vec{c}$  $=$ $\vec{v}-\vec{u}$  is the peculiar velocity vector and $\boldsymbol{I}$ is the rank-$D$ identity invariant tensor. The microscopic reduced vector associated with heat transport is 
\begin{equation} \label{df-mbgk-3}
\bar{\boldsymbol{\Upsilon}}_{q}=-\left [\frac{4+D}{2 \beta }-\frac{\mathbb{I}}{\mathbb{I}_o \beta }-c^2 \right ]\vec{c}
\end{equation} 
The  thermodynamic force associated with stress tensor, $\boldsymbol{X}_{\tau}$ and  thermal gradient vector, $\boldsymbol{X}_{q}$ is
\begin{equation} \label{df-mbgk-4}
\begin{array}{lllll}
\boldsymbol{X}_{\tau}&=\beta[ (\nabla \otimes \vec{u}) + (\nabla \otimes \vec{u})^{T} ] & , &
\boldsymbol{X}_{q}&=\nabla \beta
\end{array}
\end{equation}
\subsection{Onsager-BGK kinetic model}
Consider non-equilibrium phenomenon due to two thermodynamic forces $\boldsymbol{X}_{\tau}$  and thermodynamic forces $\boldsymbol{X}_{q}$. The particles that are in  non-equilibrium due to thermodynamic forces $\boldsymbol{X}_{\tau}$ are replaced exponentially by particles in equilibrium with characteristic time $t_{R(f,\tau)}$ and $t_{R(f_0,\tau)}$ respectively. Similarly,  particles in  non-equilibrium due to thermodynamic forces $\boldsymbol{X}_{q}$ are replaced exponentially by particles in equilibrium with characteristic time $t_{R(f,q)}$ and $t_{R(f_0,q)}$ respectively. In most cases the state of gas is not varying rapidly in the interval of relaxation time so $f$$-$$f_{0}$ is small, and $t_{R(f,\tau)}$ $=$ $t_{R(f_0,\tau)}$ $=$ $t_{R(\tau)}$ and $t_{R(f,q)}$ $=$ $t_{R(f_0,q)}$ $=$ $tr_{R(q)}$, with this the new kinetic model called \emph{Onsager-BGK model} becomes
\begin{equation} \label{slkns-kmodel} 
J_m(f,f_0)=-\left(\frac{f(\vec{x},\vec{v},\mathbb{I},t)-f_0(\vec{x},\vec{v},\mathbb{I},t)}{t_{R(\tau)}(\vec{x},t)}\right)_{\boldsymbol{X}_{q}=0}-\left(\frac{f(\vec{x},\vec{v},\mathbb{I},t)-f_0(\vec{x},\vec{v},\mathbb{I},t)}{t_{R(q)}(\vec{x},t)}\right)_{\boldsymbol{X}_{\tau}=0}
\end{equation} 
\begin{figure}\centering
{\includegraphics[scale=0.25,bb=0 0 838 718]{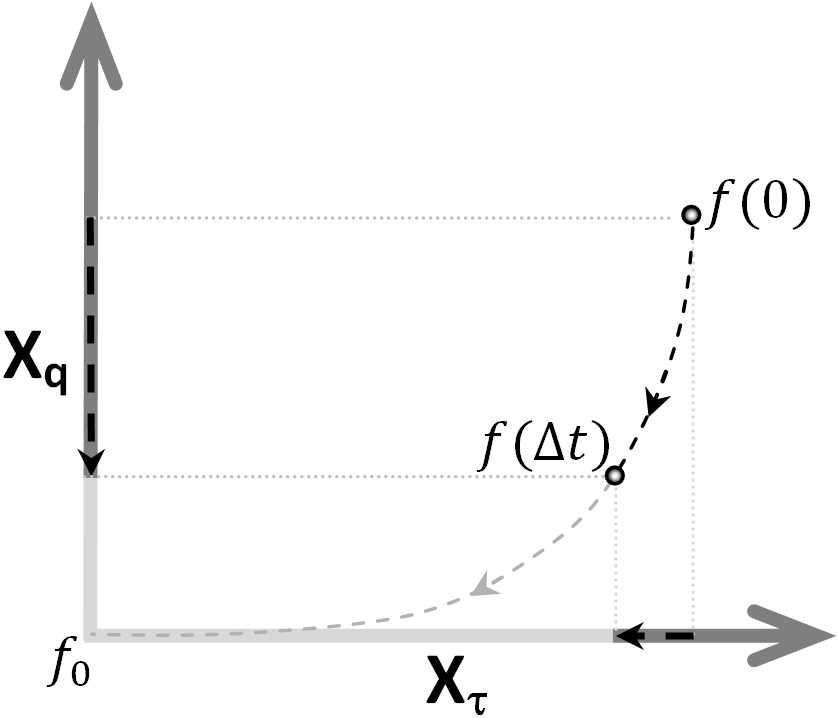}}
\caption{ \label{kmodela}Schematic of relaxation of non-equilibrium distribution $f$ to equilibrium distribution $f_{0}$ in the phase plane of thermodynamic force $\boldsymbol{X}_{\tau}$ and $\boldsymbol{X}_{q}$.}
\end{figure}
In this proposed new kinetic model only inelastic collisions are considered that are in non-equilibrium due to thermodynamic forces $\boldsymbol{X}_{\tau}$ and $\boldsymbol{X}_{q}$. The part of distribution which is in non-equilibrium due thermodynamic forces $\boldsymbol{X}_{\tau}$ first relaxes to Maxwellian $f_{0}$ in characteristic time $t_{R(\tau)}$. Simultaneously, the part of distribution which is in non-equilibrium due thermodynamic forces $\boldsymbol{X}_{q}$ relaxes to Maxwellian $f_{0}$ in characteristic time $t_{R(q)}$. Figure \ref{kmodela} shows the relaxation of non-equilibrium distribution function based on the new kinetic model in the phase plane of thermodynamic force $\boldsymbol{X}_{\tau}$ and $\boldsymbol{X}_{q}$. The relaxation step can be cast as an eigenvalue problem $(\boldsymbol{A}-\lambda \boldsymbol{I})\boldsymbol{\mathcal{X}}$ $=$ $0$ where $\boldsymbol{\mathcal{X}}$ is a tensor with components $\{\boldsymbol{X}_{\tau},\boldsymbol{X}_{q}\}$ such that positive semi-definiteness of the collision operator ensures non-negative entropy production, providing a Lyapunov criterion for the stability towards a equilibrium distribution. In a more generalized form Onsager-BGK model can be written as 
\begin{equation} \label{slkns-gmodel} 
J_m(f,f_0)=-\sum_j\left(\frac{f(\vec{x},\vec{v},\mathbb{I},t)-f_0(\vec{x},\vec{v},\mathbb{I},t)}{t_{R(j)}(\vec{x},t)}\right)_{\boldsymbol{X}_{i}=0, \forall i \neq j}
\end{equation} 
where $t_{R(j)}$ is the relaxation time for thermodynamic force $\boldsymbol{X}_{j}$. Distribution function, $f(\Delta t)$ after time interval $t=\Delta t$ relaxes as
\begin{equation} \label{slkns-kmodel-a} 
f(\Delta t) =  \sum_j\left[f(0)Exp(-\Delta t/t_{R(j)})  +f_{0}(1-Exp(-\Delta t/t_{R(j)}))\right]_{\boldsymbol{X}_{i}=0,i\ne j}
\end{equation} 
where $f(0)=f(t=0)$ is the initial non-equilibrium distribution just after streaming or convection step while $f_{0}$ is the final equilibrium distribution to be reached after sufficient collisions\footnote{It is not guaranteed that this approximate distribution function will be able to meet all the requirements of thermodynamics. The present analysis can be made more physically meaningful by using modified moment method of Eu \cite{eubc} in which the distribution function also depends on entropy derivatives called as Gibbs variables.}. Generalized Onsager-BGK model that accounts for elastic and inelastic collision is expressed as
\begin{equation} \label{slkns-gmodel} 
J_m(f,f_0)_{el,in}=-\sum_j\left(\frac{f-f_{0,el}}{t_{R(j),el}}\right)_{\boldsymbol{X}_{i}=0, \forall i \neq j}-\sum_j\left(\frac{f-f_{0}}{t_{R(j)}}\right)_{\boldsymbol{X}_{i}=0, \forall i \neq j}
\end{equation} 
Kinetic model which accounts inelastic and elastic collisions is parameterized by elastic and inelastic relaxation times  i.e. $t_{R(j),el}$ and $t_{R(j)}$ for each thermodynamic force $\boldsymbol{X}_{j}$. Elastic collisions do not contribute to equipartition of energy between translational and internal states. Due to elastic collision particles relax to equilibrium $f_{0,el}$ for translational state at a temperature which is different from the temperature at which internal state attains its equilibrium.
\subsection{Distribution function for Onsager-BGK kinetic model}
Distribution function using Chapman-Enskog method\cite{chapman} for single particle distribution function is build on two fundamental assumptions namely i) distribution function can be expanded as a power series around a local equilibrium state using Knudsen number as a parameter, ii) distribution is time independent function of locally conserved variable. On the other hand Grad's moment method follows the framework and structure of extended irreversible thermodynamics (EIT) as it expands the distribution function in terms of tensorial Hermite polynomials around a local equilibrium state. Grad's method does not have measure of order of magnitude for truncation procedure \cite{velasco}. Sharipov \cite{sharipov2011} method expands the distribution function for an arbitrary Knudsen number as a power series with respect to thermodynamic force. For a generic Onsager-BGK kinetic model, the distribution function for an arbitrary Knudsen number is generated using iterative refinement  as follows
\begin{equation} \label{dis-gmodel} 
f = f_{0} - \sum_{j}{  \boldsymbol{\Upsilon}_{j} \odot \boldsymbol{X}_{j}} + \sum_{k,j}{  \boldsymbol{\Upsilon}_{kj} \odot \boldsymbol{X}_{k}\odot \boldsymbol{X}_{j}}- \sum_{m,k,j}{  \boldsymbol{\Upsilon}_{mkj} \odot \boldsymbol{X}_{m}\odot \boldsymbol{X}_{k}\odot \boldsymbol{X}_{j}}+\cdots
\end{equation} 
where 
\begin{equation} \label{dis-gmodel} 
\begin{array}{lll}
\boldsymbol{\Upsilon}_{j} \odot \boldsymbol{X}_{j} & = & \displaystyle t_{R(j)}  \left[\frac{\partial {f}_{0} }{\partial t} +\nabla _{\vec{x}} \cdot (\vec{v}{f}_{0} )\right]_{\boldsymbol{X}_{i}=0,i\ne j} \\
\boldsymbol{\Upsilon}_{kj} \odot \boldsymbol{X}_{k} &  = & \displaystyle t_{R(k)}  \left[\frac{\partial \boldsymbol{\Upsilon}_{j} }{\partial t} +\nabla _{\vec{x}} \cdot (\vec{v}\boldsymbol{\Upsilon}_{j} )\right]_{\boldsymbol{X}_{i}=0,i\ne k} \\
\boldsymbol{\Upsilon}_{mkj} \odot \boldsymbol{X}_{m} & = & \displaystyle t_{R(m)}  \left[\frac{\partial \boldsymbol{\Upsilon}_{kj} }{\partial t} +\nabla _{\vec{x}} \cdot (\vec{v}\boldsymbol{\Upsilon}_{kj} )\right]_{\boldsymbol{X}_{i}=0,i\ne m} 
\end{array}
\end{equation} 
The higher order distribution follows Onsager's reciprocity principle and accounts for the terms due to Onsager's cross coupling e.g. Soret and Dufour effects on Burnett distribution. The first order velocity distribution function, $f_{1}$ is
\begin{equation} \label{dis-obgk-1a} 
f_{1}=\displaystyle f_{0}-\sum_j  t_{R(j)}\left[\frac{\partial f_{0} }{\partial t} +\nabla _{\vec{x}} \cdot (\vec{v}f_{0} )\right]_{\boldsymbol{X}_{i}=0,i\ne j} = f_{0}-\sum_j{  \boldsymbol{\Upsilon}_{j} \odot \boldsymbol{X}_{j}} 
\end{equation} 
The microscopic flux tensor $\boldsymbol{\Upsilon}_{j}$ in terms of reduced  microscopic flux tensor $\bar{\boldsymbol{\Upsilon}}_{j}$ is expressed as follows
\begin{equation} \label{dis-obgk-2} 
\boldsymbol{\Upsilon}_{j}=-f_{0} (\mathbb{J}^{-1} \bar{\boldsymbol{\Upsilon}})_j= -f_{0} t_{R(j)} \bar{\boldsymbol{\Upsilon}}_{j}
\end{equation} 
where linearized collision operator $\mathbb{J}$ has a dependence on the thermodynamic force i.e. it depends inversely with $t_{R(j)}$ which is the relaxation time associated with the thermodynamic force  $\boldsymbol{X}_{j}$. The perturbation terms satisfies the additive invariants property, expressed as
\begin{equation} \label{dis-obgk-5} 
\begin {array}{ll}
 \sum_{j} {\left \langle \boldsymbol{\Psi} , \boldsymbol{\Upsilon}_{j}  \odot \boldsymbol{X}_{j}\right \rangle=}& \sum_{j} \left({ \displaystyle\int _{\mathbb{R}^{+} }\int _{\mathbb{R}^{D} }\boldsymbol{\Psi} \boldsymbol{\Upsilon}_{j}  d\vec{v}d\mathbb{I} }\right)\odot \boldsymbol{X}_{j}=0
\end {array}
\end{equation} 
A generalized form of Onsager-BGK model can also be expressed as
\begin{equation} \label{slkns-gmodel-a} 
J_m(f,f_0)=\sum_j\left(\frac{ \boldsymbol{\Upsilon}_{j} \odot \boldsymbol{X}_{j}}{t_{R(j)}}\right)_{\boldsymbol{X}_{i}=0, \forall i \neq j}=-f_{0}\sum_j \bar{\boldsymbol{\Upsilon}}_{j} \odot \boldsymbol{X}_{j} 
\end{equation} 
The non-equilibrium thermodynamic flux $\boldsymbol{J}_{i}$  expressed  in a  linear phenomenological form similar to equation (\ref{b1-ons-7}) is 
\begin{equation} \label{obgk-ons-7}
\begin{array}{ll}
\boldsymbol{J}_{i}&=\displaystyle  \int_{\mathbb{R}^{+}} \int_{\mathbb{R}^{D}} \bar{\boldsymbol{\Upsilon}}_{i} f_{0} t_{R(j)}\sum_j \bar{\boldsymbol{\Upsilon}}_{j} \odot \boldsymbol{X}_{j}   d\vec{v} d\mathbb{I}={\left<\bar{\boldsymbol{\Upsilon}}_{i} , f_{0}t_{R(j)}\sum_j \bar{\boldsymbol{\Upsilon}}_{j} \odot \boldsymbol{X}_{j} \right>}\\
&=\sum_{j} \boldsymbol{L}_{ij} \odot \boldsymbol{X}_{j}
\end{array}
\end{equation} 
where $\boldsymbol{L}_{ij}$ is the phenomenological tensor of transport coefficients defined as
\begin{equation} \label{b1-ons-8}
\boldsymbol{L}_{ij}={\left<\bar{\boldsymbol{\Upsilon}}_{i}, f_{0} t_{R(j)} \bar{\boldsymbol{\Upsilon}}_{j} \right>}
\end{equation} 
Onsager-BGK kinetic model for single temperature binary mixture of non-reacting gas becomes
\begin{equation} \label{slkns1-diff} 
\begin{array}{l}
J_m(f,f_0)= -\left(\frac{f-f_{0}}{t_{R(\tau)}}\right)_{\boldsymbol{X}_q=0,\boldsymbol{X}_d=0}-\left(\frac{f-f_{0}}{t_{R(q)}}\right)_{\boldsymbol{X}_{\tau}=0,\boldsymbol{X}_d=0}-\left(\frac{f-f_{0}}{t_{R(d)}}\right)_{\boldsymbol{X}_{\tau}=0,\boldsymbol{X}_q=0}
\end{array}
\end{equation} 
where $\boldsymbol{X}_{\tau}$ , $\boldsymbol{X}_q$  and $\boldsymbol{X}_d$ are the thermodynamic force terms associated with stress tensor, thermal gradient vector and chemical potential gradient vector with $t_{R(\tau)}$, $t_{R(q)}$ and $t_{R(d)}$ as their associated relaxation time. Refer appendix \ref{kspmodel} for kinetic model of multi-temperature binary mixture of non-reacting gas. Following equation (\ref{slkns-gmodel-a}), kinetic model is written as 
\begin{equation} \label{slkns1-diff-a} 
\begin{array}{l}
J_m(f,f_0)= -f_{0}\left( \bar{\boldsymbol{\Upsilon}}_{\tau} : \boldsymbol{X}_{\tau} + \bar{\boldsymbol{\Upsilon}}_{q} \cdot \boldsymbol{X}_{q}+\bar{\boldsymbol{\Upsilon}}_{d} \cdot \boldsymbol{X}_{d} \right)
\end{array}
\end{equation} 
\begin{figure}\centering
{\includegraphics[scale=0.25,bb=0 0 1202 460]{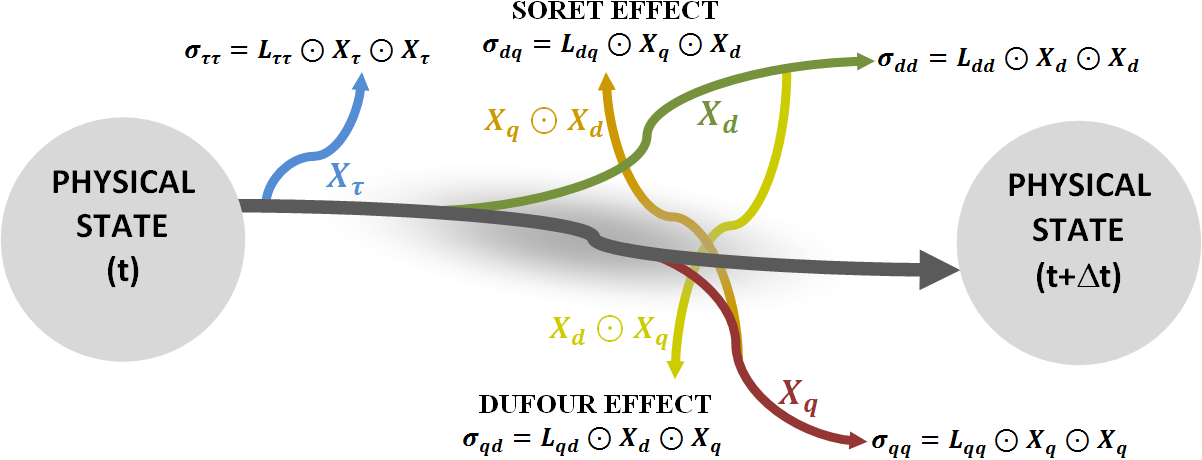}}
\caption{ \label{dufsor}Components of entropy as physical state evolves from time $t$ to time $t+\Delta t$ because of thermodynamic force due to stress tensor, thermal gradient vector and chemical potential gradient vector.}
\end{figure}
The three non-equilibrium thermodynamic fluxes namely $\boldsymbol{J}_q$, $\boldsymbol{J}_{\tau}$, and $\boldsymbol{J}_d$ are
\begin{equation} \label{slkns1-diff-flux} 
\begin{array}{l}
\boldsymbol{J}_q = \boldsymbol{L}_{qq} \cdot \boldsymbol{X}_{q} + \boldsymbol{L}_{qd} \cdot \boldsymbol{X}_{d} \\
\boldsymbol{J}_d = \boldsymbol{L}_{dq} \cdot \boldsymbol{X}_{q} + \boldsymbol{L}_{dd} \cdot \boldsymbol{X}_{q} \\
\boldsymbol{J}_{\tau} = \boldsymbol{L}_{\tau\tau} : \boldsymbol{X}_{\tau} 
\end{array}
\end{equation} 
The thermodynamic forces $\boldsymbol{X}_{\tau}$ is of different tensorial order with respect to $\boldsymbol{X}_{q}$ and $\boldsymbol{X}_{d}$ hence the phenomenological tensors of transport coefficient $\boldsymbol{L}_{\tau q}$, $\boldsymbol{L}_{\tau d}$, $\boldsymbol{L}_{d \tau}$ and $\boldsymbol{L}_{q \tau}$ do not exists. Whereas, $\boldsymbol{X}_q$ and $\boldsymbol{X}_d$ are of same tensorial order, hence Onsager's cross coupling \cite{bosch} gives the Dufour effect because of the heat flux driven by chemical potential gradient vector and Soret effect due to diffusive flux driven by thermal gradient vector. Figure \ref{dufsor} shows the components of entropy  as the physical state evolves from time $t$ to time $t+\Delta t$. The entropy generation follows Onsager's principle\cite{groot} and it is expressed as follows
\begin{equation} \label{slkns1-diff-ent} 
\begin{array}{ll}
\displaystyle \sigma =\sum_{ij}\boldsymbol{L}_{ij}\odot\boldsymbol{X}_{i}\odot\boldsymbol{X}_{j}&\left\{
{\begin{array}{l}
\underbrace{\boldsymbol{L}_{\tau \tau}:\boldsymbol{X}_{\tau}:\boldsymbol{X}_{\tau}}_{\sigma_{\tau\tau}} + \underbrace{\boldsymbol{L}_{qq}\cdot\boldsymbol{X}_{q}\cdot\boldsymbol{X}_{q}}_{\sigma_{qq}}+ \underbrace{\boldsymbol{L}_{dd}\cdot\boldsymbol{X}_{d}\cdot \boldsymbol{X}_{d}}_{\sigma_{dd}}\\
+  \underbrace{\boldsymbol{L}_{qd}\cdot \boldsymbol{X}_{d}\cdot\boldsymbol{X}_{q}}_{\sigma_{qd}}+ \underbrace{\boldsymbol{L}_{dq}\cdot\boldsymbol{X}_{q}\cdot\boldsymbol{X}_{d}}_{\sigma_{dq}}
\end{array}}\right.
\end{array}
\end{equation} 
The Dufour and Soret coefficients can be extracted from reduced matrix elements obtained from $\boldsymbol{L}_{qd}$ and $\boldsymbol{L}_{dq}$. Onsager symmetry tells us that heat flux $\boldsymbol{J}_q$ generated due to thermodynamic force $\boldsymbol{X}_d$ exactly equals the diffusion flux $\boldsymbol{J}_d$ generated due to thermodynamic force $\boldsymbol{X}_q$, so
\begin{equation} \label{slkns-srel-b} 
\begin{array}{lll}
\displaystyle \left( \frac{\boldsymbol{J}_q}{\boldsymbol{X}_d}\right)_{\boldsymbol{X}_{q}}&=&\displaystyle  \left( \frac{\boldsymbol{J}_d}{\boldsymbol{X}_q}\right)_{\boldsymbol{X}_{d}}
\end{array}
\end{equation} 
The postulate of Onsager's symmetry relationship implies equality of  phenomenological tensors given by  
\begin{equation} \label{slkns-srel-a} 
\begin{array}{lll}
\displaystyle \boldsymbol{L}_{qd} &=&\boldsymbol{L}_{dq} 
\end{array}
\end{equation} 
Since $\boldsymbol{L}_{qd}$ $=$ ${\left<\bar{\boldsymbol{\Upsilon}}_{q}, f_{0} t_{R(d)} \bar{\boldsymbol{\Upsilon}}_{d} \right>}$ and $\boldsymbol{L}_{qd}$ $=$ ${\left<\bar{\boldsymbol{\Upsilon}}_{q}, f_{0} t_{R(d)} \bar{\boldsymbol{\Upsilon}}_{d} \right>}$ as described by equation (\ref{b1-ons-8}), and equality of phenomenological tensors establishes 
\begin{equation} \label{slkns-srel-tr} 
\begin{array}{lll}
t_{R(d)} =t_{R(q)}
\end{array}
\end{equation} 
The relaxation time associated with thermal gradient vector,  $t_{R(q)}$ can be derived as
\begin{equation}\label{dis-obgk-3} 
t_{R(q)}=\frac{\kappa (\gamma -1)}{R \gamma p}
\end{equation}
where $\kappa$ is thermal conductivity. The relaxation time associated with stress tensor, $t_{R(\tau)}$ is derived as
\begin{equation}\label{dis-obgk-4} 
t_{R(\tau)}=\frac{\mu}{p}=\frac{\mu_{ref}}{p}\left(\frac{T}{T_{ref}} \right)^{\omega}
\end{equation}
where $\mu_{ref}$ is the viscosity of the gas at reference temperature $T_{ref}$, $\omega$ is the exponent of the viscosity law and $p$ is the pressure. The Prandtl number can be extracted from the ratio of  reduced matrix elements $L_{\tau \tau}/L_{qq}$ as $\text{Pr}$ can also be interpreted as the ratio of the relaxation time \cite{woods} associated with thermodynamic force $\boldsymbol{X}_{\tau}$ and $\boldsymbol{X}_{q}$, which are of different tensorial order
\begin{equation}\label{dis-obgk-4a} 
\text{Pr}=\frac{t_{R(\tau)}}{t_{R(q)}}
\end{equation}

\subsection{Statistical representation of Onsager-BGK kinetic model}
In statistics there are many varied approaches to measure divergence between two generalized probability density function $f_{g}(x)$ and $f_{h}(x)$. Kullback-Leibler divergence \cite{kld}  is one such measure which provides relative entropy \cite{villani, ghosh}.  Kullback-Leibler divergence is defined as follows 
\begin{equation} \label{s-obgk-1} 
D(f_{g}\| f_{h})=\int f_{g}(x) ln \frac{f_{g}(x)}{f_{h}(x)}dx
\end{equation} 
Kullback-Leibler divergence is asymmetric, always non-negative and becomes zero if and only if both the distributions are identical. Kullback-Leibler symmetric divergence can be written as
\begin{equation} \label{s-obgk-2}
\displaystyle D(f_{g}, f_{h})=D(f_{g}\| f_{h})+D(f_{h}\| f_{g})=\int \left(f_{g}(x)-f_{h}(x)\right) ln \frac{f_{g}(x)}{f_{h}(x)}dx
\end{equation} 
Kullback-Leibler symmetric divergence $D(f_{g}, f_{h})$ equals Mahalanobis distance \cite{mahal} when  $f_{g}(x)$ and $f_{h}(x)$ are multivariate normal distributions with common variance-covariance matrix. Mahalanobis distance uses Galilean transformation and evaluates equivalent Euclidean distance under standard normal distribution. In the kinetic theory context $ D(f, f_{ref})$ can be interpreted as Mahalanobis distance between two distributions $f$ and $f_{ref}$.
\subsubsection{Mahalanobis speed and H-function}
Consider Boltzmann H-function given by
\begin{equation} \label{s-obgk-3}
H=\int _{\mathbb{R}^{+} }\int _{\mathbb{R}^{D} }f ln f  d\vec{v} d\mathbb{I}
\end{equation} 
The time derivative of H-function can be cast as
\begin{equation} \label{s-obgk-4}
\frac{\partial H}{\partial t}=\int _{\mathbb{R}^{+} }\int _{\mathbb{R}^{D} } J_{m}(f,f_0) lnf d\vec{v} d\mathbb{I}
\end{equation} 
Based on the additive invariant property due to conservation of mass, momentum and energy we can write
\begin{equation} \label{s-obgk-5}
\int _{\mathbb{R}^{+} }\int _{\mathbb{R}^{D} } J_{m}(f,f_0) ln(f_{0})d\vec{v} d\mathbb{I}=0
\end{equation} 
Using the above relationship  the time derivative of Boltzmann H-function can also be written as
\begin{equation} \label{s-obgk-6}
\frac{\partial H}{\partial t}=\int _{\mathbb{R}^{+} }\int _{\mathbb{R}^{D} } J_{m}(f,f_0) ln\frac{f}{f_{0}} d\vec{v} d\mathbb{I}
\end{equation} 
The Boltzmann H-function can be interpreted as a summation of components of H-function belonging to each thermodynamic force i.e. each thermodynamic force will have its own H-theorem. For a generalized Onsager-BGK kinetic model described by equation (\ref{slkns-gmodel}) the time derivative of Boltzmann H-function can be written as summation of components for each thermodynamic force $\boldsymbol{X}_{i}$ as
\begin{equation} \label{s-obgk-7}
\begin{array}{ll}
\displaystyle \frac{\partial H}{\partial t}&= \displaystyle \sum_i \left(\frac{\partial H}{\partial t}\right)_{\boldsymbol{X}_{j}=0, \forall j \neq i}\\
&=\displaystyle -\sum_i \frac{1}{t_{R(i)}}\left(\int _{\mathbb{R}^{+} }\int _{\mathbb{R}^{D} } (f-f_{0}) ln\frac{f}{f_{0}} d\vec{v} d\mathbb{I}\right)_{\boldsymbol{X}_{j}=0, \forall j \neq i}\\
&\displaystyle=-\sum_i \frac{D(f,f_{0})_{\boldsymbol{X}_{j}=0, \forall j \neq i}}{t_{R(i)}} 
\end{array}
\end{equation} 
where $D(f,f_{0})_{\boldsymbol{X}_{j}=0, \forall j \neq i}$ is the Mahalanobis distance between distribution $f$ and $f_{0}$ associated with thermodynamic force $\boldsymbol{X}_{i}$. This statistical representation helps us to draw analogy with Mahalanobis distance and its positivity property  shows 
\begin{equation} \label{s-obgk-8}
\displaystyle \left(\frac{\partial H}{\partial t}\right)_{\boldsymbol{X}_{j}=0, \forall j \neq i} =-\frac{D(f,f_{0})_{\boldsymbol{X}_{j}=0, \forall j \neq i}}{t_{R(i)}} \leq 0
\end{equation} 
this establishes $\frac{\partial H}{\partial t} \leq 0$, proof of  H-theorem for the new kinetic model.
\subsubsection{Mahalanobis speed and entropy production}
The Boltzmann entropy production rate can be written as
\begin{equation} \label{s-obgk-9}
\begin{array}{ll}
\sigma(f,f_{0}) & = \displaystyle - \frac{\partial H}{\partial t}= \sum_i \frac{D(f,f_{0})_{\boldsymbol{X}_{j}=0, \forall j \neq i}}{t_{R(i)}}= \sum_i\dot{M}_{i}
\end{array}
\end{equation} 
where $\dot{M}_{i}$ is defined as Mahalanobis speed,  it gives the component of entropy production rate, $\sigma_{i}$ associated with the thermodynamic force $\boldsymbol{X}_{i}$. Mahalanobis speed $\dot{M}_{i}$ associated with  thermodynamic force $\boldsymbol{X}_{i}$ for first order distribution function can be written as
\begin{equation} \label{s-obgk-10}
\begin{array}{ll}
\displaystyle \dot{M}_{i} &\displaystyle =- \left(\frac{\partial H}{\partial t}\right)_{\boldsymbol{X}_{j}=0, \forall j \neq i} \ge 0\\
\displaystyle &\displaystyle =-\int _{\mathbb{R}^{+} }\int _{\mathbb{R}^{D} } \left(\frac{ \boldsymbol{\Upsilon}_{i} \odot \boldsymbol{X}_{i}}{t_{R(i)}}\right)_{\boldsymbol{X}_{j}=0, \forall j \neq i} lnf d\vec{v} d\mathbb{I} \ge 0\\
\displaystyle &\displaystyle = \left(\int _{\mathbb{R}^{+} }\int _{\mathbb{R}^{D} }  f_{0}\bar{\boldsymbol{\Upsilon}}_{i} lnf d\vec{v} d\mathbb{I}\right)_{\boldsymbol{X}_{j}=0, \forall j \neq i}\odot \boldsymbol{X}_{i} \ge 0\\
\displaystyle &\displaystyle = \boldsymbol{J}_{i}\odot \boldsymbol{X}_{i} = \sigma_i(f,f_0) \ge 0
\end{array}
\end{equation} 
where $\boldsymbol{J}_{i}$ is the entropy flux associated with the thermodynamic force $\boldsymbol{X}_{i}$ satisfying Onsager's relationship as the physical state evolves with Mahalanobis speed $\dot{M}_{i}$ which is Onsager's component of entropy, $ \sigma_i(f,f_0)$ associated with its conjugate thermodynamic force. Based on first order expansion the entropy generation is always positive and looks independent of velocity and temperature gradients involved in thermodynamic force $\boldsymbol{X}$. This is an incorrect interpretation as the very validity of first order Chapman-Enskog distribution is not ensured at higher gradients or when conditions described by $\| \frac{\boldsymbol{\Upsilon}_{i}\odot\boldsymbol{X}_{i}}{f_{0}}\| \leq 1$  is violated. 
\subsection{Derivation of Euler and Navier-Stokes-Fourier equations}
Consider $\boldsymbol{\Psi}$-moments of a Boltzmann equation for a  two dimensional case as follows
\begin{equation} \label{ns-obgk-1} 
\displaystyle
{\left\langle \boldsymbol{\Psi},\frac{\partial f_{1} }{\partial t} +\nabla _{\vec{x}}\cdot (\vec{v}f_{1} )=0 \right\rangle} \displaystyle \equiv\int _{\mathbb{R}^{+} }\int _{\mathbb{R}^{2} }\boldsymbol{\Psi}   \left(\frac{\partial f_{1} }{\partial t} +\nabla _{\vec{x}}\cdot (\vec{v}f_{1} )=0\right)d\vec{v}d\mathbb{I}\\\\
\end{equation} 
Substitution of the first order distribution for the polyatomic gas leads to
\begin{equation} \label{ns-obgk-2}
\begin{array}{l}
\displaystyle {\left\langle \boldsymbol{\Psi},\frac{\partial f_{0} - \sum_{j}{  \boldsymbol{\Upsilon}_{j} \odot \boldsymbol{X}_{j}} }{\partial t} \right\rangle} + {\left\langle  \frac{\partial v_{x} (f_{0} - \sum_{j}{  \boldsymbol{\Upsilon}_{j} \odot \boldsymbol{X}_{j}}) }{\partial x} \right\rangle}
\displaystyle+{\left\langle   \frac{\partial v_{y} (f_{0} - \sum_{j}{  \boldsymbol{\Upsilon}_{j} \odot \boldsymbol{X}_{j}}) }{\partial y} \right\rangle}=0
\end{array}
\end{equation} 
After solving we get Navier-Stokes equation as follows
\begin{equation} \label{ns-obgk-3}
\displaystyle{ \frac{\partial \boldsymbol{U}}{\partial t} +\frac{\partial \boldsymbol{GX_{I}}}{\partial x}+\frac{\partial \boldsymbol{GX_{V}}}{\partial x} +\frac{\partial \boldsymbol{GY_{I}} }{\partial y}+\frac{\partial \boldsymbol{GY_{V}} }{\partial y} =0 }
\end{equation} 
where $\boldsymbol{U}$$=$$\left(\rho ,\rho u_{x},\rho u_{y},\rho E\right)^{T}$ represent the conserved vector. As described earlier inviscid or Euler fluxes  $\boldsymbol{GX}_{I}$,$\boldsymbol{GY}_{I}$ are based on Maxwellian, $f_{0} $ and viscous fluxes $\boldsymbol{GX}_{V}$, $\boldsymbol{GY}_{V}$ are based on perturbation $Kn \bar{f}_{1}$ $=$ $-(\sum_{j}{ \boldsymbol{\Upsilon}_{j} \boldsymbol{\odot}\boldsymbol{X}_{j}})$. The mass, momentum and energy components of inviscid fluxes are
\begin{equation} \label{ns-obgk-4}
\begin{array}{lll}
[\boldsymbol{GX}_{I} , \boldsymbol{GY}_{I} ]& = [{\left< \boldsymbol{\Psi} v_x  f_{0} \right>},{\left<  \boldsymbol{\Psi} v_y f_{0} \right>}]
&\displaystyle \equiv \int _{\mathbb{R}^{+} } \int _{\mathbb{R}^{2} }\boldsymbol{\Psi}  \vec{v}f_{0}  d\vec{v}d\mathbb{I}
\end{array}
\end{equation} 
where $v_{x}$ and $v_{y}$  are the Cartesian components of molecular velocity $\vec{v}$. The mass, momentum and energy components of viscous fluxes are obtained as
\begin{equation} \label{ns-obgk-5}
\begin{array}{ll}
[\boldsymbol{GX}_{V}, \boldsymbol{GY}_{V}] & =[-\sum_{j} \boldsymbol{{\Lambda}}_{j}^{x,\boldsymbol{\Psi}} \boldsymbol{\odot}\boldsymbol{X}_{j},-\sum_{j} \boldsymbol{{\Lambda}}_{j}^{y,\boldsymbol{\Psi}} \boldsymbol{\odot}\boldsymbol{X}_{j}]
\end{array}
\end{equation} 
where $\boldsymbol{{\Lambda}}_{j}^{x,\boldsymbol{\Psi}}$ and $\boldsymbol{{\Lambda}}_{j}^{y,\boldsymbol{\Psi}}$ are the macroscopic tensors associated with moment function, $\boldsymbol{\Psi}$ and its conjugate thermodynamic force due to thermal gradient vector and stress tensor, expressed as
\begin{equation} \label{ns-obgk-6}
\begin{array}{lll}
[\boldsymbol{{\Lambda}}_{j}^{x,\boldsymbol{\Psi}},\boldsymbol{{\Lambda}}_{j}^{y,\boldsymbol{\Psi}}]&=[{\left< \boldsymbol{\Psi} v_x \boldsymbol{\Upsilon}_{j}  \right>},{\left< \boldsymbol{\Psi} v_y \boldsymbol{\Upsilon}_{j}  \right>}]
&\displaystyle \equiv \int _{\mathbb{R}^{+} } \int _{\mathbb{R}^{2} }   \boldsymbol{\Psi} \vec{v} \boldsymbol{\Upsilon}_{j}  d\vec{v}d\mathbb{I}
\end{array}
\end{equation} 
where $\boldsymbol{{\Lambda}}_{j}^{x,\boldsymbol{\Psi}}$ and $\boldsymbol{{\Lambda}}_{j}^{y,\boldsymbol{\Psi}}$ are expressed as
\begin{equation} \label{ns-obgk-7}
\begin{array}{lllll}
\boldsymbol{{\Lambda}}_{j}^{x,\boldsymbol{\Psi}}&=\left [\boldsymbol{{\Lambda}}_{j}^{x,\psi_1},\cdots,\boldsymbol{{\Lambda}}_{j}^{x,\psi_m}\right ]^{T}&,&
\boldsymbol{{\Lambda}}_{j}^{y,\boldsymbol{\Psi}}&=\left [\boldsymbol{{\Lambda}}_{j}^{y,\psi_1},\cdots,\boldsymbol{{\Lambda}}_{j}^{y,\psi_m}\right ]^{T}
\end{array}
\end{equation} 
where $m$ are the total number of components of $\psi_i$ $\in$ $ \boldsymbol{\Psi}$. Because of isotropy due to rotational invariance of the collision operator the macroscopic tensor associated with stress tensor follows the symmetry relationship  by satisfying 
\begin{equation}\label{ns-obgk-8}
\begin{array}{lll}
{{\Lambda}_{\tau}^{x,\psi_{i}}}({r,s})={{\Lambda}_{\tau}^{x,\psi_{i}}}({s,r})&,&
{{\Lambda}_{\tau}^{y,\psi_{i}}}({r,s})={{\Lambda}_{\tau}^{y,\psi_{i}}}({s,r})
\end{array}
\end{equation} 
where $r$, $s$ are the component index of the tensor such that $s\ne r$. The viscous fluxes are obtained as
\begin{equation} \label{ns-obgk-9} 
\begin{array}{lll}
\boldsymbol{GX}_{V} & =-\sum_{j}\boldsymbol{\Lambda}_{j}^{x,\boldsymbol{\Psi}} \boldsymbol{\odot}\boldsymbol{X}_{j}
&\equiv\left(\begin{array}{c} {\begin{array}{c} {0} \\ { -\tau _{xx} } \end{array}} \\ {-\tau _{xy} } \\  {-u_{x}^{} \tau _{xx} -u_{y}^{} \tau _{xy}+q_{x} } \end{array}\right) 
\end{array}
\end{equation} 
\begin{equation} \label{ns-obgk-10}
\begin{array}{lll}
\boldsymbol{GY}_{V} & =-\sum_{j}\boldsymbol{\Lambda}_{j}^{y,\boldsymbol{\Psi}} \boldsymbol{\odot}\boldsymbol{X}_{j}
&\equiv\left(\begin{array}{c} {0 } \\ { -\tau _{xy} } \\ {\begin{array}{l} {-\tau _{yy} } \end{array}} \\ {-u_{x}^{} \tau _{xy} -u_{y}^{} \tau _{yy}  +q_{y} } \end{array}\right)
\end{array}
\end{equation} 
Here $u_{x}^{} $ and $u_{y}$  are the macroscopic velocity components in Cartesian frame, $\rho$ is the density,  $T$ is the static temperature and $p$ is the pressure which is calculated from equation of state.

Derivation of three dimensional case \cite{mahendra-thesis} shows that the heat flux vector is $\vec{q} = -\kappa \nabla {T}$ and tensor of viscous stresses $\boldsymbol{\Pi}$ is given as $\boldsymbol{\Pi}=\mu\left[ (\nabla \otimes \vec{u}) +(\nabla \otimes \vec{u})^{T} - \frac{2}{3} \boldsymbol{I} \nabla {\cdot}  \vec{u}\right] + \zeta \boldsymbol{I} \nabla {\cdot} \vec{u}$ where $\boldsymbol{I}$ is the rank-D identity invariant tensor, $\zeta$ is the coefficient of bulk viscosity expressed as  $\zeta= \mu\left(\frac{5}{3} - \gamma \right)$. From the expression of shear stress derived using kinetic theory it is evident that \emph{Stokes hypothesis is only valid for monatomic gases} as $\zeta=0$ for $\gamma=5/3$ otherwise, $\zeta > 0$ as $1<\gamma<5/3$. For polyatomic gas the concept of bulk viscosity term will change if elastic collisions are included e.g. when elastic and inelastic collision terms are of the same order Eucken correction to heat transfer coefficient and bulk viscosity may appear.
\begin{section}
{Non-equilibrium thermodynamics based Kinetic Scheme}\label{kscheme}
\end{section}
As the state update moves from one time step to another time step it generates entropy which is the product of the thermodynamic forces and its conjugate fluxes.  All the research in the development of upwind scheme revolves around the methodology of adding the correct dissipation or entropy e.g. if the amount of dissipation is too less then the solver will fail to capture shocks and if the amount of dissipation is too high then natural viscous behavior will get overshadowed. The correct amount of dissipation and its distribution for each thermodynamic force depends on the physical process through which state update passes, hence it is difficult to have a single monolithic solver operating across the regime from rarefied flow to hypersonic continuum flow. In precise words the state update of a solver has to follow the path laid down by non-equilibrium thermodynamics and address the issue of correct distribution of entropy for each thermodynamic force associated with stress tensor and thermal gradient vector. The research in the development of such a solver will follows a rigorous procedure based on principles of kinetic theory incorporating phenomenological theory of non-equilibrium thermodynamics. The following section of the paper presents kinetic flux vector splitting based on the Onsager-BGK kinetic model. The final aim is to have a single monolithic solver that mimics the physics by naturally adding the necessary dissipation for each thermodynamic force such that it is valid across wide range of fluid regimes.
\subsection{Kinetic Flux Vector Splitting Scheme}
Pullin \cite{pullin} initiated the development of kinetic schemes for compressible Euler system based on Maxwellian distribution using Equilibrium flux method (EFM). Deshpande \cite{deshpande} pioneered Kinetic Flux Vector Splitting (KFVS) scheme which was further developed by Mandal and Deshpande \cite{mandald} for solving Euler problems. Around the same period Perthame \cite{perthame1} developed kinetic scheme and Prendergast and Xu \cite{prenxu} proposed a scheme based on BGK simplification of the Boltzmann equation. The gas kinetic scheme of Xu \cite{prenxu, xu,xuhuang} uses method of characteristics and differs from the KFVS scheme mainly in the inclusion of particle collisions in the gas evolution stage. Chou and Baganoff \cite{choub} extended KFVS for Navier-Stokes-Fourier equations  by taking moments of the upwind discretized Boltzmann equation using first order distribution function.

Non-equilibrium thermodynamics (NET) based Kinetic Flux Vector Splitting (NET-KFVS) developed in the paper involves three steps : i) in the first step the Boltzmann equation is rendered into an upwind discretized form in terms of Maxwellian distribution and its perturbation term based on microscopic tensor and its conjugate thermodynamic forces , ii) in the second step inviscid or Euler fluxes are obtained by taking $\boldsymbol{\Psi}$ moments of split Maxwellian distribution, iii) in the third step viscous fluxes are obtained by taking moments of split microscopic tensors followed by full tensor contraction with its conjugate thermodynamic force to obtain upwind scheme for macroscopic conservation equations.
\subsection{NET-KFVS based on microscopic tensor splitting}
In order to illustrate NET-KFVS, consider two-dimensional Boltzmann equation in upwind form as follows
\begin{equation} \label{ons-k-1} 
f_{1}^{t+\Delta t} =f_{1}^{t} -\Delta t\left[
 \left(\frac{\partial v_{x} f_{1}^{+\cdot}}{\partial x}\right)^{t}+\left( \frac{\partial v_{x}  f_{1}^{-\cdot}}{\partial x}\right)^{t}
+\left(\frac{\partial v_{y} f_{1}^{\cdot +}}{\partial y}\right)^{t}+\left( \frac{\partial v_{y}  f_{1}^{\cdot -}}{\partial y}\right)^{t} 
\right]
\end{equation} 
In NET-KFVS the distribution function  at time $t+\Delta t$ in a fluid domain is constructed based on half range distribution at time $t$ where $f_{1}^{+ \cdot}$ is the half-range distribution function for $0 < v_{x} <\infty$ and $-\infty < v_{y} <+\infty$ and $f_{1}^{- \cdot}$ is the half-range distribution function for $-\infty < v_{x} <0$ and $-\infty< v_{y} <+\infty$. Similarly, $f_{1}^{\cdot +}$ is the half-range distribution function for $-\infty < v_{x} <+\infty$ and $0< v_{y} <+\infty$ and $f_{1}^{\cdot -}$ is the half-range distribution function for $-\infty < v_{x} <+\infty$ and $-\infty< v_{y} <0$. The upwind Boltzmann equation after taking $\boldsymbol{\Psi}$-moments simplifies to


\begin{equation} \label{ons-k-2} 
\begin{array}{ll}
\left\langle \boldsymbol{\Psi}, f_{0}^{t+\Delta t}\right\rangle &=\left\langle \boldsymbol{\Psi}, f_{0}^{t}\right\rangle
-\Delta t\left[
\begin{array}{l}
 {\rm \; \;} \frac{\partial \left\langle \boldsymbol{\Psi},v_{x} f_{0}^{\pm\cdot}\right\rangle}{\partial x}+ \frac{\partial -\sum_{j}  \boldsymbol{\Lambda}_{j}^{x,\boldsymbol{\Psi},\pm\cdot}  \odot \boldsymbol{X}_{j}}{\partial x} \\\\ 
+\frac{\partial \left\langle \boldsymbol{\Psi},v_{y} f_{0}^{\cdot \pm}\right\rangle}{\partial y}+\frac{\partial -\sum_{j}\boldsymbol{\Lambda}_{j}^{y,\boldsymbol{\Psi},\cdot\pm}  \odot \boldsymbol{X}_{j} }{\partial y}
\end{array}
\right]^{t}
\end{array}
\end{equation}
This leads to upwind equations in macroscopic form i.e. Navier-Stokes-Fourier equations in kinetic upwind form as follows
\begin{equation} \label{kfvs-1} 
\boldsymbol{U}^{t+\Delta t} = \boldsymbol{U}^{t}-\Delta t \left[
\begin{array}{l}
\left(\frac{\partial  \boldsymbol{GX}^{+}_{I}}{\partial x}+\frac{\partial  \boldsymbol{GX}^{+}_{V}}{\partial x}\right)_{\Delta x < 0}+\left(\frac{\partial  \boldsymbol{GX}^{-}_{I}}{\partial x}+\frac{\partial  \boldsymbol{GX}^{-}_{V}}{\partial x}\right)_{\Delta x > 0}\\
+\left(\frac{\partial  \boldsymbol{GY}^{+}_{I}}{\partial y}+\frac{\partial  \boldsymbol{GY}^{+}_{V}}{\partial y}\right)_{\Delta y < 0}+\left(\frac{\partial  \boldsymbol{GY}^{-}_{I}}{\partial y}+\frac{\partial  \boldsymbol{GY}^{-}_{V}}{\partial y}\right)_{\Delta y >0}
\end{array}
\right]^{t}
\end{equation} 
The upwinding is enforced by stencil sub-division such that derivative of positive split fluxes are evaluated using negative split stencil \cite{mahslkns}. The inviscid part of the split flux is defined as
\begin{equation} \label{kfvs-2} 
\begin{array}{l}
\boldsymbol{GX}^{\pm}_{I} = \displaystyle \left\langle \boldsymbol{\Psi}, v_{x} f_{0}^{\pm\cdot}\right\rangle = \int_{\mathbb{R}^{+}}\int_{\mathbb{R}}\int_{\mathbb{R}^{\pm}} \boldsymbol{\Psi} v_{x} f_{0} dv_{x} dv_{y} d\mathbb{I}\\
\boldsymbol{GY}^{\pm}_{I} = \displaystyle \left\langle \boldsymbol{\Psi}, v_{y} f_{0}^{\cdot\pm}\right\rangle = \int_{\mathbb{R}^{+}}\int_{\mathbb{R}^{\pm}}\int_{\mathbb{R}} \boldsymbol{\Psi} v_{y} f_{0} dv_{x} dv_{y} d\mathbb{I}
\end{array}
\end{equation}
The viscous part of the split flux is defined as
\begin{equation} \label{gxv-1} 
\begin{array}{ll}
\boldsymbol{GX}_{V}^{\pm} & =-\sum_{j}\boldsymbol{\Lambda}_{j}^{x,\boldsymbol{\Psi,{\pm\cdot}}} \boldsymbol{\odot}\boldsymbol{X}_{j}=-(\boldsymbol{\Lambda}_{\tau}^{x,\boldsymbol{\Psi,{\pm\cdot}}}\boldsymbol{:}\boldsymbol{X}_{\tau}+\boldsymbol{\Lambda}_{q}^{x,\boldsymbol{\Psi,{\pm\cdot}}}\boldsymbol{\cdot} \boldsymbol{X}_{q})\\
\boldsymbol{GY}_{V}^{\pm} & =-\sum_{j}\boldsymbol{\Lambda}_{j}^{y,\boldsymbol{\Psi,{\cdot\pm}}} \boldsymbol{\odot}\boldsymbol{X}_{j}=-(\boldsymbol{\Lambda}_{\tau}^{y,\boldsymbol{\Psi,{\cdot\pm}}}\boldsymbol{:}\boldsymbol{X}_{\tau}+\boldsymbol{\Lambda}_{q}^{y,\boldsymbol{\Psi,{\cdot\pm}}}\boldsymbol{\cdot} \boldsymbol{X}_{q})
\end{array}
\end{equation} 
for example viscous split mass flux component evaluated using $\psi_{1}$ $\in$ $\boldsymbol{\Psi}$ is
\begin{equation} \label{gxv-m} 
\begin{array}{l}
GX^{\pm}_{V}(\psi_1)=\pm\frac{ Exp(-\beta u_{x}^2)}{\sqrt{\pi}}{\frac{\rho}{4 p\sqrt{\beta}}\left\{ 2 u_{x} \beta q_{x}\frac{(\gamma-1)}{\gamma}+\tau_{xx}\right\} }
\end{array}
\end{equation}
It contains features of non-equilibrium thermodynamics due to cross coupling of shear stress tensor and thermal gradient vector. Similarly, momentum and energy flux will also contain terms due to cross coupling of shear stress tensor and thermal gradient vector defined by macroscopic split tensors $\boldsymbol{\Lambda}_{j}^{x,\boldsymbol{\Psi,{\pm\cdot}}}$ and $\boldsymbol{\Lambda}_{j}^{y,\boldsymbol{\Psi,{\cdot\pm}}}$ given as
\begin{equation}\label{psi-obgk-1}
\begin{array}{l}
\boldsymbol{\Lambda}_{j}^{x,\boldsymbol{\Psi,{\pm\cdot}}}=\left\langle \boldsymbol{\Psi},v_{x}  {  \boldsymbol{\Upsilon}_{j}^{\pm\cdot} }\right\rangle
\displaystyle \equiv \int_{\mathbb{R}^{+}} \int_{\mathbb{R}}\int_{\mathbb{R}^{\pm}} \boldsymbol{\Psi} v_{x}  \boldsymbol{\Upsilon}_{j}  dv_{x} dv_{y}d\mathbb{I}\\
\boldsymbol{\Lambda}_{j}^{y,\boldsymbol{\Psi,{\cdot\pm}}}=\left\langle \boldsymbol{\Psi},v_{y}  {  \boldsymbol{\Upsilon}_{j}^{\cdot \pm} }\right\rangle
\displaystyle \equiv \int_{\mathbb{R}^{+}}\int_{\mathbb{R}^{\pm}} \int_{\mathbb{R}} \boldsymbol{\Psi} v_{y}  \boldsymbol{\Upsilon}_{j}  dv_{x} dv_{y}d\mathbb{I}
\end{array}
\end{equation}
The components of split macroscopic tensors  $\boldsymbol{{\Lambda}}_{j}^{x,\boldsymbol{\Psi,{\pm\cdot}}}$ $=$ $\left [\boldsymbol{{\Lambda}}_{j}^{x,\psi_i,{\pm\cdot}},\cdots,\boldsymbol{{\Lambda}}_{j}^{x,\psi_i,{\pm\cdot}}\right ]^{T}$  and  $\boldsymbol{{\Lambda}}_{j}^{y,\boldsymbol{\Psi,{\cdot\pm}}}$ $=$ $\left [\boldsymbol{{\Lambda}}_{j}^{y,\Psi_i,{\cdot\pm}},\cdots,\boldsymbol{{\Lambda}}_{j}^{y,\Psi_i,{\cdot\pm}}\right ]^{T}$ are defined for each $\psi_i$ $\in$ $ \boldsymbol{\Psi}$ $=$ $\left[1,\vec{v},\mathbb{I}+\frac{1}{2} v^{2}\right]^{T} $. The split macroscopic tensors for each moment component associated with shear stress tensor follow symmetry relationship because of isotropy due to rotational invariance of the collision operator by satisfying 
\begin{equation}\label{psi-obgk-2}
\begin{array}{lll}
{\Lambda}_{\tau}^{x,\psi_{i},\pm \cdot}({r,s})={\Lambda}_{j}^{x,\psi_{i},\pm \cdot}({s,r})&,&
{\Lambda}_{\tau}^{y,\psi_{i},\cdot\pm }({r,s})={\Lambda}_{j}^{y,\psi_{i},\cdot\pm }({s,r})
\end{array}
\end{equation} 
where $r$, $s$ are the component index of the tensor such that $r\ne s$. Appendix \ref{splitflux} gives the expressions of these  split macroscopic tensors. Expressions for split macroscopic tensors for three dimensional case is given in \cite{mahendra-thesis}.
\subsection{NET-KFVS in its variance reduced form}
\begin{figure}\centering
(a){\includegraphics[scale=0.2,bb=0 0 1051 497]{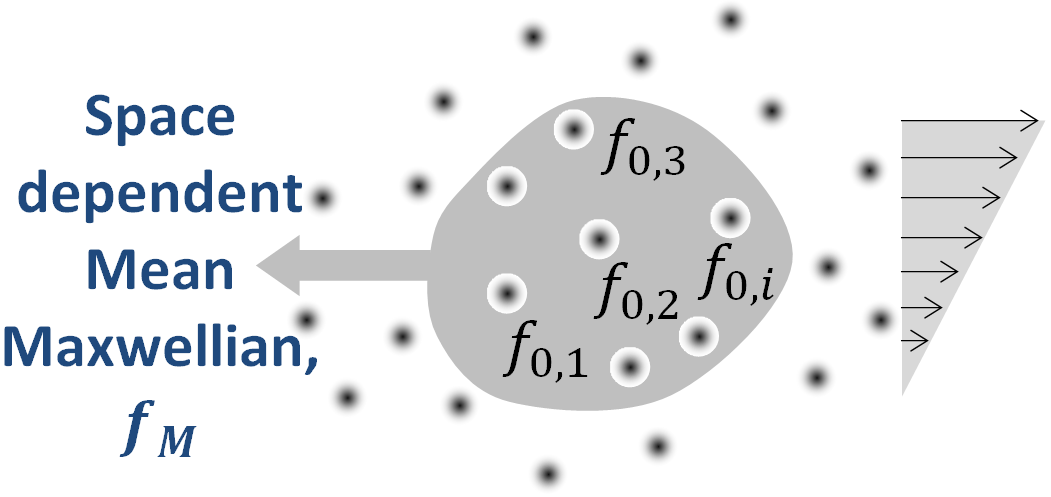}}(b){\includegraphics[scale=0.2,bb=0 0 1071 517]{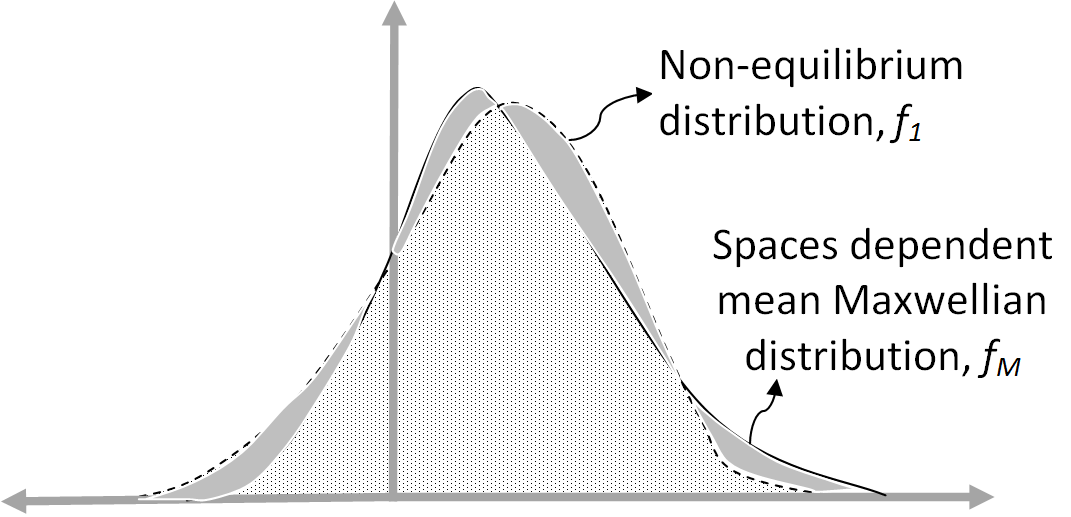}}\caption{ \label{vrkfvs} Kinetic scheme based on variance reduction approach: (a)space dependent mean Maxwellian in a fluid domain, (b) shaded portion shows perturbation of equilibrium and non-equilibrium variations over the space dependent Maxwellian.}
\end{figure}

The shear amplitude for any fluid dynamic problem should be observed in the correct frame of reference with variance reduction approach. There are flows which may have high velocity gradients but may still be governed by Maxwellian distribution for example continuum Couette flow due to flat plate moving with high speed. In most of the fluid dynamic problems the flow is a perturbation over a space dependent Maxwellian as shown in figure \ref{vrkfvs}(a), non-equilibrium first order distribution function for such a case is
\begin{equation} \label{net-obgk-1}
f_{1}=\Delta f - \sum_{j}{  \boldsymbol{\Upsilon}_{j} \odot \boldsymbol{\bar{X}}_{j}} + \mathcal{O}(\Delta t_{R(j)})
\end{equation}
where $f_{M}$ is the space dependent mean Maxwellian, $\Delta f=f_{0}-f_{M}$, $\boldsymbol{\bar{X}}_{j}$ is thermodynamic force based on velocity and temperature field  relative to  $f_{M}$ and $\Delta t_{R(j)}$ $=$ $t_{R(j),f_{0}}-t_{R(j),f_{M}}$. When non-equilibrium effects are not very dominant the term  $\Delta t_{R(j)}$ can be neglected and first order distribution approximates as $f_{1} \approx\Delta f - \sum_{j}{\boldsymbol{\Upsilon}_{j} \odot \boldsymbol{\bar{X}}_{j}}$. Boltzmann equation in this perturbative form becomes
\begin{equation}\label{net-obgk-2}
\frac{\partial \Delta f}{\partial t}+\frac{\partial \vec{v} \Delta f}{\partial \vec{x}}- \sum_{j}\frac{\partial \vec{v}\left({  \boldsymbol{\Upsilon}_{j} \odot \boldsymbol{\bar{X}}_{j}}\right)}{\partial \vec{x}}=0
\end{equation} 
Taking $\boldsymbol{\Psi}$ moments of the resulting variant of Boltzmann equation leads to Navier-Stokes-Fourier equations based on Variance Reduction Kinetic Flux Vector Splitting (VRKFVS) \cite{mahvrkfvs} as follows 
\begin{equation} \label{net-obgk-3}
\displaystyle \frac{\partial }{\partial t} \left(\Delta \boldsymbol{U}\right) +\frac{\partial }{\partial x} \left[\Delta \left(\boldsymbol{GX}_{I}^{\pm } \right)+\left(\boldsymbol{GX}_{V}^{\pm } \right)_{\Delta } \right]+\frac{\partial }{\partial y} \left[\Delta \left(\boldsymbol{GY}_{I}^{\pm } \right)+\left(\boldsymbol{GY}_{V}^{\pm } \right)_{\Delta } \right] =0 
\end{equation} 
where $\Delta \boldsymbol{U}=\boldsymbol{U}-\boldsymbol{U}_{M} $ is the deviation of the state update vector $\boldsymbol{U}$ over $\boldsymbol{U}_{M} $ based on space dependent Maxwellian distribution,$f_{M} $. The inviscid fluxes are
\begin{equation} \label{net-obgk-4} 
\begin{array}{l}
\Delta \left(\boldsymbol{GX}_{I}^{\pm } \right)= \displaystyle \left\langle \boldsymbol{\Psi}, v_{x} (f_{0}^{\pm\cdot}-f_{M}^{\pm\cdot})\right\rangle = \boldsymbol{GX}_{I}^{\pm } -\boldsymbol{GX}_{I,M}^{\pm }  \\
\Delta \left(\boldsymbol{GY}_{I}^{\pm } \right)= \displaystyle \left\langle \boldsymbol{\Psi}, v_{y} (f_{0}^{\cdot\pm}-f_{M}^{\cdot\pm})\right\rangle = \boldsymbol{GY}_{I}^{\pm } -\boldsymbol{GY}_{I,M}^{\pm } 
\end{array}
\end{equation}
where $\boldsymbol{GX}_{I,M}^{\pm }$ and $\boldsymbol{GY}_{I,M}^{\pm }$ are the inviscid split fluxes based on the Maxwellian distribution, $f_{M} $ associated with the chosen state of equilibrium. The viscous fluxes$\left(\boldsymbol{GX}_{V}^{\pm } \right)_{\Delta }$ and $\left(\boldsymbol{GY}_{V}^{\pm } \right)_{\Delta } $ are computed based on thermodynamic force $\boldsymbol{\bar{X}}$ relative to chosen Maxwellian $f_{M}$ as follows
\begin{equation} \label{net-obgk-5}
\begin{array}{lllll}
\left(\boldsymbol{GX}_{V}^{\pm } \right)_{\Delta } & =-\sum_{j}\boldsymbol{\Lambda}_{j}^{x,\boldsymbol{\Psi,{\pm\cdot}}} \boldsymbol{\odot}\boldsymbol{\bar{X}}_{j}&,&
\left(\boldsymbol{GY}_{V}^{\pm } \right)_{\Delta } & =-\sum_{j}\boldsymbol{\Lambda}_{j}^{y,\boldsymbol{\Psi,{\cdot\pm}}} \boldsymbol{\odot}\boldsymbol{\bar{X}}_{j}
\end{array}
\end{equation} 
In variance reduction approach we solve only for the perturbation of equilibrium and non-equilibrium variations over the space dependent Maxwellian as shown in figure \ref{vrkfvs}(b). This method basically evaluates the \textit{variance-reduced} form of the kinetic model similar to the variance reduction technique of Baker and Hadjiconstantinou \cite{hadjivr1} used in direct simulation Monte Carlo (DSMC). This variant of NET-KFVS based on variance reduction form of BGK-Boltzmann equation was found to be useful in capturing creeping flows, sub-sonic flows and weak secondary flow in a strong flow field environment.
\subsection{NET based Kinetic Particle Method} 
The exact solution \cite{chapman} of Boltzmann equation with BGK model $J(f,f_0)=-(f-f_0)/t_{R}$ is
\begin{equation}\label{net-obgk-6}
f = \int_{\mathbb{R}^{+}} Exp(-\acute{t}/t_R)f_{0}(\vec{v},\vec{x}-\vec{v}\acute{t},t-\acute{t})t_R^{-1}d\acute{t}
\end{equation}
Physically it  means that along the trajectory in the phase space, particles are replaced exponentially by particles in equilibrium  with characteristics time $t_{R}$. Gas kinetic scheme of Xu \cite{xuhuang} updates the distribution using this method of characteristics using BGK and Shakhov's model. Macrossan's RTSM (relaxation time simulation method) \cite{macro,wangmacro} uses method of characteristics and  BGK model to update distribution function, $f(\Delta t)$  after time $t=\Delta t$ as
\begin{equation}\label{net-obgk-7}
f(\Delta t) =  Exp(-\Delta t/t_{R}) f(0) + (1-Exp(-\Delta t/t_{R}))f_{0}
\end{equation}
where $f(0)=f(t=0)$ is the initial distribution established by the streaming or convection phase just before the simulation of collision phase. In the collision phase the distribution function relaxes towards equilibrium Maxwellian distribution, $f_{0}$  after sufficient collisions with a time constant $t_{R}$. This pseudo-collision step or the relaxation step actually carries out the step of redistribution of the particles such that the resulting distribution is a mixture of the initial distribution, $f({0})$ and the final equilibrium distribution, $f_{0}$.\\
\begin{figure}\centering
{\includegraphics[width=\textwidth,bb=0 0 1618 526]{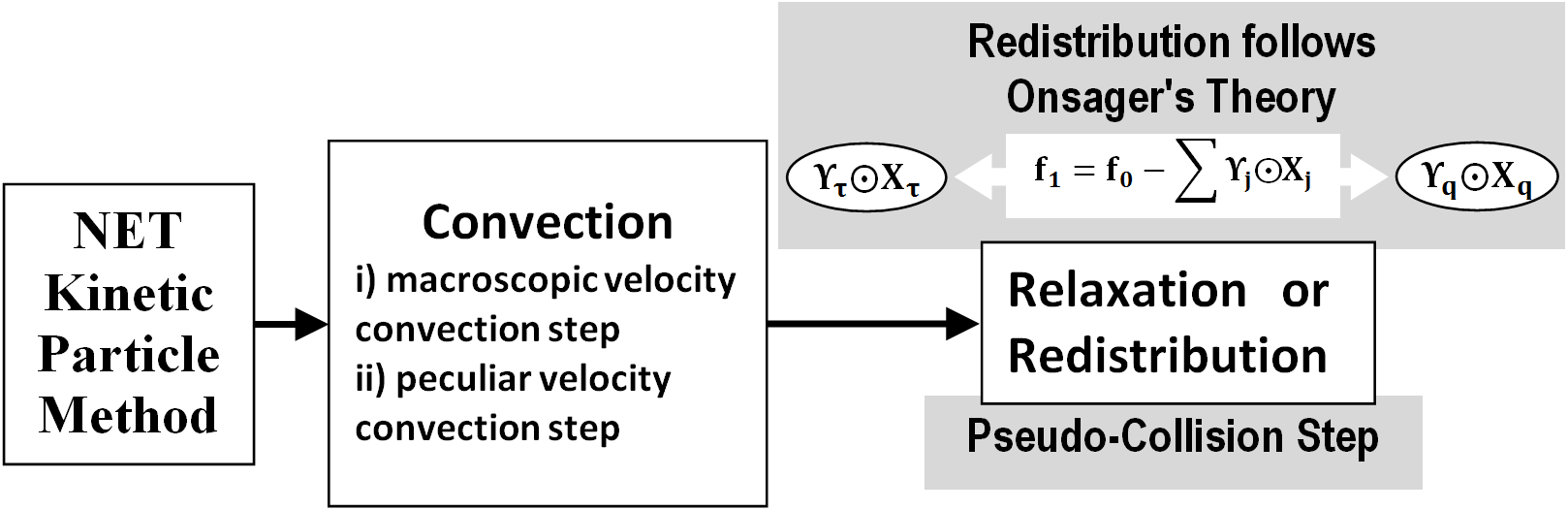}}
\caption{ \label{kpm} Kinetic particle based method.}
\end{figure}
Non-equilibrium thermodynamics based kinetic particle method like RTSM is composed of two steps i) Convection step, ii) Collision or redistribution step. The convection phase happens in two sub-steps of macroscopic velocity convection followed by peculiar velocity convection.  In the first sub-step the whole distribution travels with its orderly, deterministic macroscopic velocity $\vec{u}$ followed by the second sub-step of peculiar velocity convection where distribution is converted into particles and each particle travels in a random multi-directional path simulating peculiar velocity $\vec{c}$ $=$ $\vec{v}-\vec{u}$ following normal distribution, $N(0,1/(2 \beta))$ with zero mean and variance equal to $1/(2 \beta)$. The collision phase or redistribution phase uses the Onsager-BGK kinetic model which redistributes the particles for each thermodynamic force, using Macrossan's RTSM approach as shown in Figure \ref{kpm}. After time $t=\Delta t$ the distribution function, $f(\Delta t)$  is obtained as a mixture of the initial distribution, $f({0})$ which is the distribution obtained just after convection and the final equilibrium distribution, $f_{0}$ which is obtained by extracting a Maxwellian, this step is done for each thermodynamic force, $\boldsymbol{X}_{j}$ when all the other thermodynamic forces are absent such that 
\begin{equation}\label{net-obgk-7}
f(\Delta t) =  \sum_j\left[\alpha_j f(0)+(1-\alpha_j)(f_{0})\right]_{\boldsymbol{X}_{i}=0,i\ne j}
\end{equation}
where $\alpha_j=Exp(-\Delta t/t_{R(j)})_{X_{i}=0,i\ne j}$ and $t_{R(j)}$ is the relaxation time for $\boldsymbol{X}_{j}$. 

\begin{figure}\centering
{\includegraphics[scale=0.3,bb=0 0 820 749]{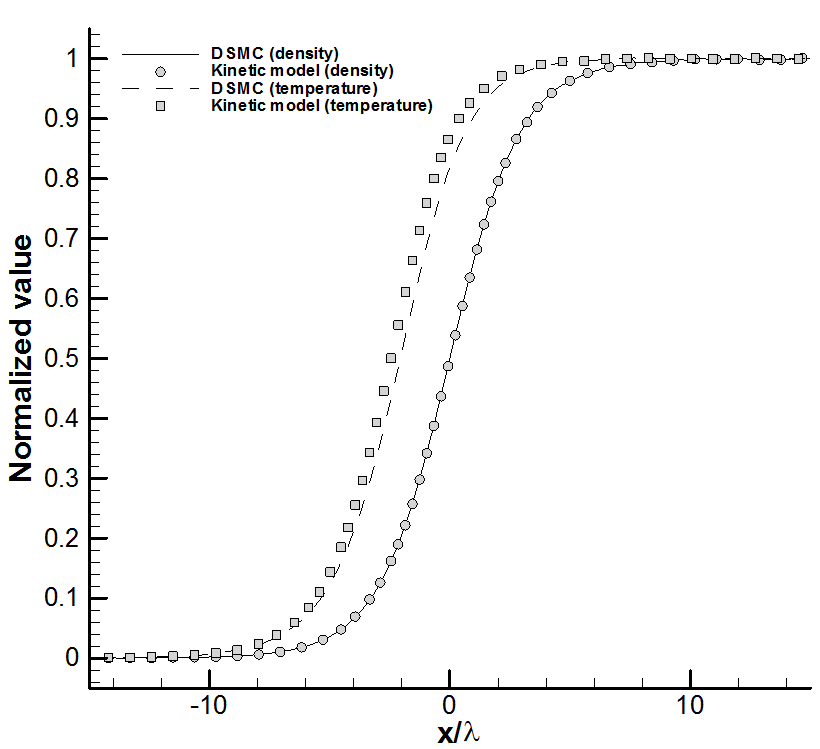}}
\caption{ \label{ssm}Argon shock structure for shock of  Mach 2.0 at $293$ K using kinetic particle method and DSMC.}
\end{figure}

Consider Argon shock structure simulation for shock of  Mach 2.0 at $293$ K with mean free path $\lambda$ $=$ $1.3442$ $\times$ $10^{-2}$ m and Prandtl number, Pr $=$ $2/3$. The computational domain spans $-0.3$ m to $0.3$ m, kinetic particle method is implemented with an initial uniform spread of $300$ nodes using the viscosity-temperature relationship $\mu=\mu_{ref}\left(\frac{T}{T_{ref}} \right)^{\omega}$  with $\omega$ $=$ $0.81$. Figure \ref{ssm}  shows the normalized density and temperature profile in argon for Mach 2.0 shock using kinetic particle method. The simulated results are compared with code \emph{DSMC1S} provided by Bird \cite{bird}.
The normalized density $\bar{\rho}$ and normalized temperature $\bar{T}$ are defined as
\begin{equation}\label{ss-1}
\begin{array}{ll}
\displaystyle \bar{\rho} = \frac{\rho - \rho_{u}}{\rho_{d}-\rho_{u}}, &\displaystyle \bar{T} = \frac{T - T_{u}}{T_{d}-T_{u}}
\end{array}
\end{equation}
\begin{figure}\centering
{\includegraphics[scale=0.4,bb=0 0  1258 528]{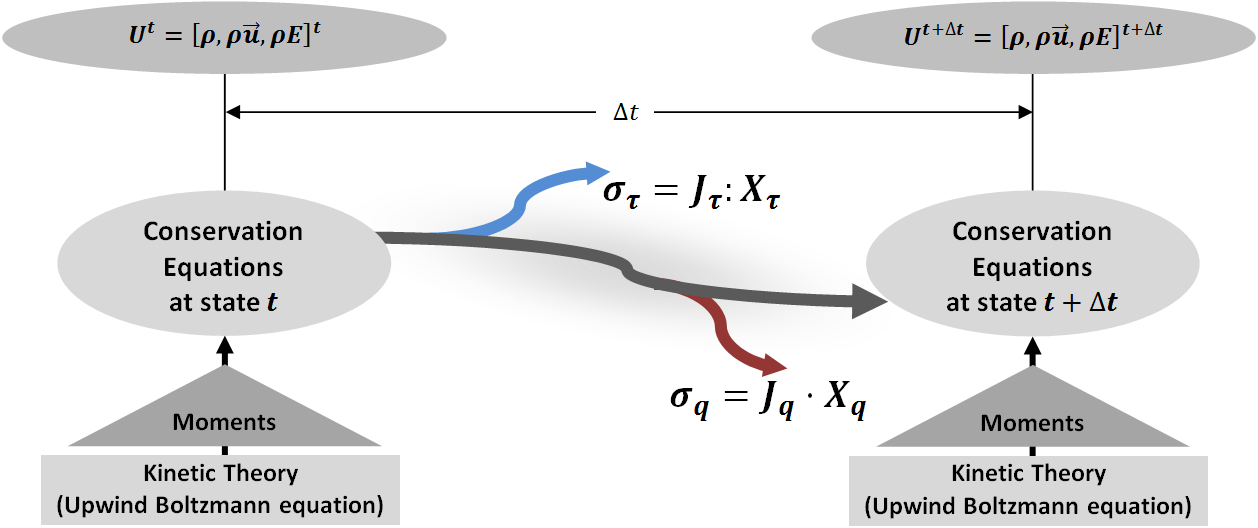}} 
\caption{ \label{solnet} State update of the kinetic method follows non-equilibrium thermodynamics.}
\end{figure}
\begin{figure}
\begin{minipage}
{.5\textwidth}\centering
{\includegraphics[scale=0.45,bb=0 0  637 784]{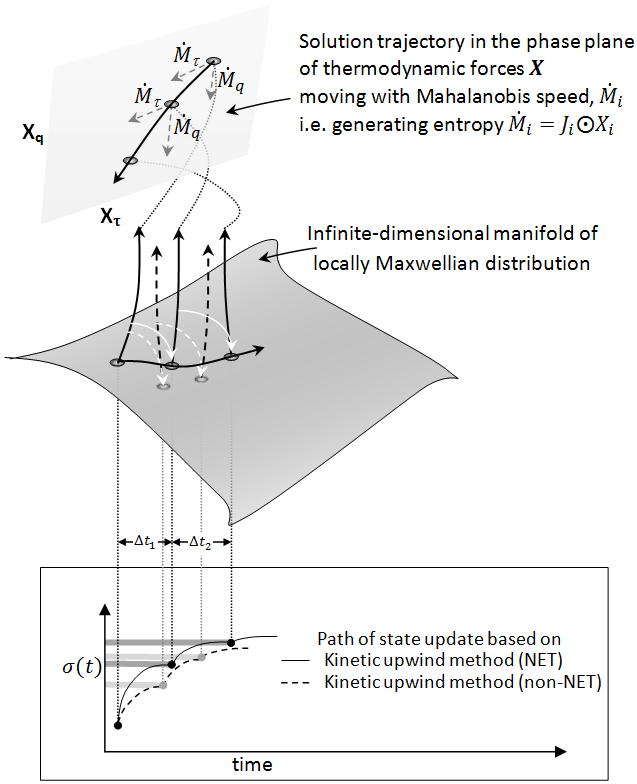}}
\caption{ \label{manmep-a}Evolution of flow and entropy generation for non-equilibrium thermodynamics (NET) and non-NET based kinetic upwind method.}
\end{minipage}
\begin{minipage}
{.05\textwidth}\centering
\end{minipage}
\begin{minipage}
{.45\textwidth}\centering
{\includegraphics[scale=0.44,bb=0 0  561 799]{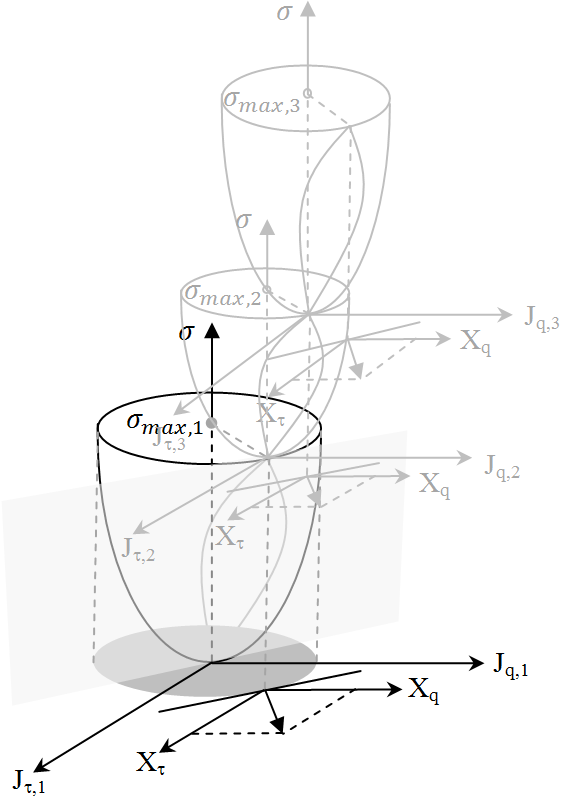}} 
\caption{ \label{manmep-b}Evolution of linear irreversible process and its dissipative surface of entropy for a case with two thermodynamic fluxes and forces.}
\end{minipage}
\end{figure}
\subsection{Linkage of non-equilibrium thermodynamics with kinetic scheme}
As the state update moves from one time step to the next, the path of evolution follows non-equilibrium thermodynamics i.e. maximizes the entropy under the constraint imposed due to conservation laws by satisfying the Onsager's variational principle. 
Kinetic model and kinetic scheme that satisfies the Onsager's variational principle will generate entropy for each thermodynamic force e.g. $\sigma_{\tau}$ $=$ $\boldsymbol{J}_{\tau}$ $\boldsymbol{:}$ $\boldsymbol{X}_{\tau}$ with respect to thermodynamic force associated with the stress tensor and  entropy, $\sigma_{q}$$=$$\boldsymbol{J}_{q}$ $\boldsymbol{\cdot}$ $\boldsymbol{X}_{q}$ with respect to thermodynamic force associated with the thermal gradient vector as shown in Figure \ref{solnet}. Linkage with non-equilibrium thermodynamics ensures \emph{ correct division of entropy generation for each thermodynamic force as the state update moves from one conservation state to another.}

Figure \ref{manmep-a}  shows the schematic picture of the evolution of flow and entropy generation for two different kinetic schemes : one following non-equilibrium thermodynamics (NET) and other non-NET based. Figure \ref{manmep-a} shows that at every time step due to thermodynamic forces the non-equilibrium flow trajectory leaves the equilibrium surface represented by infinite-dimensional manifold of locally Maxwellian distribution, as a consequence it generates irreversibility and entropy while relaxing back to equilibrium i.e. jumping back to the manifold. Each kinetic scheme will have its own path or trajectory of evolution of flow as well as entropy generation.  As shown in figure \ref{manmep-a}, this trajectory can also be observed in the phase plane of thermodynamic forces where it moves with Mahalanobis speed $\dot{M}_{i}$ $=$ $\sigma_i(f,f_0)$ i.e. generating entropy for each thermodynamic force $\boldsymbol{X}_{i}$ following Onsager's relationship. The movement of this trajectory can also be interpreted in terms of entropy production in the flux space represented as surface $\sigma(\boldsymbol{J}_{i},\boldsymbol{J}_{k})$. Figure \ref{manmep-b} shows a linear irreversible process and a dissipative surface of entropy for a case with two thermodynamic fluxes and forces \cite{mart-sele}. As the flow trajectory leaves the manifold the flux $\boldsymbol{J}_{i}$ is generated corresponding to its conjugate thermodynamic force $\boldsymbol{X}_{i}$ which is orthogonal to surface $\sigma(\boldsymbol{J}_{i},\boldsymbol{J}_{k})$ intersected by the plane $\sum_{i} \boldsymbol{J}_{i} \odot \boldsymbol{X}_{i}$.

It is the split flux not the full flux which participates in actual physical process. Split fluxes based on Onsager-BGK model also follow non-equilibrium thermodynamics, for example entropy based on the split non-equilibrium flux $\sigma^{\pm}(\boldsymbol{J}_{i},\boldsymbol{X}_{i})$ is expressed as
\begin{equation} \label{ons-split-net} 
\sigma^{\pm}(\boldsymbol{J}_{i},\boldsymbol{X}_{i})= \sum_{i} \boldsymbol{J}_{i}^{\pm} \odot \boldsymbol{X}_{i}
\end{equation} 
where non-equilibrium split flux $\boldsymbol{J}_{i}^{\pm}$ are kinetic non-equilibrium split flux conjugate to its thermodynamic force $\boldsymbol{X}_{i}$. Non-equilibrium split fluxes involve terms of higher moments $\psi_j$ $\notin$ $\boldsymbol{\Psi}$, thus it modifies the idea of extended thermodynamics, refer appendix \ref{momex} for further details.
\begin{section}
{Kinetic wall boundary condition}\label{kwallbc}
\end{section}
Wall boundary condition is an important part in simulation of fluid flow. Experimental studies as well as theoretical analysis corroborate the efficacy of a given boundary condition. No-slip and slip condition at the wall provide a realistic boundary condition used for the solution of Navier-Stokes-Fourier equations in the continuum and rarefied regime respectively.  The slip flow simulation using the continuum solver can be carried out either by using slip models or by implementing kinetic wall boundary condition. The approach of continuum solver coupled with Maxwell's velocity slip boundary condition \cite{maxwell} and von Smoluchowski's temperature jump boundary condition \cite{kennard} is the most popular as it is computationally the least expensive. Velocity slip and temperature jump can also be derived using linearized Grad's moment method which is based on expansion of distribution function around local Maxwellian in terms of Hermite tensor polynomials. Patterson \cite{patt} carried out derivation of velocity slip and temperature jump using Grad's moment method. Patterson's velocity slip condition is similar to Maxwell's velocity slip for curved surface \cite{lockerby04}. However, Patterson's temperature boundary condition \cite{patt} is approximate as it is derived under assumption of negligible tangential variation and negligible magnitude of velocity compared to thermal speed of the molecules. Accurate derivation of velocity slip and temperature jump using Grad's moment method requires at least thirteen equations \cite{struch-torr}. Li \textit{et al} \cite{lixu} have used gas kinetic upwind method to carry out slip flow modeling for hypersonic flows. Agarwal \textit{et al.} \cite{agarwalb}, Bao and Lin \cite{bao}, and Lockerby and Reese \cite{lockerbyr} have used Burnett equations coupled with slip models. Researchers \cite{mieu,kxu,liz} have used kinetic wall boundary condition obtained using the distribution function. It can be seen that there are large number of slip models existing in the literature, each with its own geometry specific slip coefficients and range of validity in Knudsen regime. Most of these slip models are for simple micro-channel flows and are valid only in slip flow regime when there are insignificant fluid dynamic variations in the tangential direction. Literature review has already revealed that the first order slip velocity not only depends on the velocity gradient in the normal direction but also on the pressure gradient in the tangential flow direction\cite{shenk}. Slip model based approaches as well as kinetic theory based methods in general have ignored the issues of non-equilibrium thermodynamics. One of the motivations of the paper is to derive a comprehensive wall boundary condition which satisfies Onsager's relationship and can simulate both continuum and rarefied slip flow within Navier-Stokes-Fourier equations in order to avoid extremely costly multi-scale simulation.
\subsection{Derivation of slip conditions for negligible tangential gradients}
The net first order distribution function at time $t$ in terms of accommodation coefficient, $\sigma$ for Maxwell gas-surface interaction model can be written as 
\begin{equation}  \label{ons-slip-1} 
\begin{array}{ll}
f_{\Sigma}(\vec{v},\mathbb{I},t) = &
\begin{array}{lll}
f_{I}(\vec{v},\mathbb{I},t) &for&   \vec{i}_n \cdot \vec{v}<0 \\
(1-\sigma )f_{R}(\vec{v},\mathbb{I},t)+\sigma f_{0,W} (\vec{v},\mathbb{I},t) &for& \vec{i}_n \cdot \vec{v}>0 
\end{array}
\end{array}
\end{equation}
where $f_{1,\Sigma } (\vec{v} ,\mathbb{I},t)$ is the total distribution resulting due to Maxwell model, $f_{1,I} (\vec{v} ,\mathbb{I},t)$ and $f_{1,R} (\vec{v} ,\mathbb{I},t)$ are the incident and  specularly reflected first order distribution respectively and $f_{0,W} (\vec{v},\mathbb{I},t)$ is the diffuse reflected Maxwellian distribution evaluated using wall conditions and mass conservation. The specularly reflected first order distribution $f_{1,R} (\vec{v} ,\mathbb{I},t)$  is written as  
\begin{equation*} \label{GrindEQ__82_} 
 f_{1,R} (\vec{v} ,\mathbb{I},t)=f_{I}(\vec{v}-2 \vec{i}_n\vec{i}_n \cdot \vec{v},\mathbb{I},t)
\end{equation*}
molecules are reflected away from the boundary ($\vec{i}_n \cdot \vec{v}$ $>$ $0$) where $\vec{i}_n$ is the surface normal. The specular reflected component of the distribution is constructed using $\vec{v}-2 \vec{i}_n\vec{i}_n \cdot \vec{v}$ i.e. with reverse sign of normal component of velocity. 

The expression of density jump, Maxwell velocity slip and von Smoluchowski's temperature jump available in the literature are derived under conditions of negligible fluid dynamical variations in the tangential directions. This subsection revisits the derivation of  density jump, Maxwell velocity slip and von Smoluchowski temperature jump using the non-equilibrium thermodynamics based distribution function obtained using Onsager-BGK kinetic model under conditions of negligible fluid dynamical variations in the tangential directions.
\subsubsection{Derivation of density jump}
Let us consider an infinitesimal area $ds$ on the surface of the wall and an elementary strip of gas extending above the wall in the $y$-direction from the elementary area $ds$ such that the strip reaches the imaginary plane set at  $y=\Delta y$ . The conservation of mass at the wall for two-dimensional geometry can be written as 
\begin{equation}  \label{ons-slip-3}
{\displaystyle
{\int_{\mathbb{R^{+}}}\int_{\mathbb{R^{-}}}\int_{\mathbb{R}}}v_{y}f_{1,I} dv_{x}dv_{y}d\mathbb{I}}
{\displaystyle + {\int_{\mathbb{R^{+}}}\int_{\mathbb{R^{+}}}\int_{\mathbb{R}}}v_{y}f_{0,W} dv_{x}dv_{y}d\mathbb{I}=0}
\end{equation}
where $f_{1,I}$ is the incident Chapman-Enskog distribution function and $f_{0,W}$ is the Maxwellian distribution function based on the wall conditions. Solving the mass conservation using first order distribution at the boundary $f_{1,I}$ $=$ $f_{0,I}$ $-$ $ \sum_{j}{   \boldsymbol{\Upsilon}_{I,j} \boldsymbol{\odot}  \boldsymbol{X_{j}}}$ leads to density jump at the wall as
\begin{equation} \label{ons-slip-4}
\rho_{w}=\frac{\rho \sqrt{\beta_{w}}}{\sqrt{\beta}}\left[1-\frac{\tau_{yy}}{2p}\right]=\rho \sqrt{\frac{T}{T_{w} } } \left(1-\frac{\tau _{yy} }{2p} \right)
\end{equation}
where $\beta_{w} = {1}/{2 R T_{w}}$ is based on wall temperature $T_{w}$.
\subsubsection{Derivation of velocity slip}
The momentum flux, $\mathbb{M}$ passing through the imaginary plane set at  $y=\Delta y$ based on linearized distribution $f_{1}$ is given as
\begin{equation} \label{ons-slip-5}
{\displaystyle \mathbb{M}_{y=\Delta y} }
{ \displaystyle =\int_{\mathbb{R^{+}}}\int_{\mathbb{R}}\int_{\mathbb{R}}(f_{0}-\sum_{j}{  \boldsymbol{\Upsilon}_{j} \boldsymbol{\odot}\boldsymbol{X}_{j}})
v_{x}v_{y}dv_{x}dv_{y}d\mathbb{I}}
{=\rho u_{x}u_{y}-\tau_{xy}}
\end{equation}
The contributing momentum component  at the wall where $y$$=$$0$ is expressed as
\begin{equation} \label{ons-slip-6}
\begin{array}{ll}
{\displaystyle \mathbb{M}_{y=0}}&
{ \displaystyle =\sigma \left[
\begin{array}{l}{
 {\displaystyle \int_{\mathbb{R^{+}}}\int_{\mathbb{R^{-}}}\int_{\mathbb{R}}}
(f_{0,I}-\sum_{j}{  \boldsymbol{\Upsilon}_{I,j} \boldsymbol{\odot}\boldsymbol{X}_{j}})v_{x}v_{y}dv_{x}dv_{y}d\mathbb{I}}\\
{\displaystyle+{\int_{\mathbb{R^{+}}}\int_{\mathbb{R^{+}}}\int_{\mathbb{R}}}f_{0,W}v_{x}v_{y}dv_{x}dv_{y}d\mathbb{I}
}
\end{array}
\right ]}
\end{array}
\end{equation}
The momentum flux on the infinitesimal surface $ds$ at the wall $y=0$ can be equated with the momentum flux on the infinitesimal surface $ds$ as $\Delta y$$\rightarrow$$0$. The conservation of momentum flux gives the velocity slip for stationary wall as
\begin{equation} \label{ons-slip-7}
\displaystyle u_{x} =\displaystyle \left[\left(\frac{2-\sigma}{\sigma}\right)\frac{\tau_{xy}}{2p}\frac{\sqrt{\pi}}{\beta}-\frac{q_{x}\eta}{2p}\right]\phi
=\displaystyle \left[\left(\frac{2-\sigma}{\sigma}\right)\tau_{xy}\frac{\lambda}{\mu}- \left(\frac{\gamma-1}{\gamma}\right)\frac{q_{x}}{2p}\right]\phi 
\end{equation}
where  $\eta=(\gamma-1)/\gamma$ and $\lambda$$=$$\frac{\mu}{2p}\sqrt{\frac{\pi}{\beta}}$ is the viscosity based mean free path. This new expression of velocity slip which is named in this paper as \emph{Onsager-Maxwell slip velocity}, differs from Maxwell's expression \cite{lockerby04} by an extra term $\phi$ $=$ $(1-\tau_{yy}/2p)^{-1}$ due to density jump.
\subsubsection{Derivation of temperature jump }
Similarly, we can carry out energy conservation by equating the energy flux on the strip $ds$ at $y=\Delta y$  and $y=0$. The energy flux, $\mathbb{Q}$  at  $y=\Delta y$ based on linearized distribution is given as
\begin{equation} \label{ons-slip-8}
\begin{array}{ll}
{\displaystyle \mathbb{Q}_{y=\Delta y} }&
{ \displaystyle =
\begin{array}{l}{
 {\displaystyle \int_{\mathbb{R^{+}}}\int_{\mathbb{R}}\int_{\mathbb{R}}}(f_{0}-\sum_{j}{  \boldsymbol{\Upsilon}_{j} \boldsymbol{\odot} \boldsymbol{X}_{j}})(\mathbb{I}+\frac{v^{2}}{2})v_{y}dv_{x}dv_{y}d\mathbb{I}}
\end{array}
}
\end{array}
\end{equation}
The energy flux at the wall where $y=0$ can be expressed as
\begin{equation}\label{ons-slip-9}
\begin{array}{ll}
{\displaystyle \mathbb{Q}_{y=0} }&
{ \displaystyle =\sigma \left[
\begin{array}{l}{
{\displaystyle \int_{\mathbb{R^{+}}}\int_{\mathbb{R^{-}}}\int_{\mathbb{R}}}(f_{0,I}-\sum_{j}{  \boldsymbol{\Upsilon}_{I,j} \boldsymbol{\odot} \boldsymbol{X}_{j}})(\mathbb{I}+\frac{v^{2}}{2})v_{y}dv_{x}dv_{y}d\mathbb{I}}\\
{\displaystyle+{\int_{\mathbb{R^{+}}}\int_{\mathbb{R^{+}}}\int_{\mathbb{R}}}f_{0,W}(\mathbb{I}+\frac{v^{2}}{2})v_{y}dv_{x}dv_{y}d\mathbb{I}
}
\end{array}
\right ]}
\end{array}
\end{equation}
For small temperature jump  we can replace pressure in terms of mean free path, $\lambda$ using relation $p= \mu \sqrt{\pi}/(2 \lambda \sqrt{\beta})$. After substituting the value of slip velocity $u_{x}$, wall density $\rho_{w}$  and carrying out the conservation such that energy flux at the wall, $y=0$ is being balanced by the energy flux as $\Delta y$$\rightarrow 0$  leading to
\begin{equation}\label{ons-slip-10}
\begin{array}{l}
2 \beta^2 (\beta_{w} \eta^{2} (\gamma-1) q_{x}^{2} + (\gamma+1)\tau_{yy}^{2})\lambda^{2} \sigma^{2} 
+ 4 \sqrt{\beta} \beta_{w} \gamma \sqrt{\pi} \tau_{yy} \lambda \mu \sigma^{2} - \beta_{w} (\gamma+1) \pi \mu^{2} \sigma^{2}\\ 
+ \beta \sqrt{\beta} \sqrt{\pi} \lambda \sigma(4 \beta_{w} \eta((\gamma-1)q_{x}\tau_{xy}-\gamma q_{y}\tau_{yy})\lambda(\sigma-2) 
-3(\gamma+1)\tau_{yy} \mu \sigma)+\beta(\gamma+1)\pi \mu^{2} \sigma^{2} \\
+ \beta\beta_{w} \lambda\left(
\begin{array}{l}
-(3\gamma-1)\tau_{yy}^{2}\lambda \sigma^{2}
+2 \pi (\sigma-2)((\gamma-1)\tau_{xy}^{2}\lambda(\sigma-2)+ 2 \eta \gamma q_{y} \mu \sigma)
\end{array}
\right)=0
\end{array}
\end{equation}
This expression can be further simplified  by neglecting  terms associated with $\tau_{yy}$ and $q_{x}$ to get
\begin{equation}\label{ons-slip-11}
T=T_{w}+ \left(\frac{2-\sigma}{\sigma}\right)^{2} \frac{\gamma  \lambda^{2} \tau_{xy}^{2}}{c_{p}(\gamma+1) \mu^{2} }-\left(\frac{2-\sigma}{\sigma}\right)\frac{2 \gamma  \lambda q_{y}}{c_{p} (\gamma+1) \mu}
\end{equation}
We get a new expression of temperature jump which is named in this paper as \emph{Onsager-von Smoluchowski's temperature jump} as it contains both the terms of heat flux vector and shear stress tensor following Onsager's reciprocity relationship. With an additional assumption that shear stress term $\tau_{xy}$ is negligible we get the von Smoluchowski's temperature jump boundary condition in terms of Prandtl number $\text{Pr}$ as 
\begin{equation}\label{ons-slip-12}
T = T_{w} - \left(\frac{2-\sigma}{\sigma}\right)\frac{2 \gamma \lambda q_{y}}{c_{p}(1+\gamma)\mu}=T_w+ \frac{(2-\sigma )}{\sigma}  \frac{2}{\text{Pr}}\frac{\gamma}{(\gamma+1)}\lambda\frac{\partial T}{\partial y}
\end{equation}
This condition is obtained with an assumption that tangential variations, terms $q_{x}$, $\tau_{yy}$ and $\tau_{xy}$ are negligible and temperature jump is mild. These expressions are valid only in slip flow regime when there are insignificant fluid dynamic variations in the tangential direction. 
\subsection{NET-KFVS based kinetic wall boundary condition}
When the variation in the tangential direction is substantial then a more generic kinetic split flux based boundary condition can be derived. The total distribution satisfies Boltzmann equation and at time $t+\Delta t$  it is constructed as follows
\begin{equation} \label{ons-kcbc-1}
f_{1,\Sigma } (\vec{v} ,\mathbb{I},t+\Delta t) =f_{1,\Sigma } (\vec{v} ,\mathbb{I},t) -\Delta t \nabla _{\vec{x}}\cdot \left(\vec{v}f_{1,\Sigma } (\vec{v} ,\mathbb{I},t)\right)
\end{equation}
The distribution at the boundary after upwind discretization for the two dimensional case is
\begin{equation}\label{ons-kcbc-2}
f_{1,\Sigma }^{t+\Delta t} =f_{1,\Sigma }^{t} -\Delta t\left[ \frac{\partial v_{x} f_{1,\Sigma}^{+ -}}{\partial x}+\frac{\partial v_{x} f_{1,\Sigma}^{- -}}{\partial x}+\frac{\partial v_{y} f_{1,\Sigma}^{\cdot -}}{\partial y} \right]^{t}
\end{equation} 
where $f_{1,\Sigma}^{+ -}$ is the half-range total distribution function for $0 < v_{x} <\infty$ and $-\infty < v_{y} <0$ and $f_{1,\Sigma}^{- -}$ is the half-range total distribution function for $-\infty < v_{x} <0$ and $-\infty< v_{y} <0$. After taking $\boldsymbol{\Psi}$ moment we can obtain state update equation expressed as
\begin{equation} \label{ons-kcbc-3}
\boldsymbol{U}^{t+\Delta t} =\boldsymbol{U}^t -\Delta t\left[
\begin{array}{l}
{\rm\;}\left(\frac{\partial \boldsymbol{\widehat{GX}}^{+ - }}{\partial x}\right)^{t}_{\Delta x < 0}+\left(\frac{\partial \boldsymbol{\widehat{GX}}^{- - }}{\partial x}\right)^{t}_{\Delta x >0}
+\left(\frac{\partial \boldsymbol{\widehat{GY}}^{\cdot -}}{\partial y}\right)^{t}_{\Delta y > 0}  
\end{array}
\right] 
\end{equation} 
where $\boldsymbol{U}$$=$$[\rho, \rho \boldsymbol{u}, \rho E]^{T}$ is the state vector  and $\Delta t$ is the time step. The $y$ component of state vector is not updated as $U_{i=3}$$=$$\rho u_{y}$$=0$. $\boldsymbol{\widehat{GX}}^{\pm - }$  represents the split flux based on half range distributions $ f_{1,\Sigma}^{\pm -}$. $\boldsymbol{\widehat{GY}}^{\cdot -}$ is the split flux resulting from half range distribution $f_{1,\Sigma}^{\cdot -}$. Derivatives of  $ \boldsymbol{\widehat{GX}}^{+ - }$, $ \boldsymbol{\widehat{GX}}^{- - }$ and $ \boldsymbol{\widehat{GY}}^{\cdot -}$ are evaluated using points on the left, right and upward side of the stencil. The mass, momentum and energy components of $x$-directional flux $\boldsymbol{\widehat{GX}}^{\pm -}$  can be written as sum of inviscid or Euler part $\boldsymbol{\widehat{GX}}^{\pm -}_{I}$ and viscous part $\boldsymbol{\widehat{GX}}^{\pm -}_{V}$ as follows
\begin{equation}\label{ons-kcbc-4}
\begin{array}{lll}
\boldsymbol{\widehat{GX}}^{\pm -}&=\boldsymbol{\widehat{GX}}^{\pm -}_{I}+\boldsymbol{\widehat{GX}}^{\pm -}_{V}&=\left< v_{x}\boldsymbol{\Psi} f_{0,\Sigma}^{\pm -} \right> -\sum_{j}\boldsymbol{\Lambda}^{x,\boldsymbol{\Psi},\pm -}_{j} \boldsymbol{ \odot} \boldsymbol{X}_{j}
\end{array}
\end{equation} 
Similarly,  components of $y$-directional flux $\boldsymbol{\widehat{GY}}^{\cdot -}$  can be written as sum of inviscid part and viscous part as follows
\begin{equation} \label{ons-kcbc-5}
\begin{array}{lll}
\boldsymbol{\widehat{GY}}^{\cdot -}&=\boldsymbol{\widehat{GY}}^{\cdot -}_{I}+\boldsymbol{\widehat{GY}}^{\cdot -}_{V} &=\left< v_{y}\boldsymbol{\Psi} f_{0,\Sigma}^{\cdot -} \right> -\sum_{j}\boldsymbol{\Lambda}^{ y,\boldsymbol{\Psi},\cdot -}_{j} \boldsymbol{ \odot} \boldsymbol{X}_{j}
\end{array}
\end{equation} 
where  macroscopic split tensors $\boldsymbol{\Lambda}_{j}^{x,\boldsymbol{\Psi,{\pm -}}}$ is defined as
\begin{equation}\label{ons-kcbc-6}
\begin{array}{l}
\boldsymbol{\Lambda}_{j}^{x,\boldsymbol{\Psi,{\pm -}}}=\left\langle \boldsymbol{\Psi},v_{x}  {  \boldsymbol{\Upsilon}_{j}^{\pm-} }\right\rangle
\displaystyle \equiv \int_{\mathbb{R}^{+}} \int_{\mathbb{R}^{-}}\int_{\mathbb{R}^{\pm}} \boldsymbol{\Psi} v_{x}  \boldsymbol{\Upsilon}_{j}  dv_{x} dv_{y}d\mathbb{I}
\end{array}
\end{equation}
The viscous fluxes are obtained using macroscopic tensors $\boldsymbol{\Lambda}^{x,\boldsymbol{\Psi},\pm -}_{j}$ and $\boldsymbol{\Lambda}^{y,\boldsymbol{\Psi}, \cdot -}_{j}$ associated with shear stress tensor and thermal gradient vector following Onsager's reciprocal relationship so as to maximize the entropy production. The fluxes $\boldsymbol{\widehat{GX}}^{\pm - }$ and $\boldsymbol{\widehat{GY}}^{-}$ can also be written in alternative form as
\begin{equation} \label{ons-kcbc-7}
\begin{array}{l}
\boldsymbol{\widehat{GX}}^{\pm - }=(2-\sigma)(\boldsymbol{GX}^{\pm - }_{I}+\boldsymbol{GX}^{\pm - }_{V}) + \sigma \boldsymbol{GX}^{\pm - }_{I,(0,W)}\\
\boldsymbol{\widehat{GY}}^{\cdot -}=(2-\sigma)(\boldsymbol{GY}^{-}_{I}+\boldsymbol{GY}^{-}_{V}) + \sigma \boldsymbol{GY}^{-}_{I,(0,W)}
\end{array}
\end{equation} 
where $\boldsymbol{GX}^{\pm - }_{I}$, $\boldsymbol{GX}^{\pm - }_{V}$, $\boldsymbol{GY}^{-}_{I}$ and $\boldsymbol{GY}^{-}_{V}$ are evaluated based on fluid conditions while $\boldsymbol{GX}^{\pm -}_{I,(0,W)}$ and $\boldsymbol{GY}^{-}_{I,(0,W)}$ are the inviscid flux corresponding to Maxwellian distribution, $f_{0,W}$ based on the wall conditions.

The kinetic wall boundary condition can also be extended for rarefied regime bordering transition flows using Onsager-BGK Knudsen layer model described in appendix \ref{kn2model}.
\begin{section}
{Results and Discussions}\label{resdis}
\end{section}
The present kinetic solver uses meshless approach to solve the NET-KFVS formulation. Consider distribution $f$ given at point $P_{o}$ and \textit{m} points surrounding it called its  connectivity $N(P_{o})$ $=$ $\left\{\forall P_{i} :d(P_{o},P_{i} )<h\; \right\}$. Finding the derivative at point $P_{o}$ is a least squares problem where error norm $\|E\|_{2}$ is to be minimized with respect to derivatives $f_{xo}$, $f_{yo}$ and $f_{zo}$ using stencil $N(P_{o})$. Least square based method using normal equations approach fails to handle highly stretched distribution of points. More-so-ever not all ill effects inherent in the normal equation approach can be avoided by using orthogonal transformation  as condition number is still relevant to some extent \cite{golub-s}. The meshless kinetic solver\cite{mahslkns} uses split least square approach by  minimizing $\|\boldsymbol{A}_{N_{x}}\boldsymbol{\mathbb{F}}_{o}-\boldsymbol{\Delta f}_{N_{x}} \|_{2}$, $\|\boldsymbol{A}_{N_{y}}\boldsymbol{\mathbb{F}}_{o}-\boldsymbol{\Delta f}_{N_{y}} \|_{2}$ and $\|\boldsymbol{A}_{N_{z}}\boldsymbol{\mathbb{F}}_{o}-\boldsymbol{\Delta f}_{N_{z}} \|_{2}$ with respect to $f_{xo} $, $f_{yo} $ and $f_{zo}$  respectively for each carefully selected sub-stencils  $N_{x} (P_{o} )$ $\in \mathbb{R}^{m_{x}}$, $N_{y} (P_{o} )$ $\in \mathbb{R}^{m_{y}}$, $N_{z}(P_{o} )$ $\in \mathbb{R}^{m_{z}}$ such that the cross-product matrix is diagonally dominant and well conditioned. Split-stencil based least squares approach retains the simplicity of normal equations while avoiding the ill-conditioning of the matrix. Based on the formulations described  split stencil least square kinetic upwind method or SLKNS in short uses NET-KFVS for fluid domain and implements NET-KFVS based kinetic wall boundary condition for non-continuum slip flow as well as continuum flow. SLKNS solver was validated for variety of test cases including continuum flows and non-continuum slip flows considering formulation based on inelastic collisions. 
\subsection{Continuum transonic flow over NACA0012 airfoil}
\begin{figure} \centering
(a){\includegraphics[width=7cm]{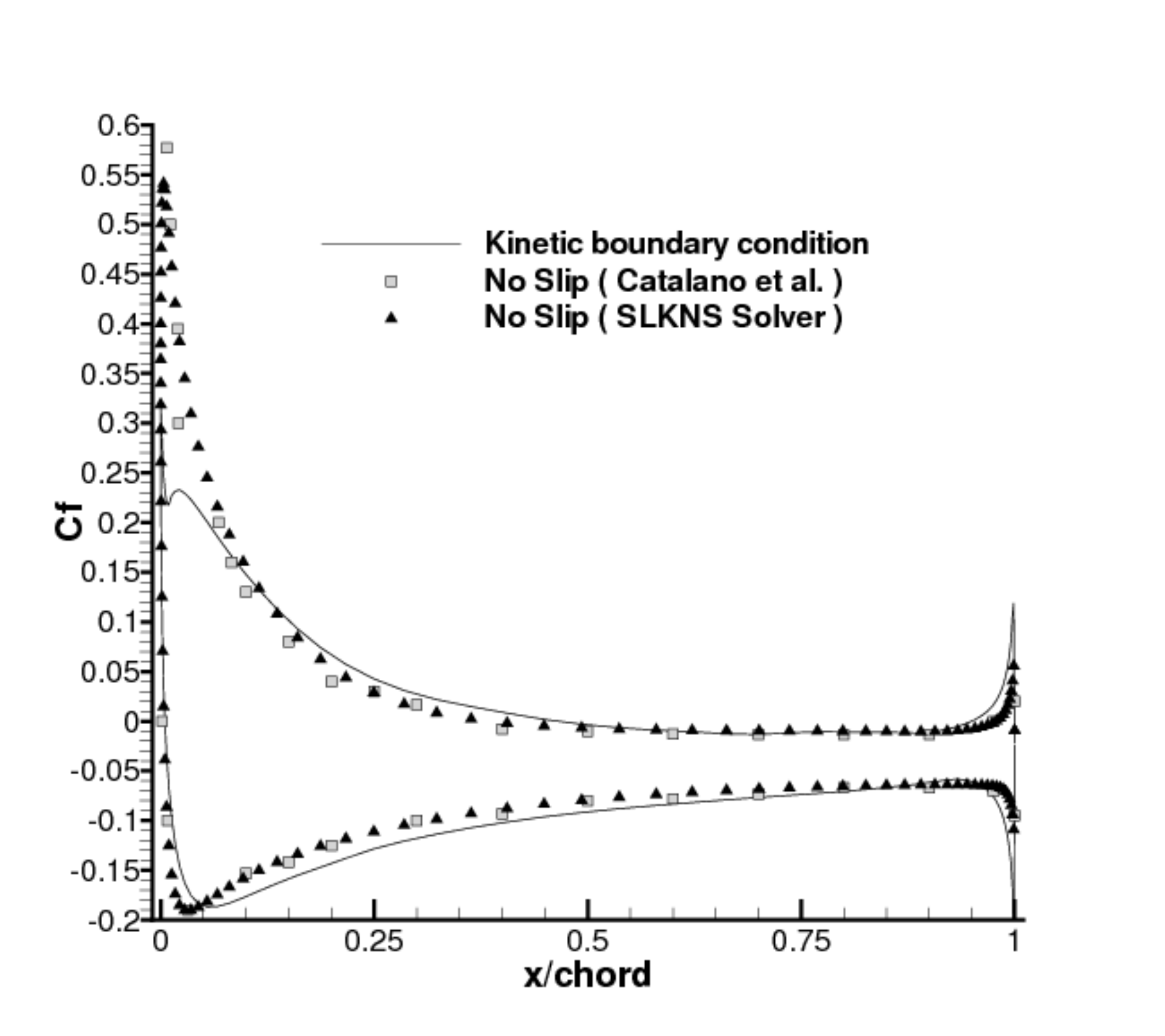}}(b){\includegraphics[width=7cm]{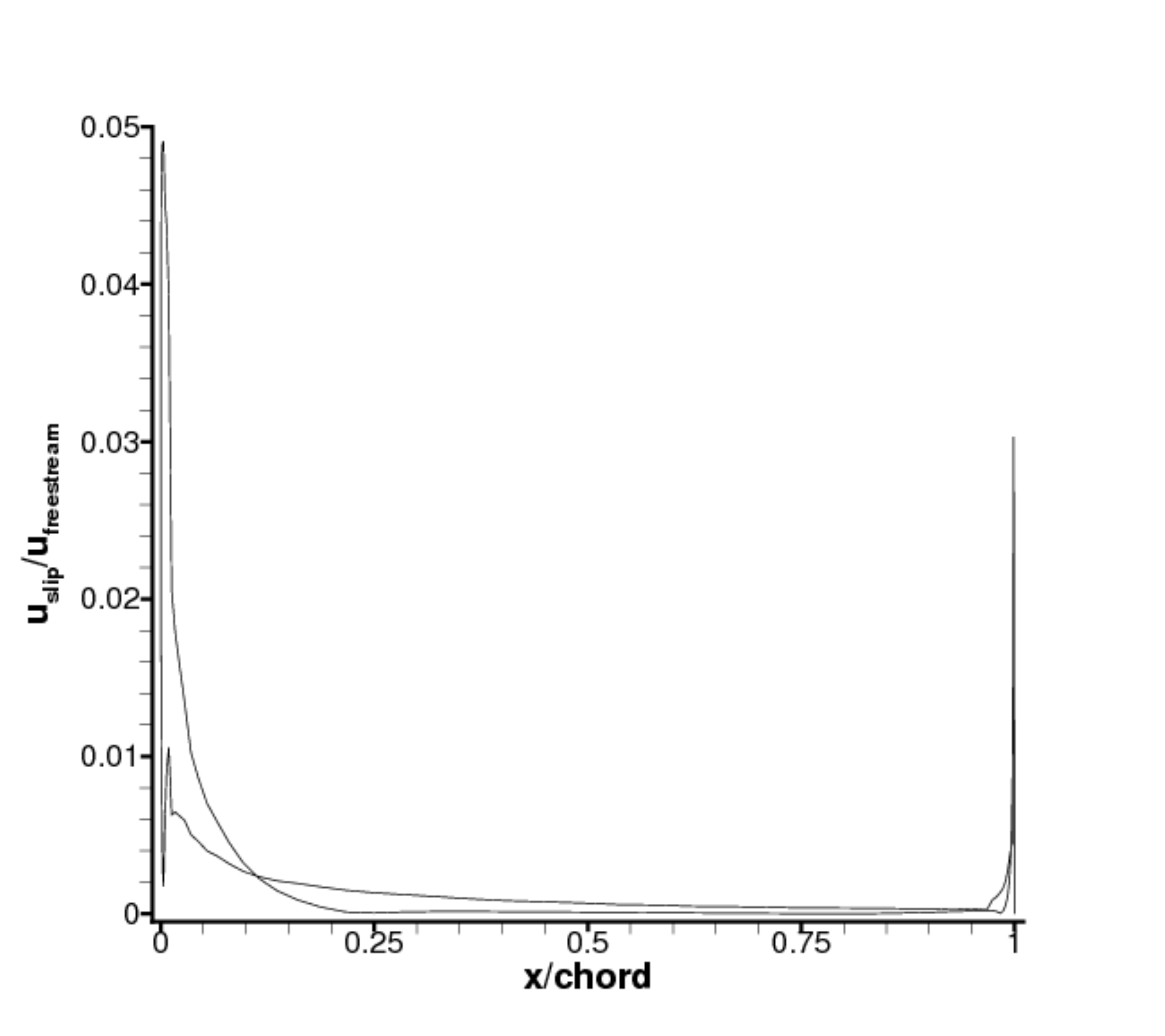}}
\caption{Continuum transonic flow past NACA0012 airfoil with kinetic wall boundary condition: (a)Coefficient of friction plot, (b) velocity slip along the surface of NACA0012 airfoil.\label{ronaca}}
\end{figure}

Consider transonic continuum flow of air at Mach $0.8$ past a NACA0012 airfoil at 10 degrees angle of attack at $\text{Re}$$=$$500$. The Knudsen number, $\text{Kn}$ $=$ $\sqrt{\frac{\pi \gamma }{2} } \frac{\text{M}}{\text{Re}}  $ evaluated for such a case is $2.4\times10^{-3}$ based on chord length.  This test case is simulated using the kinetic boundary condition with fully diffuse reflecting wall i.e. the accommodation coefficient, $\sigma$ $=$$1$. The kinetic boundary condition treats the continuum region in the same way as the non-continuum region admitting velocity slip and temperature jump which becomes negligibly small in the continuum region. Figure~\ref{ronaca} (a) shows the plot of coefficient of friction compared with SLKNS solver with kinetic wall boundary condition, SLKNS with  no-slip boundary condition and fluctuation splitting LDA scheme \cite{catalano} using no-slip boundary condition. Dip in coefficient of friction near the leading edge can be observed due to slip flow. Figure \ref{ronaca} (b) shows the small velocity slip existing on the surface. Temperature jump for this case was found to be very negligible. This continuum flow test case using kinetic wall boundary condition confirms the observation made by Struchtrup\cite{struchtrup} that temperature jump and velocity slip will be present for all dissipative walls even in continuum regime.
\begin{figure}\centering
(a){\includegraphics[scale=0.2,bb=0 0 1145 930]{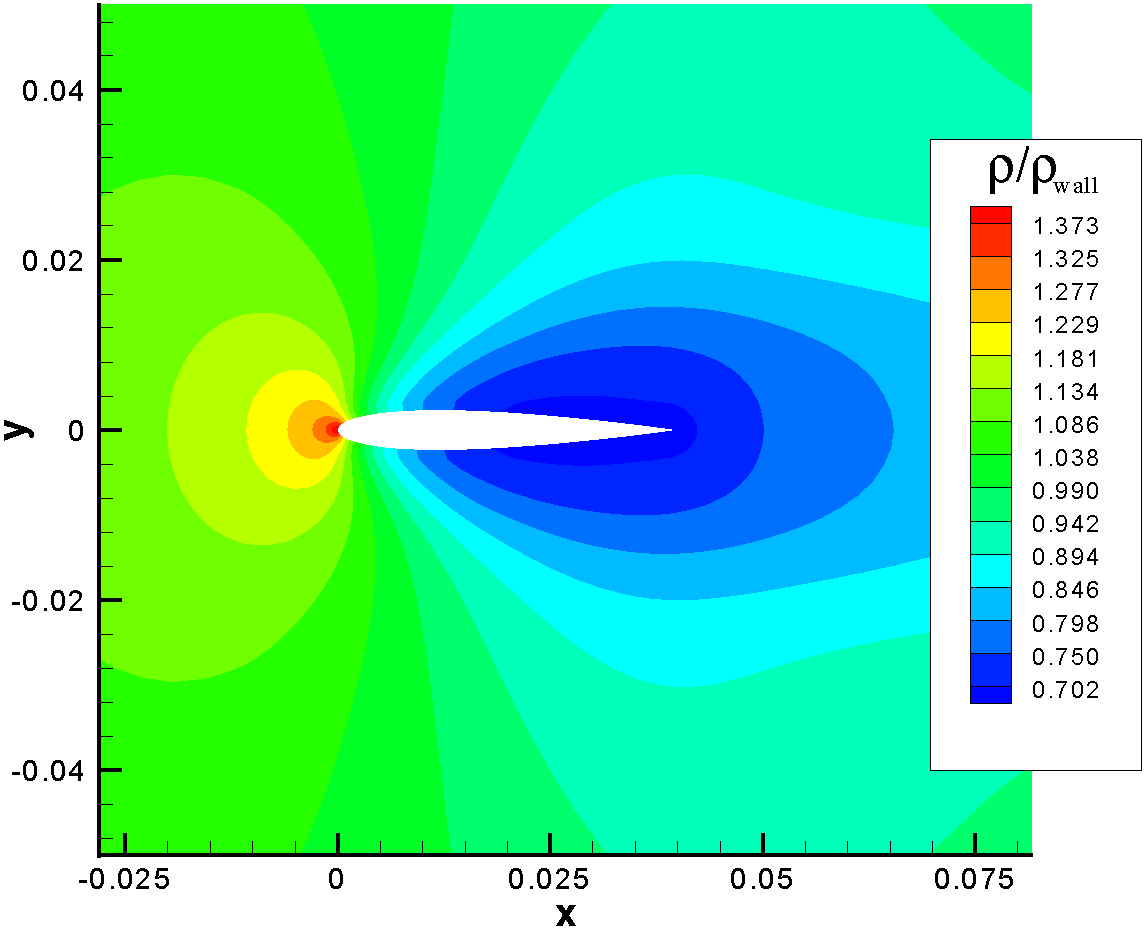}}(b){\includegraphics[width=6cm]{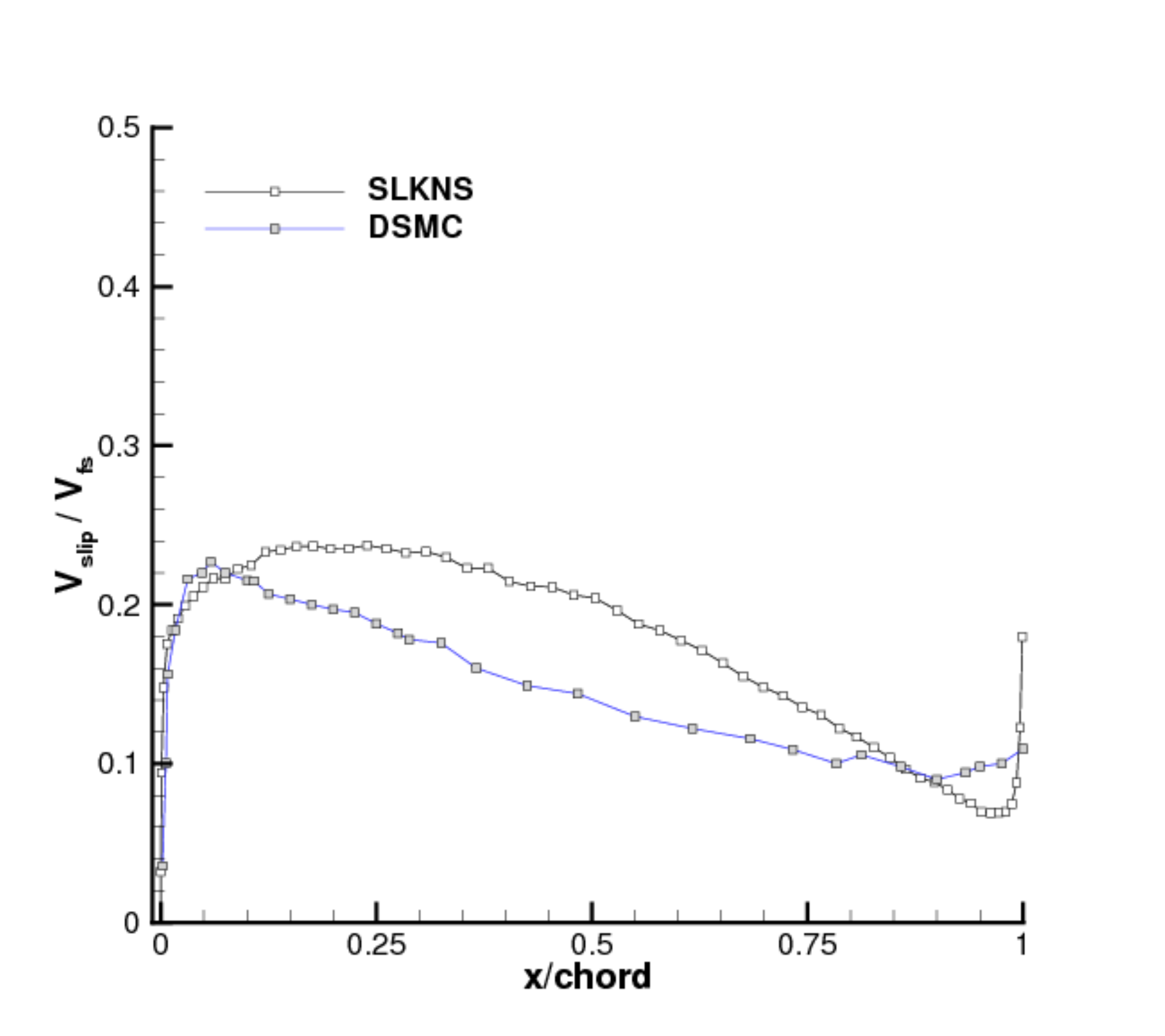}}
\caption{ \label{naca-rar}Rarefied flow past NACA0012 airfoil at Mach=0.8, Re=73., Kn=0.014 (a)Contours of $\frac{\rho}{\rho_{wall}}$, (b)Comparison of the slip velocity distribution with DSMC}
\end{figure}
\subsection{Rarefied transonic flow over NACA0012 airfoil}
Consider free stream rarefied transonic flow at Mach, $\text{M}=0.8$, density $1.116 \times 10^{-4}$  $kg/m^{3}$ and temperature $257$ K past a NACA0012 airfoil at zero angle of attack \cite{boyd-ip}. The Reynolds number based on the airfoil chord is 73 and Knudsen number is 0.014. The chord length is $0.04 m$ and wall of the airfoil is at $290$ K.  The density contours shown in figure~\ref{naca-rar}(a) reveals rise of density near stagnation point and rarefaction towards the tail where the density drops down. The viscosity based mean free path depends on the density, $\rho$ as $\lambda = \mu \sqrt{\frac{\pi}{2 \rho p}}$. The rise in the mean free path near the tail makes the slip influence more pronounced which results  in sudden rise of slip velocity. Figure~\ref{naca-rar}(b) shows the comparison of the slip velocity distribution for rarefied flow past NACA0012 airfoil based on SLKNS solver and Direct Simulation Monte Carlo (DSMC)\cite{bird} using  $120 \times 60$ points. DSMC gave better results for nearly same grid size. 
\subsection{Hypersonic rarefied flow over a flat plate}
\begin{figure} \centering
(a){\includegraphics[scale=0.15,bb=0 0 1486 719]{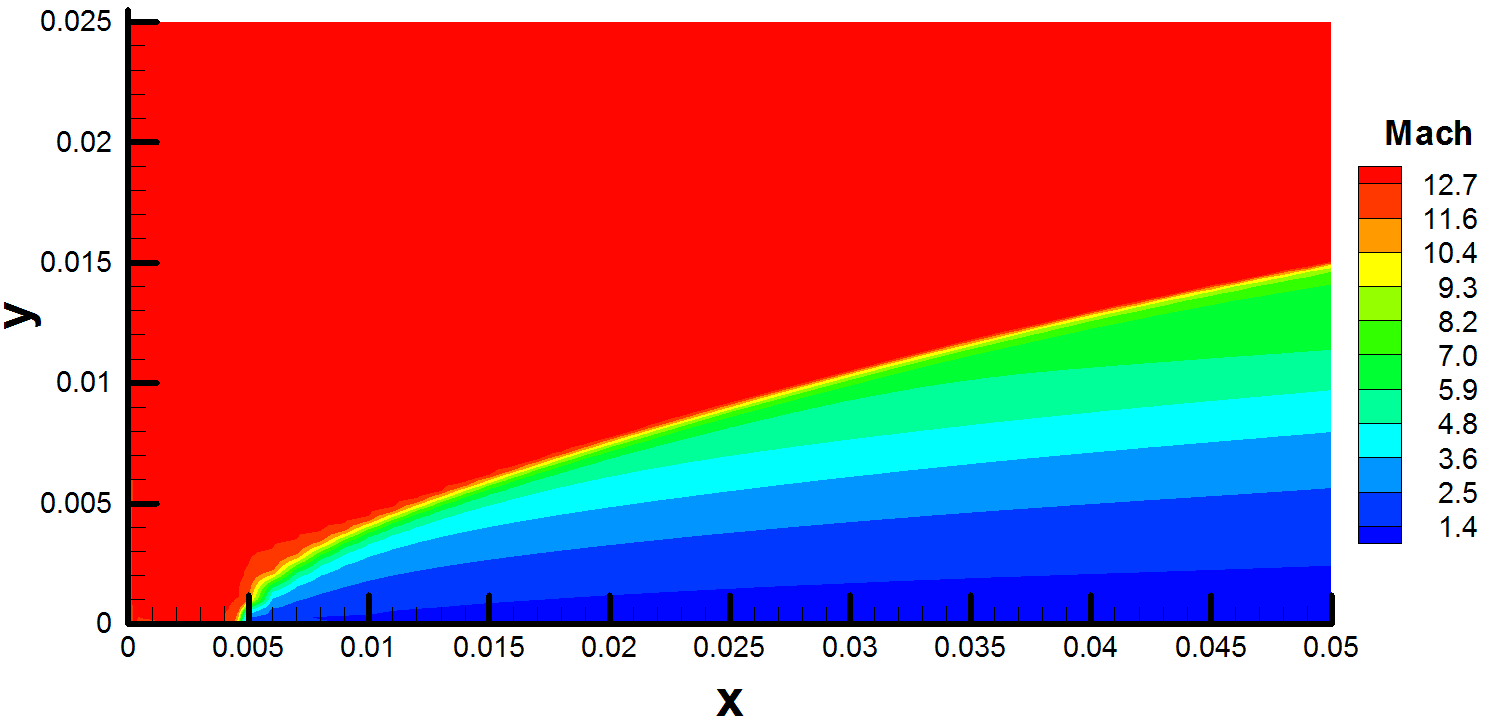}}(b){\includegraphics[width=6cm]{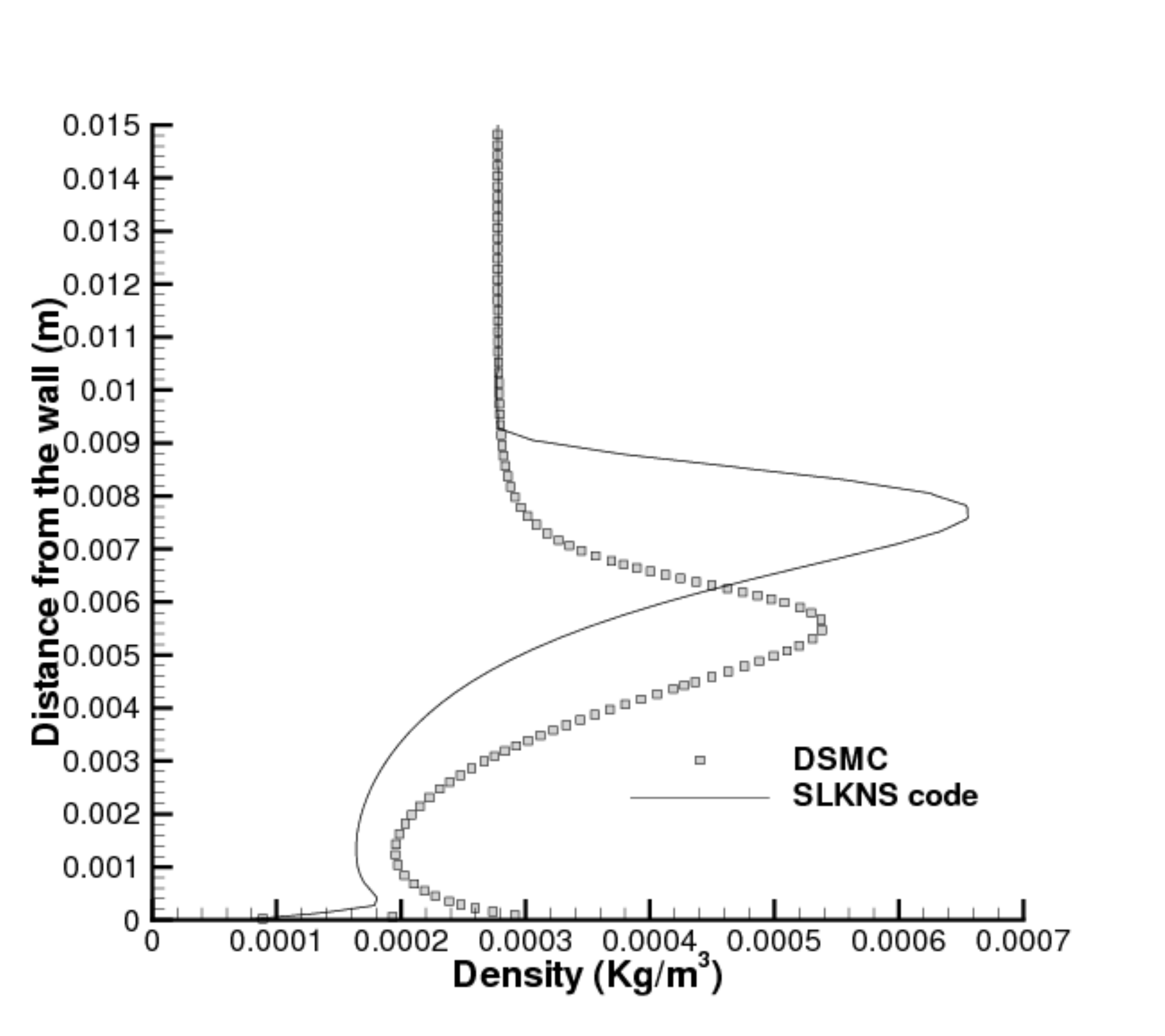}}
\caption{Hypersonic flow over a flat plate (a)Mach contours based on NET-KFVS based SLKNS solver, (b)density at cross-section $x$$=$$25$ mm from the plate tip.\label{shockro}}
\end{figure}

\begin{figure} \centering
(a){\includegraphics[width=6cm]{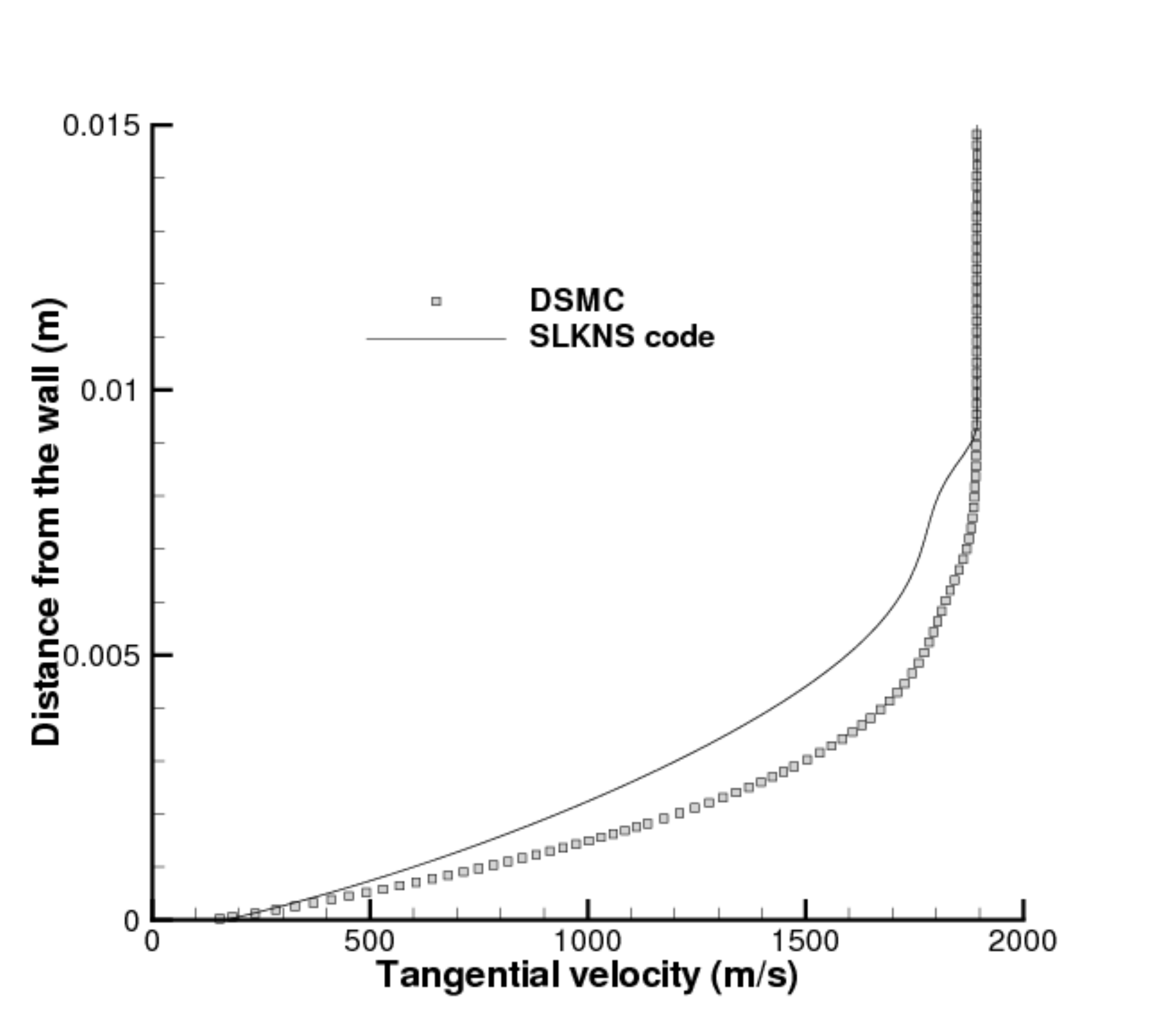}}(b){\includegraphics[width=6cm]{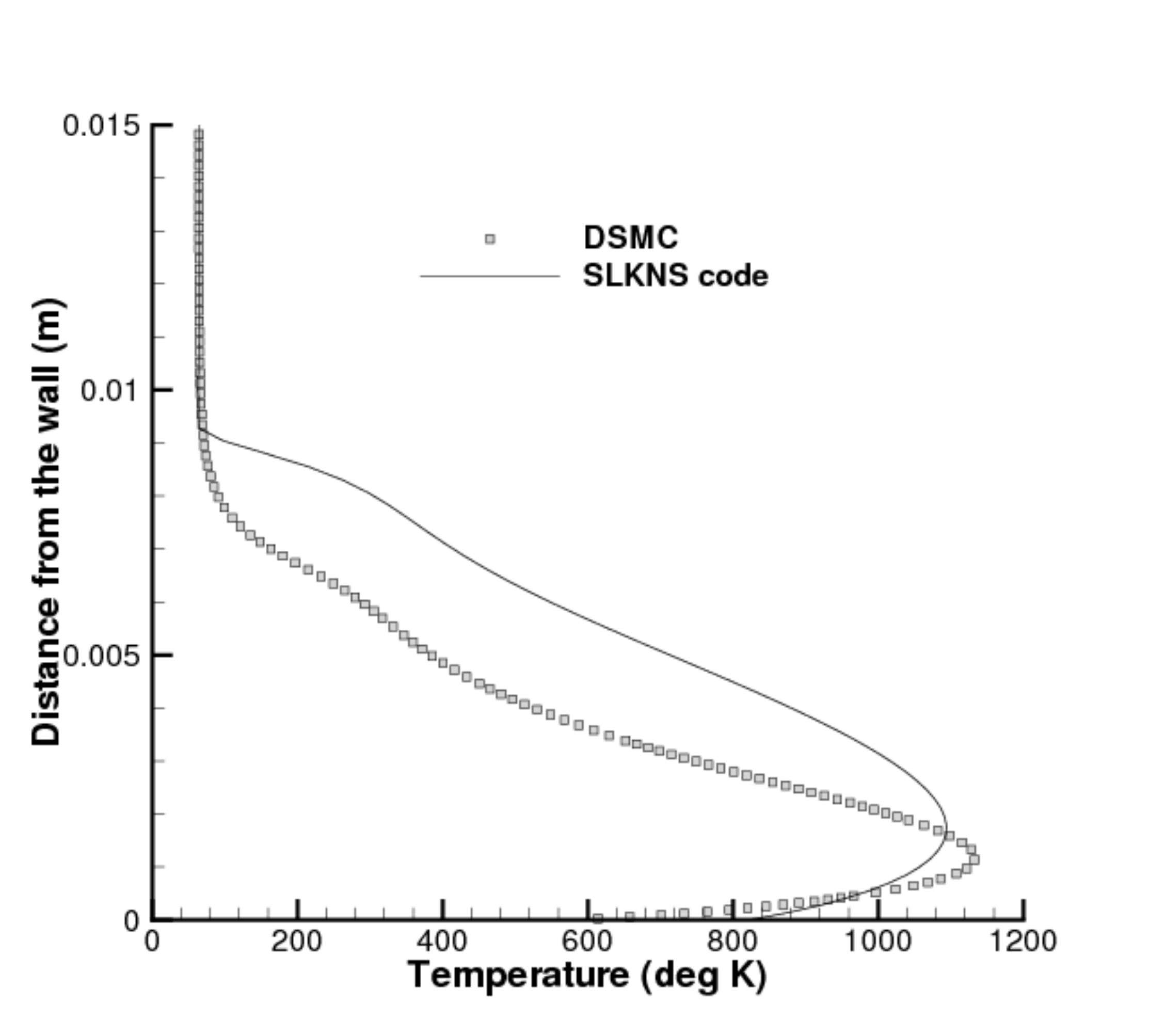}}
\caption{Hypersonic flow over a flat plate (a)Tangential velocity (b) profile of the temperature in the boundary layer at cross-section $x$$=$$25$ mm from the plate tip.\label{profileuxt}}
\end{figure}

\begin{figure} \centering
(a){\includegraphics[width=7cm]{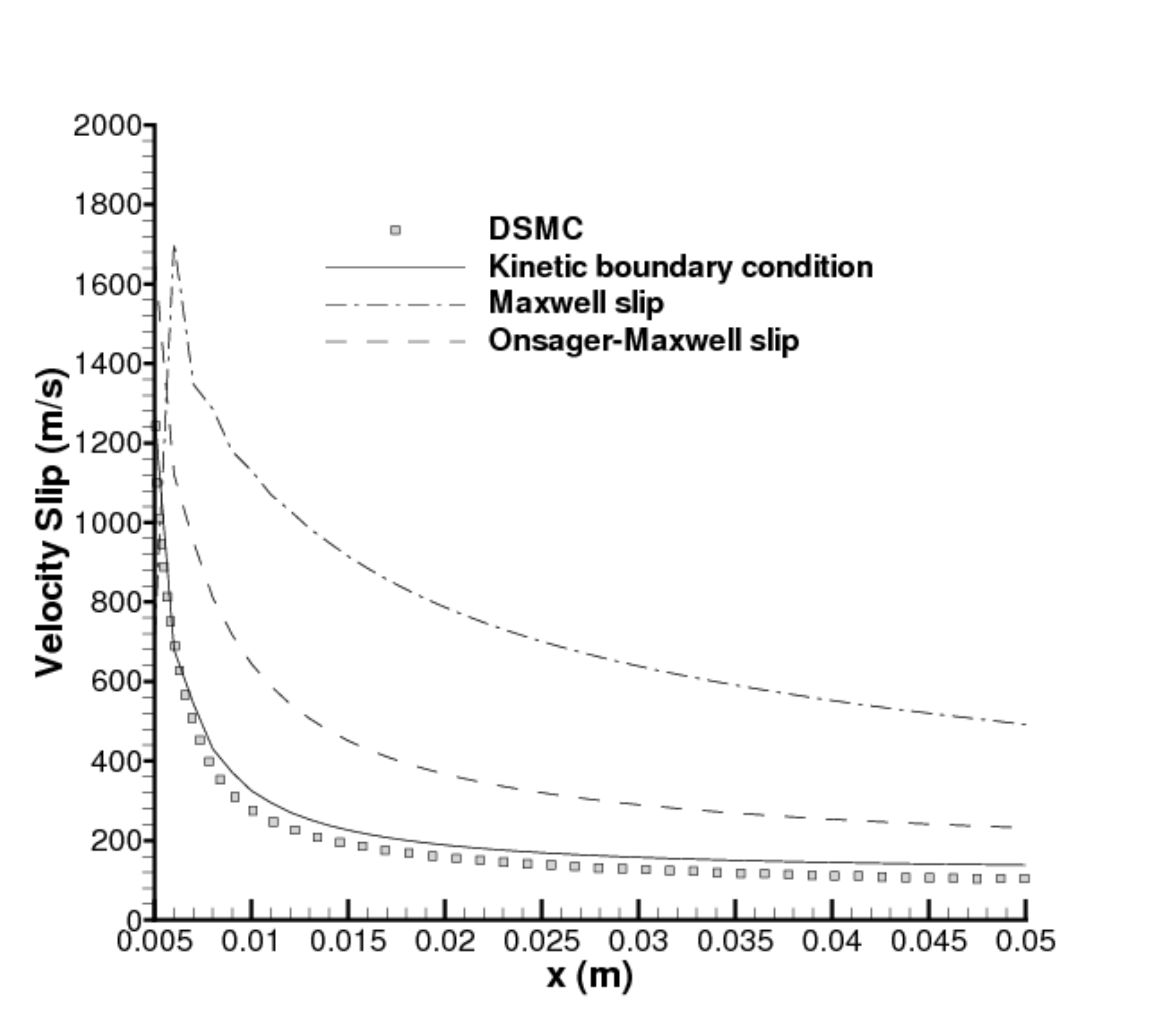}}(b){\includegraphics[width=7cm]{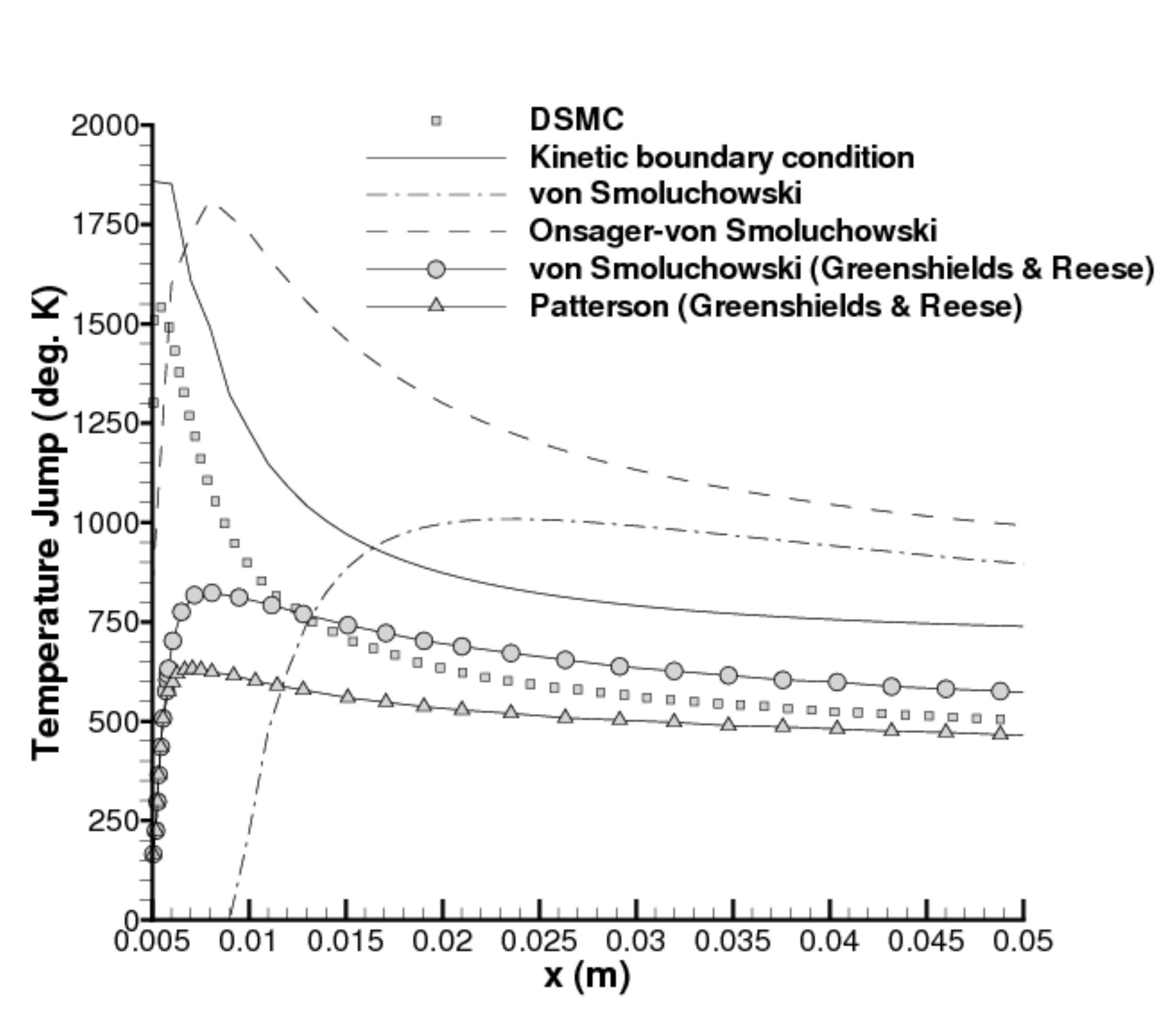}}
\caption{Variation of (a)Velocity slip and (b)Temperature jump along the surface of flat plate for various boundary conditions.\label{shockslip}}
\end{figure}

Hypersonic rarefied flow over a flat plate is one of the fundamental problem as it generates wide range of flow phenomena extending from highly non-equilibrium flow near the leading edge through the merged layer to strong and weak interaction regimes to  a classical boundary layer flow at the downstream. Kinetic flow region exists very near the leading edge caused by collisions between free stream and body reflected molecules. Near the leading edge non-continuum non-equilibrium viscous region exists where molecule-molecule and molecule-body collisions dominate the flow and as a consequence the  distribution function is far away from Maxwellian. Further downstream in the transition region molecule-molecule collisions dominate the flow, this is followed by merged layer region in which wall boundary layer merges with the a non-Rankine-Hugoniot shock \cite{pullinh}. Consider a test case of hypersonic flow of argon at free stream velocity of $(1893.7,0,0)$ m/s, with pressure of $3.73$ Pascal at temperature of $64.5$ K over a flat-plate  held at uniform temperature $T_{w}=292$ K at $0\deg$ angle of attack \cite{grs}. The test case used in this paper consists of flat plate 45 cm long placed along the x-axis in a flow domain of 25 cm $\times$ 50 cm as shown in Figure~\ref{shockro}(a). The simulation used a mesh of  $70$ $\times$ $100$  graded from $\Delta y$$\approx$$ 0.13$ mm at the plate surface to $\Delta x$$\approx$$ 0.21$ mm ahead of the plate tip. Figure \ref{shockro}(b) shows the profile of the density in the boundary layer at $x$$=$$25$ mm from the plate tip. Figure \ref{profileuxt}(a) shows the profile of the tangential velocity and  Figure \ref{profileuxt}(b) shows the profile of the temperature in the boundary layer at $x$$=$$25$ mm from the plate tip. All of these boundary conditions are compared with the flux based kinetic boundary condition and results of DSMC\cite{bird}. Figure~\ref{shockslip}(a)  shows the plot of  velocity slip for DSMC, kinetic boundary condition, Maxwell slip and Onsager-Maxwell slip.  Figure~\ref{shockslip}(b) shows the plot of temperature jump for DSMC, kinetic, von Smoluchowski, Onsager-von  Smoluchowski boundary condition and results of Greenshields and Reese\cite{grs}. Temperature boundary condition based on von Smoluchowski using NET-KFVS gave unphysical temperature jump near the leading edge, hence they were evaluated based on DSMC field data to compare it with kinetic boundary condition. Kinetic boundary condition was found to give better agreement with the results of DSMC. As also observed by Greenshields and Reese\cite{grs} that there is  discrepancy between the results of DSMC and boundary conditions of Maxwell, von Smoluchowski and Patterson. This discrepancy is because of two factors: i)missing features of non-equilibrium thermodynamics, ii)as well as due to the fact that these expressions are derived under condition of negligible tangential variations. The mass flux due to slip on the surface of the plate is governed both by the tangential as well as normal components of shear stress tensor and thermal gradient vector. In order to estimate the order of importance of this \emph{cross phenomenon due to thermodynamic forces} pertaining to shear stress tensor and thermal gradient vector a new dimensionless term  called reciprocity number, $Rp$ was derived. Dimensionless number based on the ratio  of viscous split slip fluxes due to the contribution of thermodynamic forces can be a good measure of the non-equilibrium cross phenomenon.  In this paper $Rp$ is derived as the ratio of slip mass flux due to shear stress tensor and thermal gradient vector based on half range distribution  as follows
\begin{equation} \label{ons-hyp-1}
Rp= \frac{  \boldsymbol{\Lambda}^{+ -}_{1,\tau} \boldsymbol{:}  \boldsymbol{X}_{\tau}}{ \boldsymbol{\Lambda}^{+ -}_{1, q} \boldsymbol{\cdot}  \boldsymbol{X}_{q}}=\frac{ 2 \tau_{xy} A_{1}^{+}-\tau_{xx} B_{1}^{+} }{2 \beta \eta u_{slip}( q_{y}A_{1}^{+}-q_{x} B_{1}^{+} )}
\end{equation} 
where expressions of $ A_{1}^{+}$ and $B_{1}^{+}$ are given in appendix \ref{splitflux}. The plot of $Rp$ in Figure \ref{rp-plot} shows a sudden variation in the ratio of contribution of shear stress tensor and heat flux vector near the leading edge. Near the leading edge the flow is dominated by the cross-coupling due to thermodynamic forces based on thermal gradient vector and shear stress tensor. Tangential variations become insignificant as we move away from the zone of sudden dip.
\begin{figure} \centering
{\includegraphics[width=7cm]{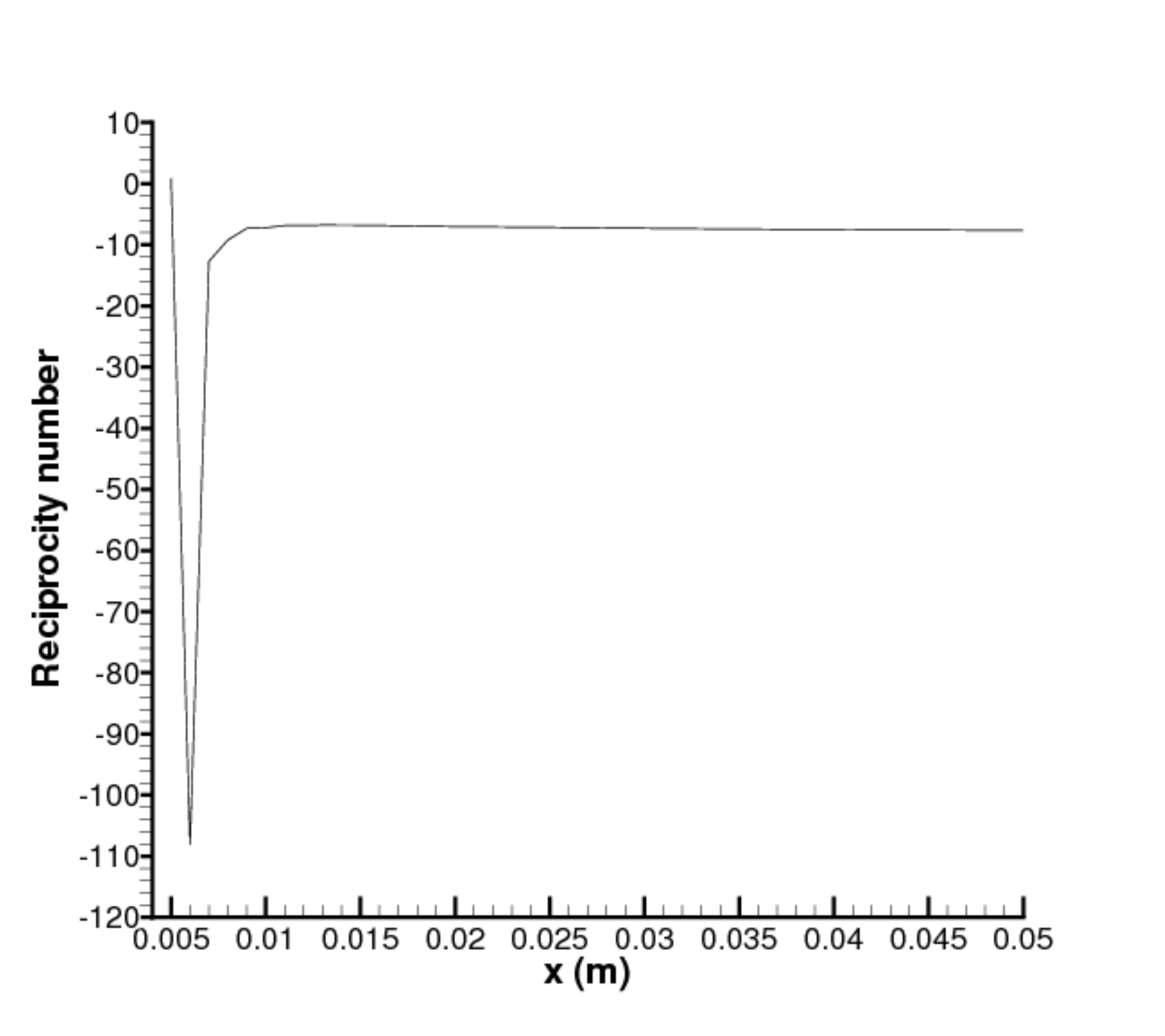}}
\caption{Variation of the $Rp$ term along the surface of the flat plate.\label{rp-plot}}
\end{figure}
\subsection{Couette rarefied flow confined in a concentric cylinder}
Couette flow between concentric inner rotating and outer stationary cylinders is one of a classical fluid dynamics problem. Consider rarefied flow with a mean free path, $\lambda$ of $0.00625$ m confined in a rotating inner cylinder of radius $3\lambda$  and stationary outer cylinder of radius $5\lambda$. The motive gas chosen is argon with initial uniform density of $1.867819 \times 10^{-5}$ $kg/m^3$ and inner cylinder held at $300$ K,  rotates at frequency of  $1000 \pi$ radians/sec.

\begin{figure}\centering
(a){\includegraphics[width=7cm]{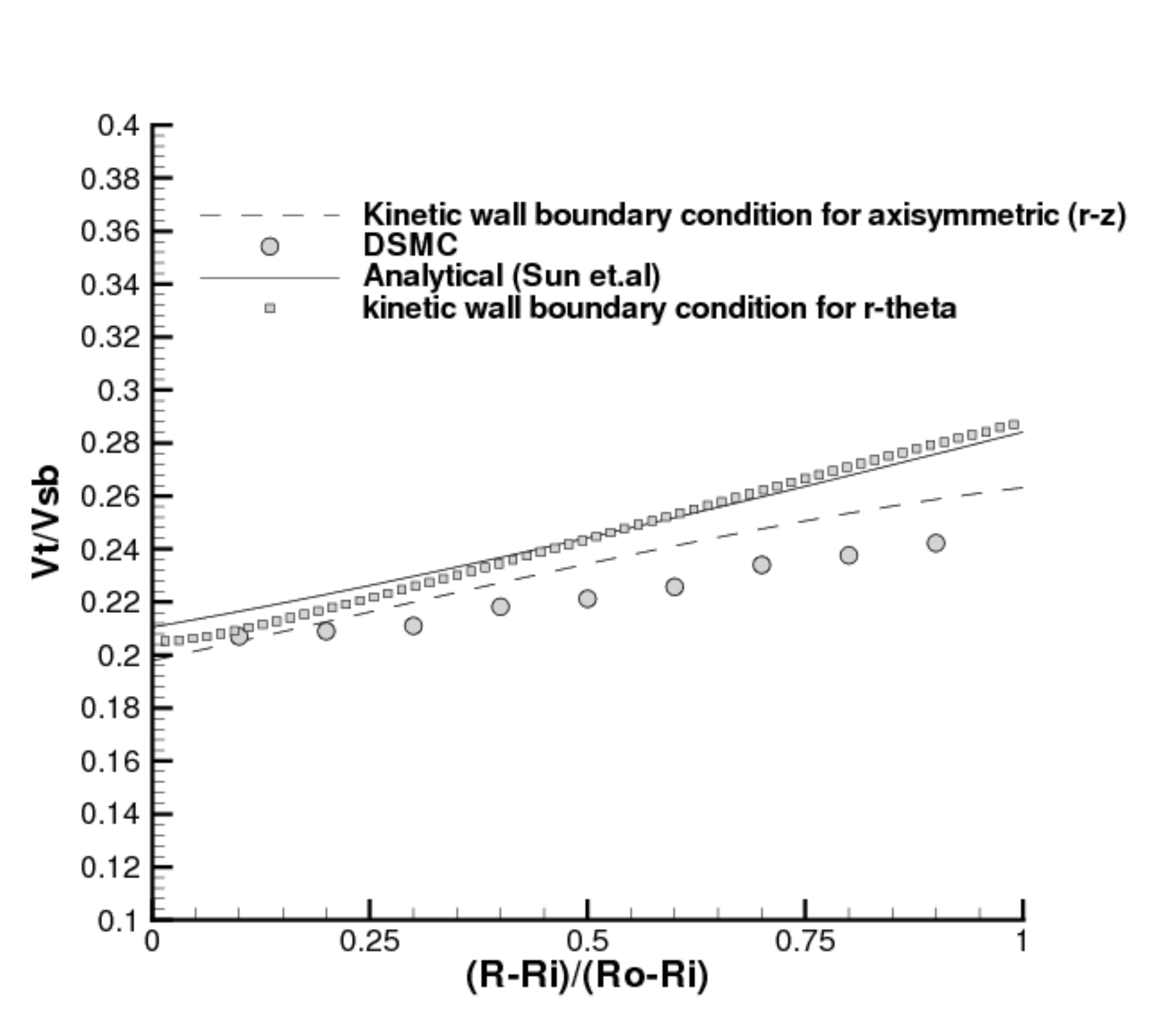}}(b){\includegraphics[scale=0.35,bb=0 0 728 599]{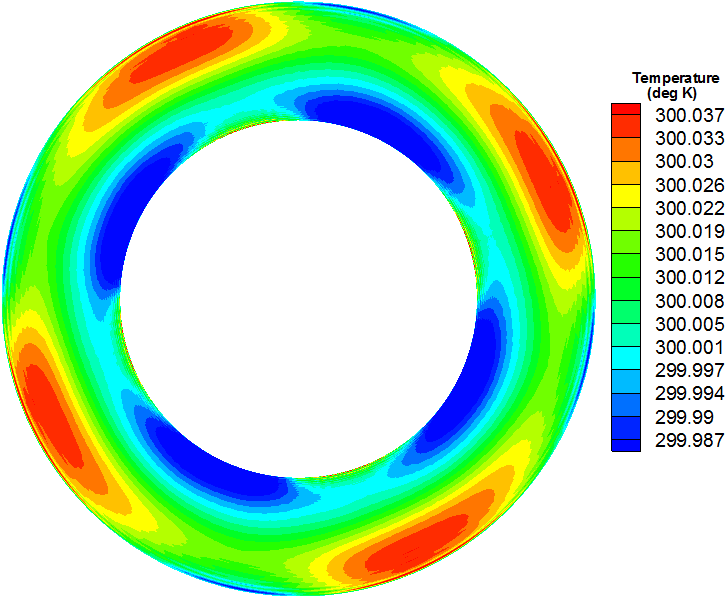}}
\caption{ \label{invslip}SLKNS  simulation on  r-$\theta$ plane (a) comparison of the non-dimensional tangential velocity with $r-z$, DSMC and analytical result \cite{sunbar}, (b) contours of temperature based on simulation on  r-$\theta$ plane.}
\end{figure}

A cloud of size $120$ $\times$ $200$ was used to carry out simulation using SLKNS with NET-KFVS based kinetic wall boundary condition for tangential momentum accommodation coefficient, $\sigma$ $=$ $0.1$. Fig.\ref{invslip}(a) shows the plot of the non-dimensional tangential velocity with respect to non-dimensional radial distance for SLKNS solver using $r$-$\theta$, and its comparison with $r-z$ formulation based on axi-symmetric Boltzmann equation \cite{mahslkns}, results of DSMC\cite{sunbar} and analytical solution\cite{sunbar} using isothermal condition and uniform density. The viscous dissipation may generate faint temperature variations which are difficult to capture using DSMC. One of the objective of the test case was to resolve such weak features associated with viscous dissipation. Figure \ref{invslip}(b) shows the contour of temperature which breaks the symmetry. It should be noted that any DSMC simulation as well as analytical derivation based on axi-symmetric approach  may no longer be accurate as symmetry breaks down due to slip flow. When the outer cylinder is specularly reflecting then no circumferential momentum is transferred to the outer cylinder \cite{aoki}. As a consequence the gas accelerates and tries to reach the stationary state of rigid body rotation (for which the distribution function is a Maxwellian), satisfying the Onsager's principle of least dissipation of energy valid for processes close to equilibrium. \emph{For non-inertial flows slip can exists even for non-dissipative specular walls.}
\begin{section}
{Conclusions and Future Recommendations}\label{conc}
\end{section}
Most of the research in kinetic theory have focused more in the issues related to entropy generation and ignored the crucial aspect of non-equilibrium thermodynamics. Kinetic models and kinetic scheme should comply with the requirements of non-equilibrium thermodynamics. Non-equilibrium thermodynamics being a phenomenological theory gives the symmetry relationship between kinetic coefficients as well as general structure of non-equilibrium phenomenon derived using kinetic theory. The Onsager's symmetry relationship is a consequence of microscopic reversibility condition due to the equality of the differential cross sections for direct and time reversed collision processes.

For a prescribed irreversible force the actual flux which satisfies Onsager's theory also maximizes the entropy production. The solution of the Boltzmann equation is in accordance with the principle of maximum entropy production (MEP). Linearized collision operator and non-equilibrium part of the distribution function can be expressed in terms of microscopic tensors and its associated thermodynamic forces. Each thermodynamic force generates non-equilibrium distribution which relaxes with its own specific relaxation time satisfying Onsager relation for entropy production. A new kinetic model called Onsager-BGK  model was formulated based on these principles of non-equilibrium thermodynamics. The Boltzmann H-function in such a case can be interpreted as a summation of components of H-function belonging to its thermodynamic force i.e. each thermodynamic force will have its own H-theorem. H-theorem can also be understood as a  ratio of Mahalanobis distance between non-equilibrium and equilibrium distribution and its associated relaxation time. The positivity property of Mahalanobis distance quickly establishes an effortless proof of H-theorem for the Onsager-BGK kinetic model. The non-equilibrium part of the distribution function resulting from the new kinetic model can also be expressed in Onsager's form i.e. as full tensor contraction of thermodynamic forces and its associated microscopic tensors. Kinetic scheme and boundary conditions should also follow the principle of non-equilibrium thermodynamics by addressing the issue of correct generation and distribution of entropy for each thermodynamic force associated with its non-equilibrium state. Velocity slip and temperature jump follow Onsager's variational principle. The simulation results based on the Onsager-BGK model and non-equilibrium thermodynamics based kinetic scheme were validated with analytical as well as the results of Direct Simulation Monte Carlo (DSMC). A new term called reciprocity number $Rp$ was also derived using the contribution of thermodynamic forces on viscous split fluxes in order to estimate the order of importance of cross phenomenon.  Non-equilibrium thermodynamics based kinetic schemes, kinetic wall boundary condition and kinetic particle method can simulate continuum as well as rarefied slip flows within Navier-Stokes-Fourier equations in order to avoid costly multi-scale simulations.

The future course of action will require validation of Onsager-BGK model for multi-component gas mixtures and further investigation on collision probability function based Onsager-BGK model for Knudsen layer. Onsager-BGK model also opens up a possibility of its metamorphosis as a lattice Boltzmann model for compressible, non-isothermal flows.
\section*{Acknowledgments}
We are thankful to Shri G. Gouthaman and Shri T.K. Bera  for help and support. First author is grateful to Prof. S.M. Deshpande for being a mentor and a source of inspiration.
\appendix
\begin{section}
{Extending Onsager-BGK model for gas-mixture}\label{kspmodel}
\end{section}
In case of two fluid approximation we will have kinetic models for each species. The kinetic model will not only depend on relaxation time associated with self-collisions but also on relaxation time associated with cross collisions of specie $i$ with specie $j$. The kinetic model $J_m(f,f_0)({i})$ for specie $i$ and $J_m(f,f_0)({j})$ for specie $j$ can be expressed as
\begin{equation} \label{ons-mix-1} 
\begin{array}{ll}
J_m(f,f_0)({i})&= -\left(\frac{f(i)-f_{0}(i)}{t_{R(\tau,ii)}}\right)_{\boldsymbol{X}_q=0}-\left(\frac{f(i)-f_{0}(i)}{t_{R(q,ii)}}\right)_{\boldsymbol{X}_{\tau}=0}\\
&-\left(\frac{f(i)-\tilde{f}_{0}(i)}{t_{R(\tau,ij)}}\right)_{\boldsymbol{X}_q=0}-\left(\frac{f(i)-\tilde{f}_{0}(i)}{t_{R(q,ij)}}\right)_{\boldsymbol{X}_{\tau}=0}
\end{array}
\end{equation} 
\begin{equation} \label{ons-mix-2} 
\begin{array}{ll}
J_m(f,f_0)({j})&= -\left(\frac{f(j)-f_{0}(j)}{t_{R(\tau,jj)}}\right)_{\boldsymbol{X}_q=0}-\left(\frac{f(j)-f_{0}(j)}{t_{R(q,jj)}}\right)_{\boldsymbol{X}_{\tau}=0}\\
&-\left(\frac{f(j)-\tilde{f}_{0}(j)}{t_{R(\tau,ji)}}\right)_{\boldsymbol{X}_q=0}-\left(\frac{f(j)-\tilde{f}_{0}(j)}{t_{R(q,ji)}}\right)_{\boldsymbol{X}_{\tau}=0}
\end{array}
\end{equation} 
where $f({i})$ and $f_{0}(i)$ are the non-equilibrium and equilibrium distribution for specie $i$. Similarly, $f({j})$ and $f_{0}(j)$ are the non-equilibrium and equilibrium distribution for specie $j$. Maxwellian distribution $\tilde{f}_{0}(i)$ and $\tilde{f}_{0}(j)$ are based on free temperature parameter\cite{mors}. Parameters $t_{R(\tau,ii)}$, $t_{R(q,ii)}$ are the relaxation time due to self collisions of specie $i$ while $t_{R(\tau,jj)}$, $t_{R(q,jj)}$ are the relaxation time due to self collisions of specie $j$.  Parameters $t_{R(\tau,ij)}$, $t_{R(q,ij)}$ are the relaxation time due to cross collisions of specie $i$ with specie $j$ while $t_{R(\tau,ji)}$, $t_{R(q,ji)}$ are the relaxation time due to self collisions of specie $j$ with specie $i$. The cross collision relaxation time $t_{R(\alpha,ij)}$ and $t_{R(\alpha,ji)}$ are related to number density as $n_j t_{R(\alpha,ij)}$ $=$ $n_i t_{R(\alpha,ji)}$
 for $\alpha$ $\in$ $\{\tau,q\}$ where $n_i$ and $n_j$ is the number density of specie $i$ and specie $j$.
\begin{section}
{Expressions of split macroscopic tensors}\label{splitflux}
\end{section}
The components of  split macroscopic tensors $\boldsymbol{{\Lambda}}_{j}^{x,\boldsymbol{\Psi,{\pm\cdot}}}$ $=$ $\left [\boldsymbol{{\Lambda}}_{j}^{x,\psi_i,{\pm\cdot}},\cdots,\boldsymbol{{\Lambda}}_{j}^{x,\psi_i,{\pm\cdot}}\right ]^{T}$ defined for each $\psi_i$ $\in$ $ \boldsymbol{\Psi}$ $=$ $\left[1,\vec{v},\mathbb{I}+\frac{1}{2} v^{2}\right]^{T} $ are
\begin{equation}\label{ons-mt-1} 
\begin{array}{ll}
\Lambda_{\tau}^{x,\psi_1,\pm\cdot}=\left\langle \psi_{1},v_{x}  {  \boldsymbol{\Upsilon}_{\tau}^{\pm\cdot} }\right\rangle=&
{\frac{\mu}{p}\left[
\begin{array}{ll}
 \frac{ { B_{1}^{\mp}} (\gamma -3)  \rho }{8  \sqrt{\pi } \beta ^{3/2}} & 0 \\
 0 & \frac{ { B_{1}^{\mp}} (\gamma-1)   \rho }{8  \sqrt{\pi } \beta ^{3/2}}
\end{array}
\right]}\end{array}
\end{equation} 
\begin{equation}\label{ons-mt-2}
\begin{array}{ll}
\Lambda_{q}^{x,\psi_1,\pm\cdot }=\left\langle \psi_{1},v_{x}  {  \boldsymbol{\Upsilon}_{q}^{\pm\cdot} }\right\rangle=&
{\frac{ (\gamma -1) }{ \gamma } \frac{\kappa }{p  R}\left[
\begin{array}{ll}
 -\frac{ { B_{1}^{\mp}}  {u_{x}}  \rho }{4  \sqrt{\pi } \beta ^{3/2} } &
 0
\end{array}
\right]}\end{array}
\end{equation} 
\begin{equation}\label{ons-mt-3}
\begin{array}{ll}
\Lambda_{\tau}^{x,\psi_2,\pm\cdot}=\left\langle \psi_{2},v_{x}  {  \boldsymbol{\Upsilon}_{\tau}^{\pm\cdot} }\right\rangle=&
{\frac{\mu}{p}\left[
\begin{array}{ll}
 -\frac{ {A_{1}^{\pm}}  (\gamma-3)  \rho }{8  \beta ^2} & 0 \\
 0 & -\frac{ {A_{1}^{\pm}} (\gamma-1) \rho }{8  \beta ^2}
\end{array}
\right]}
\end{array}
\end{equation} 
\begin{equation}\label{ons-mt-4}
\begin{array}{ll}
\Lambda_{q}^{x,\psi_2,\pm\cdot }=\left\langle \psi_{2},v_{x}  {  \boldsymbol{\Upsilon}_{q}^{\pm\cdot} }\right\rangle=&
{\frac{ (\gamma -1) }{ \gamma } \frac{\kappa }{p  R}\left[
\begin{array}{ll}
 \frac{ { B_{1}^{\mp}} \rho }{4 \sqrt{\pi } \beta ^{5/2} } &
 0
\end{array}
\right]}
\end{array}
\end{equation} 
\begin{equation}\label{ons-mt-5}
\begin{array}{ll}
\Lambda_{\tau}^{x,\psi_3,\pm\cdot}=\left\langle \psi_{3},v_{x}  {  \boldsymbol{\Upsilon}_{\tau}^{\pm\cdot} }\right\rangle=&
{\frac{\mu}{p}\left[
\begin{array}{ll}
 \frac{ { B_{1}^{\mp}}  {u_{y}}  (\gamma-3) \rho }{8 \sqrt{\pi } \beta ^{3/2}} & \frac{ {A_{1}^{\pm}}  \rho }{8 \beta ^2} \\
 \frac{ {A_{1}^{\pm}}  \rho }{8 \beta ^2} & \frac{ { B_{1}^{\mp}}  {u_{y}} (\gamma-1)  \rho }{8 \sqrt{\pi } \beta ^{3/2}}
\end{array}
\right]}
\end{array}
\end{equation} 
\begin{equation}\label{ons-mt-6}
\begin{array}{ll}
\Lambda_{q}^{x,\psi_3,\pm\cdot }=\left\langle \psi_{3},v_{x}  {  \boldsymbol{\Upsilon}_{q}^{\pm\cdot} }\right\rangle=&
{\frac{ (\gamma -1) }{ \gamma } \frac{\kappa }{p  R}\left[
\begin{array}{ll}
 -\frac{ { B_{1}^{\mp}}  {u_{x}}  {u_{y}}  \rho }{4  \sqrt{\pi }  \beta ^{3/2}} &
 \frac{ { B_{1}^{\mp}} \rho }{8  \sqrt{\pi }  \beta ^{5/2} }
\end{array}
\right]}
\end{array}
\end{equation} 
\begin{equation}\label{ons-mt-7}
\begin{array}{ll}
\Lambda_{\tau}^{x,\psi_4,\pm\cdot}=\left\langle \psi_{4},v_{x}  {  \boldsymbol{\Upsilon}_{\tau}^{\pm\cdot} }\right\rangle=&
{\frac{\mu}{p}\left[
\begin{array}{ll}
 -\frac{ (\gamma-3) C_{1}^{\pm}  \rho }{32 \sqrt{\pi } \beta ^{5/2} (\gamma-1)} & \frac{ {A_{1}^{\pm}}  {u_{y}}  \rho }{8 \beta ^2} \\
 \frac{ {A_{1}^{\pm}}  {u_{y}}  \rho }{8  \beta ^2} & -\frac{ C_{1}^{\pm} \rho }{32 \sqrt{\pi } \beta ^{5/2}}
\end{array}
\right]}
\end{array}
\end{equation} 
\begin{equation}\label{ons-mt-8}
\begin{array}{ll}
\Lambda_{q}^{x,\psi_4,\pm\cdot }=\left\langle \psi_{4},v_{x}  {  \boldsymbol{\Upsilon}_{q}^{\pm\cdot} }\right\rangle=&
{\frac{ (\gamma -1) }{ \gamma } \frac{\kappa }{p  R}\left[
\begin{array}{ll}
 -\frac{({ B_{1}^{\mp}}  {\sqrt{\beta } u_{x}} (3+2  {u_{y}}^2 \beta  (\gamma-1)-\gamma )+2 {A_{1}^{\pm}} \sqrt{\pi } \gamma ) \rho }{16  \sqrt{\pi }  \beta ^{3} (\gamma-1)} &
 \frac{ { B_{1}^{\mp}}  {u_{y}}  \rho }{8  \sqrt{\pi } \beta ^{5/2}  }
\end{array}
\right]}
\end{array}
\end{equation} 
\noindent Similarly components of $\boldsymbol{{\Lambda}}_{j}^{y,\boldsymbol{\Psi,{\cdot\pm}}}$ $=$ $\left [\boldsymbol{{\Lambda}}_{j}^{y,\Psi_i,{\cdot\pm}},\cdots,\boldsymbol{{\Lambda}}_{j}^{y,\Psi_i,{\cdot\pm}}\right ]^{T}$  are
\begin{equation}\label{ons-mt-9}
\begin{array}{ll}
\Lambda_{\tau}^{y,\psi_1,\cdot \pm}=\left\langle \psi_{1},v_{y}  {  \boldsymbol{\Upsilon}_{\tau}^{\cdot \pm} }\right\rangle=&
{\frac{\mu}{p}\left[
\begin{array}{ll}
 \frac{ { B_{2}^{\mp}} (\gamma-1) \rho }{8 \sqrt{\pi } \beta ^{3/2}} & 0 \\
 0 & \frac{ { B_{2}^{\mp}}  (\gamma-3) \rho }{8  \sqrt{\pi } \beta ^{3/2}}
\end{array}
\right]}
\end{array}
\end{equation} 
\begin{equation}\label{ons-mt-10}
\begin{array}{ll}
\Lambda_{q}^{y,\psi_1,\cdot \pm }=\left\langle \psi_{1},v_{y}  {  \boldsymbol{\Upsilon}_{q}^{\cdot \pm} }\right\rangle=&
{\frac{ (\gamma -1) }{ \gamma } \frac{\kappa }{p  R}\left[
\begin{array}{ll}
 0 &
 -\frac{ { B_{2}^{\mp}}  {u_{y}}   \rho }{4 \sqrt{\pi } \beta ^{3/2} }
\end{array}
\right]}
\end{array}
\end{equation} 
\begin{equation}\label{ons-mt-11}
\begin{array}{ll}
\Lambda_{\tau}^{y,\psi_2,\cdot \pm}=\left\langle \psi_{2},v_{y}  {  \boldsymbol{\Upsilon}_{\tau}^{\cdot \pm} }\right\rangle=&
{\frac{\mu}{p}\left[
\begin{array}{ll}
 \frac{ { B_{2}^{\mp}}  {u_{x}} (\gamma-1)  \rho }{8 \sqrt{\pi } \beta ^{3/2}} & \frac{ {A_{2}^{\pm}}  \rho }{8 \beta ^2} \\
 \frac{ {A_{2}^{\pm}} \rho }{8 \beta ^2} & \frac{ { B_{2}^{\mp}}  {u_{x}}  (\gamma-3) \rho }{8 \sqrt{\pi } \beta ^{3/2}}
\end{array}
\right]}
\end{array}
\end{equation} 
\begin{equation}\label{ons-mt-12}
\begin{array}{ll}
\Lambda_{q}^{y,\psi_2,\cdot \pm }=\left\langle \psi_{2},v_{y}  {  \boldsymbol{\Upsilon}_{q}^{\cdot \pm} }\right\rangle=&
{\frac{ (\gamma -1) }{ \gamma } \frac{\kappa }{p  R}\left[
\begin{array}{ll}
 \frac{ { B_{2}^{\mp}}  \rho }{8 \sqrt{\pi } \beta ^{5/2} } &
 -\frac{ { B_{2}^{\mp}}  {u_{x}}  {u_{y}}  \rho }{4 \sqrt{\pi } \beta ^{3/2}  }
\end{array}
\right]}
\end{array}
\end{equation} 
\begin{equation}\label{ons-mt-13}
\begin{array}{ll}
\Lambda_{\tau}^{y,\psi_3,\cdot \pm}=\left\langle \psi_{3},v_{y}  {  \boldsymbol{\Upsilon}_{\tau}^{\cdot \pm} }\right\rangle=&
{\frac{\mu}{p}\left[
\begin{array}{ll}
- \frac{ {A_{2}^{\pm}} (\gamma-1) \rho }{8 \beta ^2} & 0 \\
 0 & -\frac{ {A_{2}^{\pm}}  (\gamma-3) \rho }{8 \beta ^2}
\end{array}
\right]}
\end{array}
\end{equation} 
\begin{equation}\label{ons-mt-14}
\begin{array}{ll}
\Lambda_{q}^{y,\psi_3,\cdot \pm }=\left\langle \psi_{3},v_{y}  {  \boldsymbol{\Upsilon}_{q}^{\cdot \pm} }\right\rangle=&
{\frac{ (\gamma -1) }{ \gamma } \frac{\kappa }{p  R}\left[
\begin{array}{ll}
 0 &
 \frac{ { B_{2}^{\mp}} \rho }{4 p \sqrt{\pi }\beta ^{5/2} }
\end{array}
\right]}
\end{array}
\end{equation} 
\begin{equation}\label{ons-mt-15}
\begin{array}{ll}
\Lambda_{\tau}^{y,\psi_4,\cdot \pm}=\left\langle \psi_{4},v_{y}  {  \boldsymbol{\Upsilon}_{\tau}^{\cdot \pm} }\right\rangle=&
{\frac{\mu}{p}\left[
\begin{array}{ll}
 -\frac{ C_{2}^{\pm}  \rho}{32  \sqrt{\pi } \beta ^{5/2}} & \frac{ {A_{2}^{\pm}}  {u_{x}}  \rho }{8 \beta ^2} \\
 \frac{ {A_{2}^{\pm}}  {u_{x}}  \rho }{8 \beta ^2} & 
-\frac{  (\gamma-3)C_{2}^{\pm}  \rho}{32 \sqrt{\pi } \beta ^{5/2}}
\end{array}
\right]}
\end{array}
\end{equation} 
\begin{equation}\label{ons-mt-16}
\begin{array}{ll}
\Lambda_{q}^{y,\psi_4,\cdot \pm }=\left\langle \psi_{4},v_{y}  {  \boldsymbol{\Upsilon}_{q}^{\cdot \pm} }\right\rangle=&
{\frac{ (\gamma -1) }{ \gamma } \frac{\kappa }{p  R}\left[
\begin{array}{ll}
 \frac{ { B_{2}^{\mp}}  {u_{x}}   \rho }{8 p \sqrt{\pi }  \beta ^{5/2} } &
 \frac{  (-{ B_{2}^{\mp}} {u_{y}} \sqrt{\beta } (3+2  {u_{x}}^2 \beta  (\gamma-1)-\gamma )+2  {A_{2}^{\pm}} \sqrt{\pi } \gamma ) 
 \rho }{16 \sqrt{\pi }  \beta ^3 (\gamma-1) }
\end{array}
\right]}
\end{array}
\end{equation} 
where $B_{1}^{\pm}= \pm Exp[-\beta u_x^2]$, $B_{2}^{\pm}= \pm Exp[-\beta u_y^2]$, $A_{1}^{\pm}=1 \pm Erf[\sqrt{\beta} u_x]$, $A_{2}^{\pm}=1 \pm Erf[\sqrt{\beta} u_y]$, $C_{1}^{\pm}=4  {A_{1}^{\pm}}  {u_{x}} \sqrt{\beta }\sqrt{\pi } (\gamma-1)+{ B_{1}^{\mp}} (1-2  u_{y}^2\beta  (\gamma-1)-3 \gamma ) $ and $C_{2}^{\pm}=4  {A_{2}^{\pm}}  {u_{y}} \sqrt{\beta }\sqrt{\pi } (\gamma-1)+ { B_{2}^{\mp}} (1-2  u_{x}^2 \beta  (\gamma-1)-3 \gamma )$
\begin{section}
{Non-equilibrium split fluxes and extended thermodynamics}\label{momex}
\end{section}
The moments of the Boltzmann equation satisfy an infinite hierarchy of balance laws, the macroscopic state vector at the ${k+1}^{th}$ hierarchy is based on the flux vector $\boldsymbol{J}^{k}_{i}$ $=$ $\sum_j \boldsymbol{L}_{ij}^{k} \odot \boldsymbol{X}_{j}$ in the $k^{th}$ hierarchy where Onsager's phenomenological tensor at $k^{th}$ hierarchy, $\boldsymbol{L}_{ij}^{k}$ is obtained using $\boldsymbol{L}_{ij}^{k-1}$ as $\boldsymbol{L}_{ij}^{k}$ $=$ $\left \langle\boldsymbol{\Psi} , \vec{v} \boldsymbol{L}_{ij}^{k-1} \right \rangle$. Higher order transport equations can be seen as a fractal entity based on Onsager's phenomenological tensor emerging through iteration of extended thermodynamics. Navier-Stokes equations is a limiting case of extended thermodynamics (ET) when relaxation times of diffusive fluxes are neglected\cite{mullerruggeri}. \\
In actual physical process it is split fluxes which participates in the conservation hence the moments of the Boltzmann equation should satisfy an infinite hierarchy of balance laws in terms of split fluxes. At the $k^{th}$ hierarchy the macroscopic state vector $\boldsymbol{U}_{m_{1}m_{2}\cdots m_{k}}$ and split flux vector $\boldsymbol{G}_{m_{1}m_{2}\cdots m_{k}}^{\pm}$ are based on the $k^{th}$ component of the vector of collision invariant, $\boldsymbol{\Psi}_{m_{1}m_{2}\cdots m_{k}}$ given as
\begin{equation} \label{momex-1} 
\boldsymbol{U}_{m_{1}m_{2}\cdots m_{k}} ={ \int _{\mathbb{R}^{+} }\int _{\mathbb{R}^{D} }\boldsymbol{\Psi}_{m_{1}m_{2}\cdots m_{k}}  f_{} d\vec{v}d\mathbb{I}  }
\end{equation} 
\begin{equation} \label{momex-2} 
\boldsymbol{G}_{m_{1}m_{2}\cdots m_{k}}^{\pm} ={ \int _{\mathbb{R}^{+} }\int _{\mathbb{R}^{{D}^{\pm}} }\boldsymbol{\Psi}_{m_{1}m_{2}\cdots m_{k}} \vec{v} f  d\vec{v}d\mathbb{I}  }
\end{equation} 
Unlike the idea of rational extended thermodynamics\cite{mullerruggeri}  it is the split flux terms in an equation which becomes the density in the next one as follows
\begin{equation}\label{momex-3} 
\begin{array}{c} 
\frac{\partial \boldsymbol{U}_{m_{1}}}{\partial t} + \frac{\partial \boldsymbol{G}_{m_{1}}^{\pm}}{\partial \vec{x}}=0\\
{\swarrow} \\
\frac{\partial \boldsymbol{U}_{m_{1}m_{2}}}{\partial t} + \frac{\partial \boldsymbol{G}_{m_{1}m_{2}}^{\pm}}{\partial \vec{x}}=0\\
{\swarrow} \\
{\vdots}
\end{array}
\end{equation}
The state vector $\boldsymbol{U}_{m_{1}m_{2}\cdots m_{k}m_{k+1}}$ in the $(k+1)^{th}$ hierarchy is based on the split flux $\boldsymbol{G}_{m_{1}m_{2}\cdots m_{k}}^{\pm}$ based on half range distribution function defined in $k^{th}$ hierarchy. For example, consider one dimensional case where split mass flux composed of inviscid and viscous contribution, $GX^{\pm}(\psi_{1})$ evaluated using $\psi_{1}$ $\in$ $\boldsymbol{\Psi}$ is given as
\begin{equation} \label{momex-4}  
\begin{array}{l}
GX^{\pm}(\psi_{1})=\frac{1}{2}\rho \left(A^{\pm} u_{x} + \frac{B^{\pm}}{\sqrt{\pi \beta}}\right)-
{\frac{\rho}{4 p\sqrt{\beta}}\frac{B^{\pm}}{\sqrt{\pi}}\left( 2 u_{x} \beta q_{x}\frac{(\gamma-1)}{\gamma}+\tau_{xx}\right) }
\end{array}
\end{equation}
where $A^{\pm}$ $=$ $1\pm Erf(u_{x}\sqrt{\beta})$ and $B^{\pm}$ $=$ $\pm Exp(-\beta u_{x}^2)$. The split mass flux contain terms of momentum, shear stress tensor and heat flux vector. These terms involve i)$\vec{v}$, ii) $\vec{v} \otimes \vec{v}$, and iii) $\left(\mathbb{I}+\frac{1}{2}\vec{v}^{T} \vec{v}\right)\vec{v}$  components of collision invariants. Thus, 5 moments based on $\boldsymbol{\Psi}$$=$$\left[1,\vec{v},\mathbb{I}+\frac{1}{2} \vec{v}^{T} \vec{v}\right]^{T}$  are inadequate as vector of collision invariant should include $\boldsymbol{\Psi}$$=$$\left[1,\vec{v},\mathbb{I}+\frac{1}{2} \vec{v}^{T} \vec{v},\vec{v} \otimes \vec{v},\left(\mathbb{I}+\frac{1}{2} \vec{v}^{T} \vec{v}\right)\vec{v}\right]^{T} $ giving rise to at-least 13 moment equations such that split mass flux becomes density in the second step and split momentum flux becomes density in the third step. This set of 13-moment Grad like system\cite{grad}  includes evolution of pressure tensor and heat flux vector. However, the present 5 moments based formulation is adequate for the simulation of most of the engineering slip flow problems which lie in the regime of  linear irreversible thermodynamics. The present approach will not be adequate for cases that involve large Mach number in shock waves, high frequencies for sound waves, etc. It should also be noted that the approach based on non-equilibrium thermodynamics may also modify Levermore \cite{levermorem} procedure which generates hierarchy of closed systems of moment equations that ensures every member of the hierarchy is symmetric hyperbolic with an entropy, and formally recovers to Euler limit. The finite dimensional linear subspace $\Xi $ of functions of $\vec{v}$ in Levermore procedure should ensure that entropy generation follows Onsager's reciprocity principle.

In real media it is the split flux which participates in any physical process and for non-equilibrium flows the split fluxes contain dissipative terms. For example split flux associated with mass flow  contains contribution of viscous terms. It is also interesting to note that the presence of dissipative terms due to thermodynamic forces in the split fluxes brings out its relationship and difference with the hydrodynamic theory of Brenner \cite{brenner} and Quasi-gas dynamics \cite{elizarova} where dissipative terms were introduced in un-split flux terms such that time-spatial averages are invariant under Galileo transform. 
\begin{section}
{Onsager-BGK model for the Knudsen layer}\label{kn2model}
\end{section}
Near the wall at a normal distance of order $\mathcal{O}(\lambda)$ $=$ $\lambda_{e}$ there exists a Knudsen layer parametrized by $\lambda_{e}$ which is the effective mean free path depending on the effective viscosity and wall conditions. In the Knudsen layer some molecules may collide more with the wall and may not suffer as much collisions with the molecules as compared to the molecules above the Knudsen layer. 
 For modeling slip near transition regime ideally we require an approach which is computationally cheap includes higher moments thereby terms of order ${\rm Kn}^{2}$.  It should be noted that validity of Chapman-Enskog expansion procedure can only be said for ${\rm Kn}$ $\leq$ $1$, the more correct way to obtain non-linear distribution has to be based on extended irreversible thermodynamics (EIT) by expanding the distribution function in terms of microscopic tensors and thermodynamic forces. From non-equilibrium thermodynamics point of view  linear irreversible thermodynamics (LIT) is no longer valid in the Knudsen layer as  fluxes are no longer linear functions of its conjugate force, regime shifts to extended irreversible thermodynamics (EIT) described by $\boldsymbol{J}_{i}=\sum_{j} \boldsymbol{L}_{ij} \odot \boldsymbol{X}_{j}+\frac{1}{2}\sum_{jk} \boldsymbol{L}_{ijk} \odot \boldsymbol{X}_{j}\odot \boldsymbol{X}_{k}+\cdots$. Simplest approach may be to approximate EIT based flux  as a function of  LIT based flux by using suitable scaling function. Consider a function  $\mathbb{P}(\bar{y})$\footnote{For two dimensional symmetry problems collision probability function, $\mathbb{P}(\bar{y})$ will also depend on the curvature of the surface as it is a volume dependent parameter related to Onsager's dissipation function.} as a measure of probability of collision at any normal distance $\bar{y}$$=$$\frac{y}{\lambda_{e}}$ $\leq$$1$ such that non-equilibrium EIT based flux, $\boldsymbol{J}_{i}$ can be approximated in terms of $\mathbb{P}(\bar{y})$ for curvature free surface as $\boldsymbol{J}_{i}\approx\mathbb{P}(\bar{y})\sum_{j} \boldsymbol{L}_{ij} \odot \boldsymbol{X}_{j}$. In such a case the single particle velocity distribution $f_{1}(\bar{y})$ in the Knudsen layer at any normal dimensionless distance $\bar{y}$$=$$\frac{y}{\lambda_{e}}$ $\leq$$1$  can be expressed as
\begin{equation} \label{knons-3} 
f_{1}(\bar{y})= f_{0} - \mathbb{P}(\bar{y})\sum_{j}{  \boldsymbol{\Upsilon}_{j} \odot \boldsymbol{X}_{j}} 
\end{equation} 
where $f_{0}$ is the Maxwellian distribution. The total distribution function, $f_{1,\Sigma }$ at the wall based on Maxwell gas-interaction model in terms of accommodation coefficient $\sigma $ can be written as 
\begin{equation}  \label{knons-4} 
\begin{array}{ll}
f_{1,\Sigma }(\vec{v},\mathbb{I},t,\bar{y}=0) = &
\begin{array}{lll}
f_{1,I} (\vec{v},\mathbb{I},t,\bar{y}=0) &\text{for}&   \vec{i}_n \cdot \vec{v}<0 \\
(1-\sigma )f_{1,R}(\vec{v},\mathbb{I},t,\bar{y}=0) +\sigma f_{0,W}(\vec{v},\mathbb{I},t,\bar{y}=0) &\text{for}& \vec{i}_n \cdot \vec{v}>0 
\end{array}
\end{array}
\end{equation}
where $f_{1,\Sigma }(\vec{v},\mathbb{I},t,\bar{y}=0))$ is the total, $f_{1,I}(\vec{v},\mathbb{I},t,\bar{y}=0)$ is the incident Knudsen layer Chapman-Enskog distribution, $f_{1,R}(\vec{v},\mathbb{I},t,\bar{y}=0)$ $=$ $f_{1,I}(\vec{v}-2 \vec{i}_n\vec{i}_n \cdot \vec{v},\mathbb{I},t,\bar{y}=0)$ is the specular reflected Knudsen layer Chapman-Enskog distribution and $f_{0,W}(\vec{v},\mathbb{I},t,\bar{y}=0)$ is the diffuse reflected Maxwellian distribution based on wall conditions and conservation. At the wall $\mathbb{P}(\bar{y})$ $=0$ so the non-equilibrium part of the incident Chapman-Enskog distribution vanishes and the temperature as well as velocity gradients  are singular  similar to the findings of Lilley and Sader \cite{lilley}. This can also be interpreted as a Onsager-BGK Knudsen layer model, $J(f,f_0)_{\text{Kn}}$ valid in the Knudsen layer with varying relaxation time expressed as
\begin{equation}\label{knons-5} 
J(f,f_0)_{\text{Kn}}=-\left(\frac{f -f_{0}^{}}{ \tilde{t}_{R(\tau)}(\bar{y}) }  \right)_{\boldsymbol{X}_{q}=0}-\left(\frac{f-f_{0}^{}}{ \tilde{t}_{R(q)}(\bar{y})}  \right)_{\boldsymbol{X}_{\tau}=0}
\end{equation} 
The relaxation time $\tilde{t}_{R(j)}= \mathbb{P}(\bar{y}) t_{R(j)}$ varies with the normal distance from the wall based on the collision probability function, $\mathbb{P}(\bar{y})$. The state update equations in the macroscopic form in the Knudsen layer becomes
\begin{equation} \label{knons-6} 
\displaystyle \boldsymbol{U}(t+\Delta t) = \displaystyle \boldsymbol{U}(t)-\Delta t \displaystyle \left[
\displaystyle \frac{\partial  \boldsymbol{GX}^{\pm}_{I}(t)}{\partial x}+\frac{\partial \boldsymbol{\bar{GX}}^{\pm}_{V}(t)}{\partial x}
\displaystyle+\frac{\partial  \boldsymbol{GY}^{\pm}_{I}(t)}{\partial y}+\frac{\partial \boldsymbol{\bar{GY}}^{\pm}_{V}(t)}{\partial y}
\right]
\end{equation}
The viscous part of the flux component, $\boldsymbol{\bar{GX}}^{\pm}_{V}$ and $\boldsymbol{\bar{GY}}^{\pm}_{V}$  are obtained as
\begin{equation} \label{knons-7} 
\begin{array}{ll}
\boldsymbol{\bar{GX}}_{V}^{\pm} & =\mathbb{P}(\bar{y})\boldsymbol{GX}_{V}^{\pm}=-\mathbb{P}(\bar{y})\sum_{j}\boldsymbol{\Lambda}_{j}^{x,\boldsymbol{\Psi},\pm \cdot} \boldsymbol{\odot}\boldsymbol{X}_{j}\\
\boldsymbol{\bar{GY}}_{V}^{\pm} & =\mathbb{P}(\bar{y})\boldsymbol{GY}_{V}^{\pm}=-\mathbb{P}(\bar{y})\sum_{j}\boldsymbol{\Lambda}_{j}^{y,\boldsymbol{\Psi},\cdot\pm} \boldsymbol{\odot}\boldsymbol{X}_{j}
\end{array}
\end{equation} 
This kinetic model is quite easy to implement as the viscous fluxes are just multiplied by the collision probability function, $\mathbb{P}(\bar{y})$.

\end{document}